\documentclass[preprint, 3p]{elsarticle}
\makeatletter
\def\ps@pprintTitle{%
	\let\@oddhead\@empty
	\let\@evenhead\@empty
	\def\@oddfoot{\centerline{\thepage}}%
	\let\@evenfoot\@oddfoot}
\makeatother

\usepackage{hyperref}
\usepackage{amsmath}
\usepackage{bm}

\DeclareMathOperator*{\argmin}{arg\,min}
\usepackage[official]{eurosym}
\usepackage[skip=0pt]{caption}
\usepackage[inline]{enumitem}
\usepackage{subcaption}
\usepackage{multicol}
\begin{document}
	
	\begin{frontmatter}
		
		\title{Value of information from vibration-based structural health monitoring extracted via Bayesian model updating}
		
		\author[mymainaddress]{Antonios Kamariotis\corref{mycorrespondingauthor}}
		\cortext[mycorrespondingauthor]{Corresponding author}
		\ead{antonis.kamariotis@tum.de}
		\author[mysecondaryaddress]{Eleni Chatzi}
		\ead{chatzi@ibk.baug.ethz.ch}
		\author[mymainaddress]{Daniel Straub}
		\ead{straub@tum.de}
		
		\address[mymainaddress]{Engineering Risk Analysis Group, Technical University of Munich, Theresienstrasse 90, 80333 Munich, Germany}
		\address[mysecondaryaddress]{Institute of Structural Engineering, ETH Zurich, Stefano-Franscini-Platz 5, 8093 Zurich, Switzerland}
		
		\begin{abstract}
			Quantifying the value of the information extracted from a structural health monitoring (SHM) system is an important step towards convincing decision makers to implement these systems.
			We quantify this value by adaptation of the Bayesian decision analysis framework. In contrast to previous works, we model in detail the entire process of data generation to processing, model updating and reliability calculation, and investigate it on a deteriorating bridge system. The framework assumes that dynamic response data are obtained in a sequential fashion from deployed accelerometers, subsequently processed by an output-only operational modal analysis scheme for identifying the system's modal characteristics. We employ a classical Bayesian model updating methodology to sequentially learn the deterioration and estimate the structural damage evolution over time. This leads to sequential updating of the structural reliability, which constitutes the basis for a preposterior Bayesian decision analysis. Alternative actions are defined and a heuristic-based approach is employed for the life-cycle optimization. By solving the preposterior Bayesian decision analysis, one is able to quantify the benefit of the availability of long-term SHM vibrational data. Numerical investigations show that this framework can provide quantitative measures on the optimality of an SHM system in a specific decision context. 
		\end{abstract}
		
		
	\end{frontmatter}
	
	%\linenumbers
	
	\section{Introduction}
	
	The advancements in the development of reliable and low-cost sensors, capable of measuring different structural response quantities (e.g. accelerations, displacements, strains, temperatures, loads, etc.) have led to vast scientific and practical developments in the field of Structural Health Monitoring (SHM) over the last four decades \cite{Worden_introduction}. Techniques for processing the raw measurement data and obtaining indicators of structural ``health" have been made readily available \cite{Limongelli}. However, despite the advancements in the field, SHM still remains predominantly applied within the research community \cite{Worden_book} and has not yet translated to extensive application on real-world structures and infrastructure systems. One main reason for this is that the effect and the potential benefit from the use of SHM systems can only be appraised on the basis of the decisions that are triggered by monitoring data. Key open-ended questions include \cite{VOI_ID}: How can information obtained from an SHM system provide optimal decision support? What is the Value of Information (VoI) from SHM systems? How can it be maximized?
	
	Preposterior Bayesian decision analysis can be employed as a formal framework for quantifying the VoI \cite{Raiffa}, which adequately incorporates the uncertainties related to the structural performance and the associated costs, the monitoring measurements, etc. A VoI analysis provides the necessary mathematical framework for quantifying the benefit of an SHM system prior to its installation. In the civil and infrastructure engineering context, the computation of the VoI has been considered mainly related to optimal inspection planning for deteriorating structural systems \cite{Straub_Faber, Jesus, Vereecken}. Recent works \cite{Pozzi, Zonta, Straub_VOI, Thons, Konakli, Andriotis, Zhang, Iannacone} use the VoI concept in an attempt to quantify the value of SHM on idealized structural systems within a Bayesian framework. All works to date, however, adopt rather simplified assumptions regarding the type of information offered by the SHM system. They thus largely rely on hypothetical likelihood functions or observation models, which render these demonstrations, although insightful, not easily transferable to realistic applications. A first attempt towards modeling the entire SHM process and the monitoring information has been made by the authors in \cite{Kamariotis}, which is formalized and extended herein.
	
	Installation of a continuous monitoring system on a structure allows for continuous measurement of the dynamic response of the structure (e.g. accelerations, strain, etc.). In an in-operation regime, a precise measurement of the acting loads, which are usually distributed along a system (e.g. wind, traffic), is a challenging task. Output-only operational modal analysis (OMA) \cite{SSI, Au_OMA} techniques have been developed to alleviate the burden of the absence of acting load measurements. Using an OMA procedure one can identify the system eigenfrequencies and mode shapes of typical structures excited by unmeasured ambient (broadband) loads. This is beneficial, since the operation of the structure is not obstructed, as it would be in the case of forced vibration testing. 
	
	Further to data acquisition and system identification, model updating forms a popular subsequent step toward modeling the system performance on the basis of the monitoring information. This process is also referred to as the process of establishing a digital twin via model updating \cite{Wright}. Bayesian model updating (BMU) using identified modal data has proved successful in identifying damage on a global or local level within a structure \cite{Vanik, Papadimitriou, Beck_MCMC, Simoen, Yuen, Behmanesh}. These methods hold significant promise for application with actual full-scale structures \cite{Ntotsios, Moaveni, Argyris}. The vast majority of studies are focused on investigating how the BMU framework performs in detecting, localizing and quantifying different types of artificially created damage given some fixed set of modal data. A few recent studies are concerned with BMU using vibrational data obtained in a continuous fashion from SHM systems \cite{Behmanesh,Simoen_progressive, Ierimenti}. However, no studies are available that systematically quantify the benefit of BMU using continuous SHM data towards driving optimal informed maintenance decision making.
	
	This work embeds a sequential implementation of the BMU framework within a preposterior Bayesian decision analysis, to quantify the VoI from long-term vibrational data obtained from an SHM system. We employ a numerical benchmark for continuous monitoring under operational variability \cite{bench19} to test and demonstrate the approach. The numerical benchmark serves as a tool to create continuous reference monitoring data from a two-span bridge system subject to different types (scour, corrosion) of deterioration at specific hotspots over its lifespan. The benchmark is used as a simulator for extracting dynamic response data, i.e. simulated measurements (accelerations), corresponding to a typical deployment of accelerometers on the structure. Acceleration measurements are provided as input to an output-only OMA algorithm, which identifies the system's modal characteristics. We implement Bayesian model and structural reliability updating methods in a sequential setting for incorporating the continuous OMA-identified modal data within a decision making framework. This proposed procedure follows the roadmap to quantifying the benefit of SHM presented in \cite{VOI_ID}. We employ a simple heuristic-based approach for the solution of the life-cycle optimization problem in the preposterior Bayesian decision analysis. The resulting optimal expected total life-cycle costs are computed in the preposterior case, and compared against the optimal expected total life-cycle costs obtained in the case of only prior knowledge, thus enabling the quantification of the VoI of SHM.
	
	\section{VoI from SHM analysis}
	\label{sec:VoI_workflow}
\begin{figure}[ht]
		\centerline{
			\includegraphics[width=0.9\textwidth]{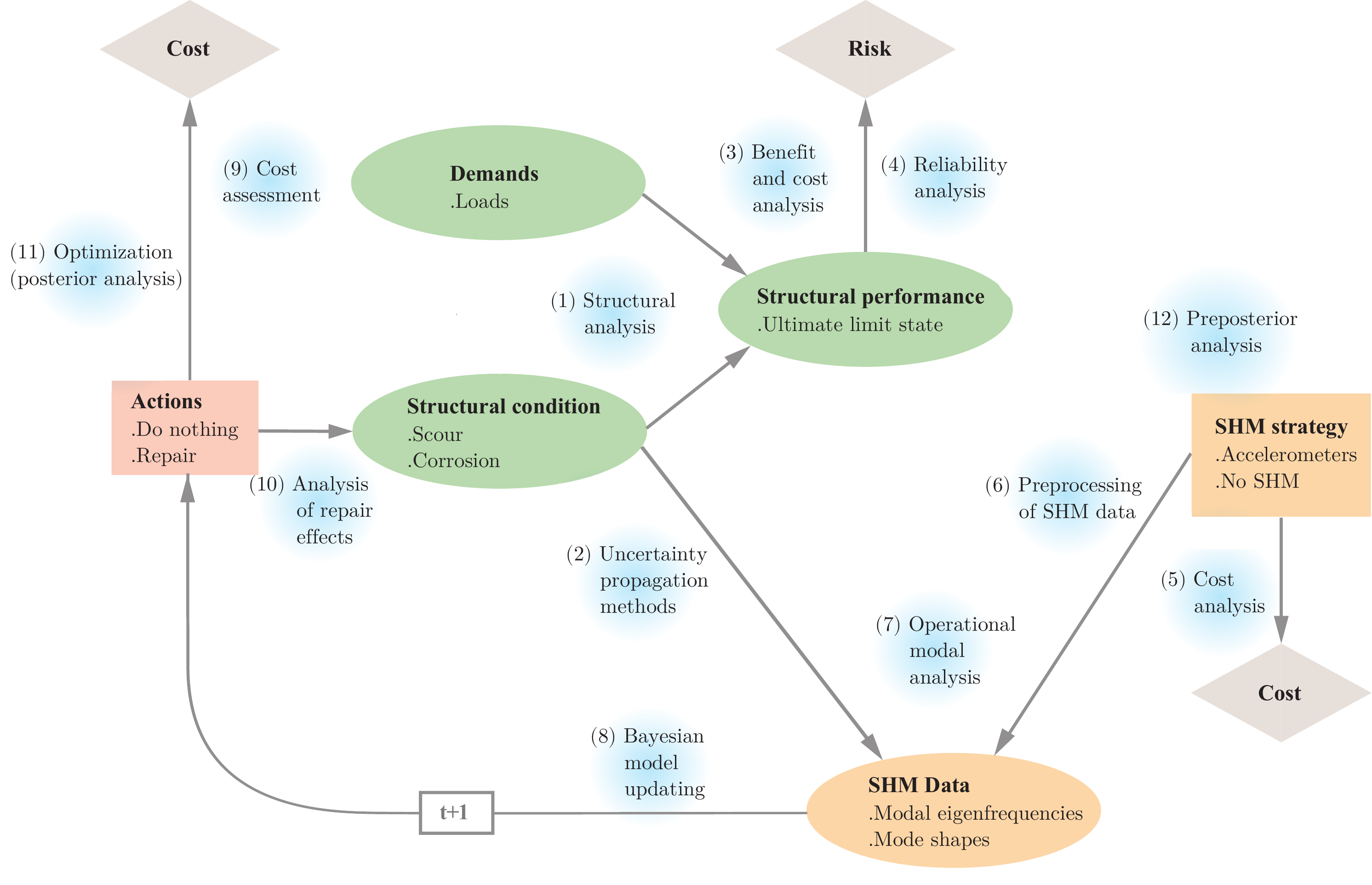}
		}
		\caption{Influence diagram of the SHM process for a preposterior Bayesian decision analysis to quantify the VoI.}
		\label{fig: framework}
	\end{figure}

	The monitoring of a structural system through deployment of an appropriately designed SHM system is a viable means to support decision-making related to infrastructure maintenance actions. But is gathering this information worth it? Preposterior Bayesian decision analysis provides the necessary formal mathematical framework for quantifying the VoI of an SHM system. A concise representation of a such an analysis with the use of an influence diagram (ID) has been introduced in \cite{VOI_ID}. An adaptation of this ID for the purposes of the VoI analysis that we propose and apply on a simulated SHM benchmark study in this paper is offered in Figure \ref{fig: framework}.
	
	Influence diagrams build upon Bayesian networks (BN), which offer a concise graphical tool to model Bayesian probabilistic inference problems, and extend these through the addition of decision and utility nodes to model decision-making under uncertainty \cite{Jensen}. 
	In the ID of Figure \ref{fig: framework}, green oval nodes model uncertain parameters and models/processes related to the structural system, the orange square node models the decision on the SHM system, while the orange oval node models the monitoring data that is extracted via use of a specific SHM system. This data can be used to learn the structural condition via Bayesian updating to then inform the decision on maintenance/repair actions (red square node). Finally, the grey diamond-shaped nodes represent the different costs that enter into the process. The box [t+1] shows that this ID represents a decision process over the lifetime of the structure. The blue text bubbles introduce the different computational methods that are incorporated in the different parts of the process. The large number of these bubbles highlights the modeling and computational challenges associated with a full VoI analysis.
	
	In this paper, for the first time in existing literature, we avoid overly simplifying assumptions in some parts of the modeling of the preposterior Bayesian decision analysis for quantifying the VoI from SHM, but we still model some parts of the process in a simplistic way. The main contribution lies in the modeling of the SHM data. As can be seen in Figure \ref{fig: framework}, we employ continuous SHM information over the lifetime of a deteriorating structural system in the form of acceleration time series, which are subsequently processed by an OMA procedure that identifies the modal eigenfrequencies and mode shapes. These SHM modal data are then used within a BMU procedure to sequentially identify the structural condition (see Section \ref{sec:Bayes}). The way in which the SHM data sets are sampled within a preposterior Bayesian decision analysis with the use of the benchmark structural model is described in detail in Section \ref{subsec: Synthetic}. We treat the modeling of the structural performance node of the ID, as well as the incorporation of the monitoring information within a reliability updating, in a realistic and computationally efficient approach (see Section \ref{sec: SR}). To provide a computationally viable solution to the VoI analysis, we adopt a rather simplified modeling of the action decision node, and we perform the life cycle optimization with the use of heuristics (see Section \ref{sec:LCC}).
	
	The solution of the preposterior Bayesian decision analysis leads to monitoring-informed optimization of the repair action, which in turn leads to the computation of the optimal expected total life-cycle cost in the case of having an SHM system installed. If the adopted SHM strategy is to implement no SHM system, then life-cycle optimization is conducted on the basis of prior information only. By comparing the optimal expected total life-cycle costs in the prior and preposterior cases, the VoI is implicitly quantified as the difference between the two.
	
	\section{Bayesian model updating}
	\label{sec:Bayes}
	In this section, the Bayesian model updating framework with the use of OMA-identified modal data is presented. The Bayesian formulation presented here corresponds to the state-of-the-art formulation \cite{Simoen, Yuen, Moaveni}.
	
	\subsection{Bayesian formulation}
	We consider deterioration that leads to local stiffness reductions. The random variables (RVs) describing the uncertainty within the employed deterioration models are $\boldsymbol{\theta} \in {\rm I\!R}^d$, with $d$ being the total number of RVs. The goal of the Bayesian inverse problem is to infer the deterioration model parameters $\boldsymbol{\theta}$ given noisy OMA-identified modal data. These are the modal eigenvalues $\widetilde{\lambda}_m = (2\pi \widetilde{f}_m)^2$, which can be identified quite accurately, and/or mode shape vector components $\boldsymbol{\widetilde{\Phi}}_{m} \in {\rm I\!R}^{N_{s}}$ at the $N_s$ degrees of freedom (DOF) that correspond to the sensor locations, where $m=1,...,N_m$ is the number of identified modes. An accurate identification of the mode shape displacements requires the deployment of a relatively large number of sensors. Conditional on a fairly good representation of the mode shape vector, one can then derive other modal characteristics, such as the mode shape curvatures $\boldsymbol{\widetilde{K}}_{m} \in {\rm I\!R}^{N_{s}}$, which are shown to be more sensitive to local damage \cite{Pandey}. If only the eigenvalue data is at hand, damage can be detected on a global level, while damage localization requires the existence of spatial information, in the form of mode shape (or mode shape curvature) data.

	Consider a linear finite element (FE) model, which is parameterized through the parameters $\boldsymbol{\theta}$ of the deterioration models. The goal of the Bayesian probabilistic framework is to estimate the parameters $\boldsymbol{\theta}$, and their uncertainty, such that the FE model predicted modal eigenvalues $\lambda_{m}(\boldsymbol{\theta})$  and mode shapes $\boldsymbol{\Phi}_{m}(\boldsymbol{\theta})$, or mode shape curvatures $\boldsymbol{K}_{m}(\boldsymbol{\theta})$, best match the corresponding SHM modal data.
	
	Using Bayes' theorem, the posterior probability density function $\pi_{\text{pos}}$ of the deterioration model parameters $\boldsymbol{\theta}$ given an identified modal data set $[\boldsymbol{\widetilde{\lambda}},\boldsymbol{\widetilde{\Phi}}]$ is computed via equation (\ref{Bayes}); it is proportional to the likelihood function $L(\boldsymbol{\theta}; \boldsymbol{\widetilde{\lambda}},\boldsymbol{\widetilde{\Phi}})$ multiplied with the prior PDF of the model parameters $\pi_{pr}(\boldsymbol{\theta})$. The proportionality constant is the so-called model evidence $Z$ and requires the solution of a $d$-dimensional integral, shown in equation (\ref{Evidence}).
	\begin{equation}
		\pi_{\text{pos}}(\boldsymbol{\theta} \mid \boldsymbol{\widetilde{\lambda}},\boldsymbol{\widetilde{\Phi}}) \propto  L(\boldsymbol{\theta}; \boldsymbol{\widetilde{\lambda}},\boldsymbol{\widetilde{\Phi}}) \pi_{\text{pr}}(\boldsymbol{\theta})
		\label{Bayes}
		\end{equation}
	\begin{equation}
		Z=  \int_{\Omega_{\boldsymbol{\theta}}}L(\boldsymbol{\theta}; \boldsymbol{\widetilde{\lambda}},\boldsymbol{\widetilde{\Phi}}) \pi_{\text{pr}}(\boldsymbol{\theta}) d\boldsymbol{\theta}
		\label{Evidence}
		\end{equation}
	
	The model updating procedure contains significant uncertainties, which should be taken into account within the Bayesian framework. According to \cite{Simoen}, these are classified into i) measurement uncertainty, including random measurement noise and variance or bias errors induced in the SSI procedure, and ii) model uncertainty. In \cite{Behmanesh} the existence of inherent variability emerging from changing environmental conditions is highlighted. The combination of all the above uncertainties is called the total prediction error in literature \cite{Simoen, Behmanesh}. In order to construct the likelihood function, the eigenvalue and mode shape (similarly for mode shape curvature) prediction errors for a specific mode $m$ are defined as in equations (\ref{eigenvalue_error}) and (\ref{modeshape_error}).
	\begin{equation}
		\eta_{\lambda_{m}} = \widetilde{\lambda}_{m} - \lambda_{m}(\boldsymbol{\theta}) \in {\rm I\!R}
		\label{eigenvalue_error}
		\end{equation}
	\begin{equation}
		\boldsymbol{\eta}_{\boldsymbol{\Phi}_m} = \gamma _{m} \boldsymbol{\widetilde{\Phi}}_{m} - \boldsymbol{\Phi}_{m}(\boldsymbol{\theta}) \in {\rm I\!R}^{N_{s}}
		\label{modeshape_error}
		\end{equation}
	where $\gamma_{m}$ is a normalization constant, which is computed as in equation (\ref{gamma}). $\boldsymbol{\Gamma}$ is a binary matrix for selecting the FE degrees of freedom that correspond to the sensor locations.
	\begin{equation}
		\gamma_{m} = \frac{{\boldsymbol{\widetilde{\Phi}}_{m}}^T \boldsymbol{\Gamma}\boldsymbol{\Phi}_{m}}{\left \| \boldsymbol{\widetilde{\Phi}}_{m} \right \|^2}
		\label{gamma}
		\end{equation}
	The probabilistic model of the eigenvalue prediction error is a zero-mean Gaussian random variable with standard deviation assumed to be proportional to the measured eigenvalues:
	\begin{equation}
		\eta_{\lambda_{m}} \sim \mathcal{N}\left(0, c_{\lambda m}^2 \widetilde{\lambda}_{m}^2 \right)
		\label{eigenvalue_error_std}
		\end{equation}
	All the $N_s$ mode shape prediction error components in the vector $\boldsymbol{\eta}_{\boldsymbol{\Phi}_m}$ are assigned a zero-mean Gaussian random variable with the same standard deviation, assumed proportional to the $L_2$-norm of the measured mode shape vector. A multivariate Gaussian distribution is used to model this error:
	\begin{equation}
		\begin{split}
		&\boldsymbol{\eta}_{\boldsymbol{\Phi}_m}  \sim \mathcal{N}(\boldsymbol{0}, \boldsymbol{\Sigma}_{\boldsymbol{\Phi}_{m}}) \\
		\boldsymbol{\Sigma}_{\boldsymbol{\Phi}_{m}} &= \text{diag}\left(c_{\Phi m}^2 \left \| \gamma_{m} \boldsymbol{\widetilde{\Phi}}_{m} \right \|^2\right)
		\end{split}
		\end{equation}
	The factors $c_{\lambda m}$ and $c_{\Phi m}$ can be regarded as assigned coefficients of variation, and their chosen values reflect the total prediction error. In practical applications, usually very little (if anything) is known about the structure or the magnitude of the total prediction error. At the same time, even if the assumption of an uncorrelated zero mean Gaussian model for the errors has computational advantages and can be justified by the maximum entropy principle, the choice of the magnitude of the factors $c_{\lambda m}$ and $c_{\Phi m}$ clearly affects the results of the Bayesian updating procedure. It appears that most published works do not properly justify this particular choice of the magnitude of the error. 
	
	Assuming statistical independence among the $N_m$ identified modes, the likelihood function for a given modal data set can be written as in equation (\ref{likelihood}).
	\begin{equation}
		L\left(\boldsymbol{\theta}; \boldsymbol{\widetilde{\lambda}}, \boldsymbol{\widetilde{\Phi}}\right)= \prod_{m=1}^{N_m}N\left(\eta_{\lambda_{m}}; 0, c_{\lambda m}^2 \widetilde{\lambda}_{m}^2 \right)
		N(\boldsymbol{\eta}_{\boldsymbol{\Phi}_m} ; \boldsymbol{0},  \boldsymbol{\Sigma}_{\boldsymbol{\Phi}_{m}})
		\label{likelihood}
		\end{equation}
	The benefit of SHM is that the sensors can provide data in a continuous fashion, therefore resulting in an abundance of measurements received almost continually. Assuming independence among $N_t$ modal data sets obtained at different time instances, the likelihood can be expressed as: 
	\begin{equation}
		L\left(\boldsymbol{\theta}; \boldsymbol{\widetilde{\lambda}}_{1}...\boldsymbol{\widetilde{\lambda}}_{N_t},\boldsymbol{\widetilde{\Phi}}_{1}...\boldsymbol{\widetilde{\Phi}}_{N_t} \right) = \prod_{t=1}^{N_t}\prod_{m=1}^{N_m}N\left(\widetilde{\lambda}_{t_m}- \lambda_{t_m}(\boldsymbol{\theta}); 0, c_{\lambda m}^2 \widetilde{\lambda}_{t_m}^2\right)
		N\left(\gamma _{t_m} \boldsymbol{\widetilde{\Phi}}_{t_m} - \boldsymbol{\Phi}_{t_m}(\boldsymbol{\theta}); \boldsymbol{0}, \boldsymbol{\Sigma}_{\boldsymbol{\Phi}_{t_m}} \right)
		\label{time_likelihood}
		\end{equation}
	where the index $t_m$ indicates the modal data of mode $m$ identified at time instance $t$. The formulation in equation \eqref{time_likelihood} allows for sequential implementation of the Bayesian updating process.
	At any time step $t_i$ when new data becomes available, the distribution of the parameters given all the data up to time $t_i$, $\pi_{\text{pos}}(\boldsymbol{\theta} \mid \boldsymbol{\widetilde{\lambda}}_{1:i},\boldsymbol{\widetilde{\Phi}}_{1:i})$ or the one step ahead predictive distributions for time $t_{i+1}$ can be obtained. The inclusion of data in a continuous fashion can increase the level of accuracy of the Bayesian model updating procedure. However, one should be aware that the assumption of independence in equation \eqref{time_likelihood} typically does not hold. This could be addressed by a hierarchical modeling of $\boldsymbol{\theta}$ \cite{ Behmanesh}.
	
	\subsection{Solution methods}
	\label{subsec: Bayesian solution}
	The solution of the Bayesian updating problem in the general case involves the solution of the $d$-dimensional integral for the computation of the model evidence. Analytic solutions to this integral are available only in special cases, otherwise numerical integration or sampling methods are deployed. The two solution methods that we employ within this work are the Laplace asymptotic approximation and an adaptive Markov Chain Monte Carlo (MCMC) algorithm.
	\subsubsection{Laplace approximation}
	\label{subsubsec: Laplace}
	A detailed presentation of this method can be found in \cite{Katafygiotis, Papadimitriou}. The main idea is that for globally identifiable cases \cite{Katafygiotis}, and for large enough number of experimental data, the posterior distribution can be approximated by a multivariate Gaussian distribution $N(\boldsymbol{\mu}, \boldsymbol{\Sigma})$. The mean vector $\boldsymbol{\mu}$ is set equal to the most probable value, or maximum aposteriori (MAP) estimate, of the parameter vector, which is obtained by minimizing the negative logposterior:
	\begin{equation}
		\boldsymbol{\mu} = \boldsymbol{\theta}_{MAP} = \underset{\boldsymbol{\theta}}{\argmin}(-\operatorname{ln}\pi_{\text{pos}}(\boldsymbol{\theta} \mid\boldsymbol{\widetilde{\lambda}},\boldsymbol{\widetilde{\Phi}})) = \underset{\boldsymbol{\theta}}{\argmin}(-\operatorname{ln}L(\boldsymbol{\theta}; \boldsymbol{\widetilde{\lambda}},\boldsymbol{\widetilde{\Phi}}) -\operatorname{ln}\pi_{\text{pr}}(\boldsymbol{\theta}))
		\label{MAP}
		\end{equation}
	and the covariance matrix $\boldsymbol{\Sigma}$ is equal to the inverse of the Hessian of the log-posterior evaluated at the MAP estimate. When new data becomes available, the new posterior distribution has to be approximated. The MAP estimate of the previous time step is used as the initial point for the optimization at the current time step, to facilitate a faster convergence of the optimization algorithm.
	
	\subsubsection{MCMC sampling}
	For more accurate estimates of the posterior distributions than the one obtained by using the Laplace approximation, one can resort to MCMC sampling methods. Among the multiple available MCMC algorithms, here we employ the adaptive MCMC algorithm from \cite{Haario}, in which the adaptation is performed on the covariance matrix of the proposal PDF. Whenever new data becomes available, the MCMC algorithm has to be rerun to obtain the new posterior distribution. The posterior mean of the parameters estimated via MCMC at the previous time step is used as seed of the new Markov chain, which allows the chain to converge faster.
	
	\section{Structural reliability of a deteriorating structural system and its updating}
	\label{sec: SR}
	Estimation of the structural reliability, and the use of vibrational data to update this, is instrumental for the framework that we are presenting here. A detailed review of the ideas presented in this section can be found in \cite{Melchers, Straub1}.
	
	\subsection{Structural reliability analysis for a deteriorating structural system}
	\label{subsec: SR_prior}
	In its simplest form, a failure event at time $t$ can be described in terms of a structural system capacity $R(t)$ and a demand $S(t)$. Both $R$ and $S$ are random variables. With $D(\boldsymbol{\theta},t)$ we define a parametric stochastic deterioration model. Herein we assume that the structural capacity $R(t)$ can be separated from the demand $S(t)$, and that the capacity is deterministic and known for a given deterioration $D(\boldsymbol{\theta},t)$, hence we write $R\left(D(\boldsymbol{\theta},t)\right)$. More details on how this deterministic curve can be obtained for specific cases are given in Section 6, which contains the numerical examples. Therefore, at a time $t$ the structural capacity includes the effect of the deterioration process. The uncertain demand acting on the structure is here modeled by the distribution of the maximum load in a one-year time interval. The cumulative distribution function (CDF) of this distribution is denoted $F_{s_{max}}$. Such a modeling choice simplifies the estimation of the structural reliability, as will be made clear in what follows, which is vital within a computationally expensive VoI analysis framework.
	
	We discretize time in annual intervals $j=1,..,T$, where the $j$-th interval represents $t \in (t_{j-1}, t_j]$. For the type of problems that we are considering, the time-variant reliability problem can be replaced by a series of time-invariant reliability problems \cite{Straub1}. $F_j^*$ is defined as the event of failure in interval $(t_{j-1}, t_j]$. For a given value of the deterioration model parameters $\boldsymbol{\theta}$ and time $t_j$, the capacity $R\left(D(\boldsymbol{\theta},t_j)\right)$ is fixed, and the conditional interval probability of failure is defined as:
	\begin{equation}
		\text{Pr}(F_j^* \mid \boldsymbol{\theta}, t_j)= 1 -F_{s_{max}}\left(R\left(D(\boldsymbol{\theta},t_j)\right)\right)
		\end{equation}
	We define $\text{Pr}[F(t_i)] = \text{Pr}(F_1^*\cup F_2^*\cup...F_i^*)$ as the accumulated probability of failure up to time $t_i$.
	One can compute $Pr[F(t_i)]$ through the conditional interval probabilities $Pr(F_j^*|R(\boldsymbol{\theta}, t_j))$ as:
	\begin{equation}
		\text{Pr}[F(t_i)\mid \boldsymbol{\theta}] =1-\prod_{j=1}^{i}[1 - \text{Pr}(F_j^*\mid \boldsymbol{\theta}, t_j)]
		\label{conditional_accumulated}
		\end{equation}
	Following the total probability theorem, the unconditional accumulated probability of failure is:
	\begin{equation}
		\text{Pr}[F(t_i)] = \int_{\Omega_{\boldsymbol{\theta}}}\text{Pr}[F(t_i)\mid \boldsymbol{\theta}] \pi_{\text{pr}}(\boldsymbol{\theta})d\boldsymbol{\theta}
		\label{accumulated}
		\end{equation}
	The solution to the above integral is approximated using Monte Carlo simulation (MCS). We draw samples from the prior distribution $\pi_{\text{pr}}(\boldsymbol{\theta}) $ of the uncertain deterioration model parameters and the integral in (\ref{accumulated}) is approximated by:
	\begin{equation}
		\text{Pr}[F(t_i)] \approx \frac{1}{n_{\text{MCS}}} \sum_{k=1}^{n_{\text{MCS}}} \text{Pr}[F(t_i)\mid \boldsymbol{\theta}^{(k)}]
		\label{MCS}
		\end{equation}
	Having computed the probabilities $Pr[F(t_i)]$, one can compute the hazard function $h(t_i)$ for the different time intervals $t_i$, which expresses the failure rate of the structure conditional on survival up to time $t_{i-1}$:
	\begin{equation}
		h(t_i) = \frac{\text{Pr}[F(t_i)] - \text{Pr}[F(t_{i-1})]}{1 - \text{Pr}[F(t_{i-1})]}
		\label{hazard_prior}
		\end{equation}
	
	\subsection{Structural reliability updating using SHM modal data}
	\label{subsec:SR_updating}
	The goal of SHM is to identify structural damage. Monitoring data can be employed in order to identify the parameters $\boldsymbol{\theta}$ of the deterioration models and obtain their posterior distribution, as shown in Section \ref{sec:Bayes}. Consequently this leads to the updating of the accumulated probability of failure at time $t_i$, which can now be conditioned on data $\boldsymbol{Z}_{1:i-1}$ obtained up to time $t_{i-1}$. 
	\begin{equation}
		\text{Pr}[F(t_i)\mid \boldsymbol{Z}_{1:i-1}] = \text{Pr}(F_1^*\cup F_2^*\cup...F_i^*\mid \boldsymbol{Z}_{1:i-1})
		\label{conditional}
		\end{equation}
	The accumulated probability of failure up to time $t_i$ conditional on modal data obtained up to time $t_{i-1}$ is:
	\begin{equation}
		\text{Pr}[F(t_i)\mid \boldsymbol{Z}_{1:i-1}] =\\
		\int_{\Omega_{\boldsymbol{\theta}}}\text{Pr}[F(t_i)\mid \boldsymbol{\theta}] \pi_{\text{pos}}(\boldsymbol{\theta}| \boldsymbol{\widetilde{\lambda}}_{1:i-1}, \boldsymbol{\widetilde{\Phi}}_{1:i-1})d\boldsymbol{\theta}
		\label{conditional_integral}
		\end{equation}
	In (\ref{conditional_integral}), one needs to integrate over the posterior distribution of the parameters $\boldsymbol{\theta}$. As described in Section (\ref{subsec: Bayesian solution}), two different methods for obtaining samples from this posterior distribution at each time step are implemented. In the case that an adaptive MCMC algorithm is used, at every step of the sequential updating we obtain the desired posterior distribution of the parameters in the form of correlated MCMC samples. In the case that the posterior distributions are approximated by multivariate Gaussian distributions using the Laplace approximation, independent posterior samples can be drawn from this approximate posterior density. Using $n_{\text{pos}}$ samples $\boldsymbol{\theta}^{(k)}$ from either MCMC or the asymptotic approximation , the integral in equation (\ref{conditional_integral}) can be approximated:
	\begin{equation}
		\text{Pr}[F(t_i)\mid \boldsymbol{Z}_{1:i-1}]
		\approx \frac{1}{n_{\text{pos}}} \sum_{k=1}^{n_{\text{pos}}} \text{Pr}[F(t_i)\mid \boldsymbol{\theta}^{(k)}]
		\label{accumulated_posterior}
		\end{equation}
	The hazard function conditional on the monitoring data can then be obtained as:
	\begin{equation}
		h(t_i \mid  \boldsymbol{Z}_{1:i-1}) = \frac{\text{Pr}[F(t_i) \mid  \boldsymbol{Z}_{1:i-1}] - \text{Pr}[F(t_{i-1}) \mid  \boldsymbol{Z}_{1:i-1}]}{1 - \text{Pr}[F(t_{i-1})\mid  \boldsymbol{Z}_{1:i-1}]}
		\label{hazard_posterior}
		\end{equation}
	
	\section{Life-cycle cost with SHM}
	\label{sec:LCC}
	
	\subsection{Life-cycle optimization based on heuristics}
	The VoI is the difference in life-cycle cost between the cases with and without SHM system. To calculate the life-cycle cost we optimize the maintenance strategy. A strategy $S$ is a set of policies that determine which action to take at any time step $t_i$, conditional on all the information at hand up to that time \cite{Jensen}, \cite{Elizabeth}. One may define policies based on simple decision rules, also called heuristics, which may emerge from basic engineering understanding. 
	
	A detailed presentation of the use of heuristics in optimal inspection and maintenance planning can be found in \cite{Jesus, Elizabeth}. With the use of heuristics, the space of solutions to the decision problem is drastically reduced, but the problem is solved only approximately. Here, we utilize a simple heuristic for maintenance decisions. The simple heuristic chosen in this work is the following: Perform a repair action whenever the estimate of the hazard function (the conditional failure rate) is larger than a predefined threshold $h_{thres}$. The use of the hazard function as a decision criteria for condition assessment and maintenance planning is a popular choice in literature \cite{Elingwood}. The parameter $w = h_{thres}$ describing the heuristic is a parameter of the strategy $S$. For simplicity, we assume herein that performing a repair action results in replacing the damaged components and bringing them back to the initial state, and that no failure will occur once a repair action has been performed. In this way, after a repair action, the computation of the total life cycle cost stops. This modeling choice is simplifying, but does allow for a viable computation of the VoI herein.
	
	The total life-cycle cost $C_{\text{tot}}$ is here taken as the total cost of maintenance and the risk of failure costs over the lifetime of the structure. The initial cost is not included in $C_{\text{tot}}$, because it is the same with or without SHM, therefore it cancels out when caclulating the VoI.
	
	With the use of heuristics, solving the decision problem boils down to finding the optimal value of the heuristic parameter $w$ which minimizes the expected total cost, i.e. to the solution of the optimization problem: 
	\begin{equation}
		w^* = \underset{w}{\argmin} \boldsymbol{\text{E}}[C_{\text{tot}} \mid w]
		\label{optimization}
		\end{equation}
	
	\subsection{Computation of the expected total life-cycle cost in the prior case}
	\label{subsec: Prior_LCC}
	In the prior case, where only the prior deterioration model is available, the expectation in equation (\ref{optimization}) is with respect to the system state, i.e. the deterioration model parameters $\boldsymbol{\theta}$. 
	The total cost of maintenance and risk is the sum of the repair costs and the risk of failure costs over the lifetime of the bridge, $C_{\text{tot}}(w, \boldsymbol{\theta}) = C_{\text{R}}(w) + C_{\text{F}}(w, \boldsymbol{\theta})$, therefore the expected total life-cycle cost for a given heuristic parameter $w$ is: 
	\begin{equation}
		\boldsymbol{\text{E}}_{\boldsymbol{\theta}}[C_{\text{tot}} \mid w] = \boldsymbol{\text{E}}_{\boldsymbol{\theta}}[C_{\text{R}}(w) \mid w] + 
		\boldsymbol{\text{E}}_{\boldsymbol{\theta}}[C_{\text{F}}(w, \boldsymbol{\theta}) \mid w] 
		\label{cost_breakdown}
		\end{equation}
	
	The first part of the right hand side of equation (\ref{cost_breakdown}) can be computed in the following way. We draw samples $\boldsymbol{\theta}^{(k)}, k=1,..,n_{\text{MCS}}$, from the prior distribution $\pi_{\text{pr}}(\boldsymbol{\theta})$ and use them to compute the accumulated probability of failure via equation (\ref{MCS}), and subsequently compute the hazard function with equation (\ref{hazard_prior}). When the hazard function exceeds the threshold, i.e. when $h(t_i) \geq w$, then we define $t_{\text{repair}}(w) = t_{i-1}$ as the time that the repair takes place. The time of repair is thus a function of our chosen heuristic. Hence the expected total cost of repair over the lifetime is given as:
	\begin{equation}
		\boldsymbol{\text{E}}_{\boldsymbol{\theta}}[C_{\text{R}}(w) \mid w] = \hat{c}_R\gamma(t_{\text{repair}}(w))
		\label{cost_repair}
		\end{equation}
	where $\hat{c}_R$ is the fixed cost of the repair, and $\gamma(t) = \frac{1}{(1+r)^t}$ is the discounting function, $r$ being the annually compounded discount rate.
	
	The risk of failure over the lifetime can be computed via MCS, using the samples $\boldsymbol{\theta}^{(k)}$, $k=1,..,n_{\text{MCS}}$, that were drawn from the prior distribution $\pi_{\text{pr}}(\boldsymbol{\theta})$, with the following formula:
	\begin{equation}
		\boldsymbol{\text{E}}_{\boldsymbol{\theta}}[C_{\text{F}}(w, \boldsymbol{\theta}) \mid w] \approx \frac{1}{n_{\text{MCS}}} \sum_{k=1}^{n_{\text{MCS}}} C_{\text{F}}(w, \boldsymbol{\theta}^{(k)})
		\label{MC_for_risk}
		\end{equation}
	where:
	\begin{equation}
		C_{\text{F}}(w, \boldsymbol{\theta}^{(k)}) = \sum_{i=1}^{t_{repair}(w)}\hat{c}_F\gamma(t_i) \{\text{Pr}[F(t_i)\mid \boldsymbol{\theta}^{(k)}]- \text{Pr}[F(t_{i-1})\mid \boldsymbol{\theta}^{(k)}]\}
		\label{cost_failure}
		\end{equation}
	and $\hat{c}_F$ is the fixed cost of the failure event.
	
	Following the solution of the optimization problem in (\ref{optimization}), the expected total life-cycle cost associated with the optimal decision in the prior case without any monitoring data is $\boldsymbol{\text{E}}_{\boldsymbol{\theta}}[C_{\text{tot}} \mid w_0^*]$.
	
	\subsection{Computation of the expected total life-cycle cost in the preposterior case}
	\label{subsec: Preposterior_cost}
	The goal of a preposterior analysis is to act as a decision tool on whether collecting SHM data is beneficial, and to quantify the VoI of an SHM system, prior to its installation. Therefore this type of analysis is performed before any actual SHM data are obtained. Instead, the SHM monitoring data histories must be sampled over the lifetime from the prior distribution of the uncertain deterioration model parameters $\boldsymbol{\theta}$, as will be explained shortly. A sampled monitoring data history vector $\boldsymbol{Z}= [\boldsymbol{Z}_1,...,\boldsymbol{Z}_{n_T}]$ contains the OMA identified modal data at fixed time instances over the structure lifetime. The rate at which this data is sampled should be chosen on the basis of the problem at hand. In this case, we explore a slow evolving deterioration process, and further ignore dependence on temperature effects. Therefore, for this investigation we employ one set of OMA-identified modal data per year. Under the currently assumed independence on environmental and operational conditions (EOCs) this sparse assumption is further justified from the observation that increasing the number of data sets used in the BMU process seemingly decreases the parameter estimation uncertainty, albeit not properly reflecting the full variability of the updated parameters \cite{Behmanesh, Vanik}.
	
	In a preposterior analysis, the expectation in equation (\ref{optimization}) is operating over both the system state $\boldsymbol{\theta}$ and on the monitoring outcomes $\boldsymbol{Z}$.
	\begin{equation}
		\boldsymbol{\text{E}}_{\boldsymbol{\theta}, \boldsymbol{Z}}[C_{\text{tot}} \mid w] = \int_{\Omega_{\boldsymbol{\theta}}} \int_{\Omega_{\boldsymbol{Z}}} C_{\text{tot}}(w, \boldsymbol{\theta}, \boldsymbol{z})f_{\boldsymbol{\Theta}, \boldsymbol{Z}}(\boldsymbol{\theta},\boldsymbol{z} )d\boldsymbol{z}d\boldsymbol{\theta}
		\label{posterior_case}
		\end{equation}
	The total cost of maintenance and risk is again the sum of the repair cost and the risk of failure cost over the lifetime of the structure, which now both depend also on the monitoring outcomes $\boldsymbol{Z}$, $C_{\text{tot}}(w, \boldsymbol{\theta},  \boldsymbol{Z}) = C_{\text{R}}(w, \boldsymbol{Z}) + C_{\text{F}}(w, \boldsymbol{\theta}, \boldsymbol{Z})$.
	
	The integral in equation (\ref{posterior_case}) is computed with crude MCS. We draw samples from the uncertain deterioration model parameters $\boldsymbol{\theta}$, which correspond to a deterioration history over the lifetime, as given by the deterioration model equation  $D(\boldsymbol{\theta},t)$. For each of these histories, we generate noisy acceleration measurements every year, feed them into an SSI algorithm, and obtain one vector of monitoring modal data $\boldsymbol{Z}$ (one identified modal data set per year). In this way we are jointly sampling the system state space and monitoring data space, and equation (\ref{posterior_case}) is approximated as:
	\begin{equation}
		\boldsymbol{\text{E}}_{\boldsymbol{\theta}, \boldsymbol{Z}}[C_{\text{tot}} \mid w] =\frac{1}{n_{MCS}}\sum_{k = 1}^{n_{MCS}}[C_{\text{R}}(w, \boldsymbol{z}^{(k)}) + C_{\text{F}}(w, \boldsymbol{\theta}^{(k)}, \boldsymbol{z}^{(k)})]
		\label{post_case_MCS}
		\end{equation}
	For each of the sampled system states and corresponding monitoring data, we compute the updated hazard rate as given by equation (\ref{hazard_posterior}), and when $h(t_i \mid  \boldsymbol{z}^{(k)}_{1:i-1}) \geq w$, then $t_{\text{repair}}(w, \boldsymbol{z}^{(k)}) = t_{i-1}$.
	
	The cost of repair is:
	\begin{equation}
		C_{\text{R}}(w, \boldsymbol{z}^{(k)}) = \hat{c}_R\gamma(t_{\text{repair}}(w,\boldsymbol{z}^{(k)}))
		\label{cost_repair_post}
		\end{equation}
	
	The risk of failure is:
	\begin{equation}
		C_{\text{F}}(w, \boldsymbol{\theta}^{(k)},\boldsymbol{z}^{(k)} ) = \sum_{i=1}^{t_\text{repair}(w, \boldsymbol{z}^{(k)})}\hat{c}_F\gamma(t_i) \{\text{Pr}[F(t_i)\mid \boldsymbol{\theta}^{(k)}]- 
		\text{Pr}[F(t_{i-1})\mid \boldsymbol{\theta}^{(k)}]\}
		\label{cost_failure_post}
		\end{equation}
	Comparing equations (\ref{cost_failure}) and (\ref{cost_failure_post}) it is evident that adoption of the same samples of $\boldsymbol{\theta}$ in both prior and preposterior analysis, leads to an identical estimate of the risk of failure for the two analyses up to the time of the repair. The only difference between prior and preposterior case is the resulting $t_\text{repair}(w, \boldsymbol{z}^{(k)})$.
	
	Solving equation (\ref{optimization}), we obtain the optimal expected total life-cycle cost given the monitoring data, $\boldsymbol{\text{E}}_{\boldsymbol{\theta, Z}}[C_{\text{tot}} \mid w_{mon}^*]$.
		
	\subsection{Summary of the proposed methodology to calculate the VoI}
	\label{subsec: VOI}
	The proposed procedure for the VoI analysis consists of the following steps:
	\begin{enumerate}
	    \item Choose a prior stochastic deterioration model describing the structural condition over the lifetime of the structure. Define a decision analysis time discretization, maintenance/repair actions, costs of actions and cost of failure event. Choose a heuristic parameter $w$ (threshold on hazard rate) for a heuristic-based solution of the decision problem.
	    \item Draw Monte Carlo samples $\boldsymbol{\theta}$ of the stochastic deterioration model parameters. 
	    \item Perform a prior decision analysis:
	    \begin{itemize}
	        \item Use the prior $\boldsymbol{\theta}$ samples to estimate the lifetime accumulated probability of failure $Pr[F(t_i)]$ and the corresponding hazard rate $h(t_i)$.
	        \item Solve the LCC optimization problem to obtain the optimal value of the heuristic parameter $w_0^*$ and the corresponding optimal $t_{repair}$. Obtain the optimal expected LCC in the prior case:  $\boldsymbol{\text{E}}_{\boldsymbol{\theta}}[C_{\text{tot}}(\boldsymbol{\theta},w) \mid w_0^*]$.
	   \end{itemize}
	   \item Perform a preposterior decision analysis:
	   \begin{itemize}
	       \item For each individual prior sample $\boldsymbol{\theta}$ realization and given value of the heuristic parameter $w$ do the following:
	       \begin{enumerate}[label=(\alph*)]
	           \item Sample the corresponding noisy acceleration time series data for every year over the lifetime of the structure. Feed the accelerations into the SSI algorithm to identify the structure's modal data vectors $\boldsymbol{Z}$.
	           \item Perform a posterior Bayesian analysis: BMU to sequentially learn the posterior distributions of $\boldsymbol{\theta}$ and subsequently obtain an updated estimate of the accumulated probability of failure $\text{Pr}[F(t_i)\mid \boldsymbol{Z}_{1:i-1}]$ and the hazard rate $h(t_i \mid  \boldsymbol{Z}_{1:i-1})$.
	           %\item Solve the LCC optimization to obtain the optimal time to perform the repair action conditional on this specific deterioration and monitoring data realization.
	       %\end{enumerate}
	       %\item Average over all the posterior optimal solutions to obtain the optimal expected LCC in the preposterior case $\boldsymbol{\text{E}}_{\boldsymbol{\theta, Z}}[C_{\text{tot}}(\boldsymbol{\theta},\boldsymbol{Z},w) \mid w_{mon}^*]$.
	       	   \item Find the time to perform the repair action for this specific deterioration and monitoring data realization, conditional on a value of the heuristic parameter $w$.  
	       \end{enumerate}
	       \item Solve the LCC optimization problem to obtain the optimal value of the heuristic parameter $w_{mon}^*$ which minimizes the expected LCC in the preposterior case $\boldsymbol{\text{E}}_{\boldsymbol{\theta, Z}}[C_{\text{tot}}(\boldsymbol{\theta},\boldsymbol{Z},w) \mid w_{mon}^*]$.
	   \end{itemize}
	   \item Compute the VoI.
	   \begin{equation}
	       VoI= \boldsymbol{\text{E}}_{\boldsymbol{\theta}}[C_{\text{tot}}(\boldsymbol{\theta},w) \mid w_0^*] - \boldsymbol{\text{E}}_{\boldsymbol{\theta, Z}}[C_{\text{tot}}(\boldsymbol{\theta},\boldsymbol{Z},w) \mid w_{mon}^*]
	       \label{VOI}
	   \end{equation}
	\end{enumerate}
	
	\subsection{Value of Partial Perfect Information}
	\label{subsec: PI}
	The case of partial perfect information is related to a hypothetical situation, in which the SHM system provides perfect information on the condition of the structure. This means that there is no uncertainty on the parameters $\boldsymbol{\theta}$ of the deterioration model, and the optimal decision is found conditional on this perfect knowledge of $\boldsymbol{\theta}$. Because the SHM system is not able to provide any information about the load acting on the structure, which here is modeled by an uncertain Gumbel random variable, therefore one uses the term ``partial''.
	
	Estimation of the value of partial perfect information is given by:
	\begin{equation}
		VPPI= \underset{w}{\min} \boldsymbol{\text{E}}_{\boldsymbol{\theta}}[C_{\text{tot}}(\boldsymbol{\theta},w)] -  \boldsymbol{\text{E}}_{\boldsymbol{\theta}}\{\underset{w}{\min}[C_{\text{tot}}(\boldsymbol{\theta},w)\mid \boldsymbol{\theta}]\}
		\label{VPPI}
		\end{equation}
	
	The left hand side of equation (\ref{VPPI}) is the optimal expected total life-cycle cost in the prior case, exactly as presented in Section \ref{subsec: Prior_LCC}. On the right hand side, first the optimal heuristic is found conditional on exact knowledge of $\boldsymbol{\theta}$, then the expected value of the total life-cycle costs associated with optimal decisions is computed. This quantity corresponds to the value of information that one would obtain in the case of perfect monitoring and perfect decision making with the chosen heuristic.
	
	The VPPI provides an upper limit on the value that the VoI can obtain. Since it can be computed much easier than the VoI, the VPPI can provide a first estimate on the maximum investment that should be made for SHM systems. Therefore, we motivate the idea that a VPPI computation should always be performed first.
	
	\section{Numerical investigations}
	\label{sec: Numerical_Investigations}
	\subsection{Numerical benchmark: Continuously monitored bridge system subject to deterioration}
\begin{figure}[ht]
		\centerline{
			\includegraphics[width=\textwidth]{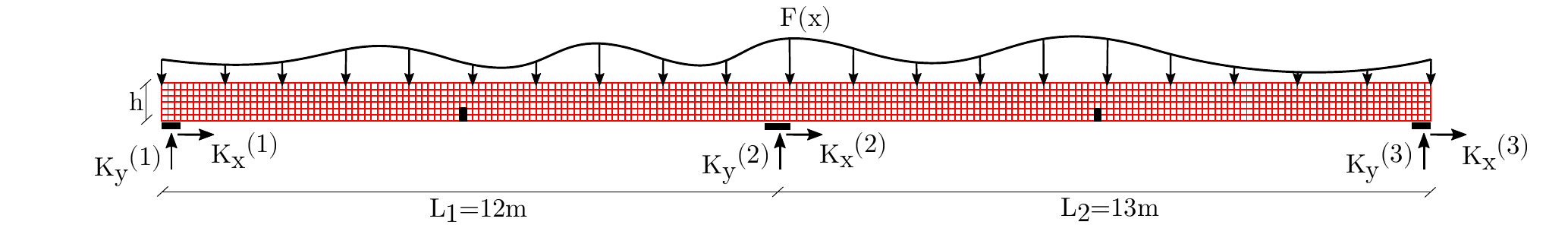}
		}
		\caption{Benchmark model}
		\label{f:benchmark}
	\end{figure}

	We consider the two-span bridge model of Figure~\ref{f:benchmark}, with its reference behavior \cite{bench19} simulated by a FE model of isoparametric plane stress quadrilateral elements. This benchmark structure has been developed as part of the TU1402 COST Action and serves for verification of analysis methods and tools for SHM. 200 elements are employed to mesh the $x$ direction, and 6 elements are assumed per height ($y$ direction). The beam dimensions form configurable parameters of the benchmark and are set as: height $h$ = 0.6m, width $w$ = 0.1m, while the lengths are $L_1$ = 12m for 
	the first span and $L_2$ = 13m for the second span. A linear elastic material with Young's modulus $E$ = 30GPa, Poisson ratio $\nu$ = 0.2, and material density $\rho$ = 2000 kg/m\textsuperscript{3} is assigned. Elastic boundaries in both directions are assumed for all three support points, in the form of translational springs with $K_x$ = 10\textsuperscript{8} N/m and $K_y$ = 10\textsuperscript{7} N/m.
	
	It is assumed that the simulated two-span bridge is continuously monitored using a set of sensors measuring vertical acceleration, whose locations correspond to predefined FE nodes. A distributed Gaussian white noise excitation $F(x)$ is used as the load acting on the bridge, to simulate the unknown ambient excitation. A dynamic time history analysis of the model, for a given realization of the load, results in the measured vertical acceleration signals at the assigned sensor locations. 
	%\vspace{-10pt}
	\subsection{Deterioration modeling}
	A prior model describing structural deterioration is a prerequisite for a VoI analysis. A detailed presentation of probabilistic deterioration models for life-cycle performance assessment of structures can be found in \cite{Elingwood, Frangopol, Biondini}. For time-dependent reliability assessment purposes, the use of simple empirical models, which are still flexible enough to model different kinds of deterioration mechanisms, can be adopted \cite{Elingwood}.
	Within this work, we use a simple rate equation of the form:
	\begin{equation}
		D(t) = A t^B
		\label{deterioration}
		\end{equation}
	to model structural deterioration, where $D(t)$ is the unit-less deterioration parameter (loss of stiffness) entering in the assumed damage model, and $A, B$ are random variables driving the uncertainty in this model. Parameter $A$ models the deterioration rate, while parameter $B$ is related to the nonlinearity effect in terms of a power law in time. We consider herein the following two case studies related to structural deterioration of the bridge structure.
	
	\subsubsection{Bridge system subject to scour}
	We assume that the middle elastic support (pier) of the bridge structure is subjected to gradual deterioration, simulating the case of scour \cite{Prendergast}. Damage is introduced as a progressive reduction of the stiffness in $y$-direction of the spring $K_y^{(2)}$ at the middle elastic support of the bridge (Figure \ref{f:benchmark}). The evolution of the stiffness reduction of the vertical spring support over the lifespan of the bridge is described by employing the damage model of equation (\ref{stifness_reduction}), where $K_{y,0}^{(2)}$ is the initial undamaged value, and $D(t)$ is the stiffness reduction described by equation (\ref{deterioration}). We consider a lifespan of $T$=50 years for the structure. The uncertain parameters of the deterioration model are summarized in Table \ref{table:Parameters_scour}. The mean and coefficient of variation of the parameters $A$ and $B$ are chosen to reflect a significant a-priori uncertainty. They result in a 10\% probability that $D(t=50)>9$ at the end of the lifespan.
	\begin{equation}
	K_y^{(2)}(t) = \frac{K_{y,0}^{(2)}}{(1+D(t))} = \frac{K_{y,0}^{(2)}}{(1+At^{B})}
	\label{stifness_reduction}
	\end{equation}
	
	\begin{table}[!ht]
		\caption{Parameters of the stochastic deterioration model for scour.}
		\footnotesize
		\centering
		\begin{tabular}{cccc}\hline
			Parameter&Distribution&Mean&CV\\\hline
			A&Lognormal&7.955$\times$10\textsuperscript{-4}&0.5\\
			B&Normal&2.0&0.15\\\hline
		\end{tabular}
		\label{table:Parameters_scour} 
	\end{table}
	
	\subsubsection{Bridge system subject to corrosion deterioration}
	As a second separate case study, we assume that the bridge structure is subjected to gradual deterioration from corrosion in the middle of both midspans (elements in black in Figure \ref{f:benchmark}). At both locations, damage is introduced as a progressive reduction of the stiffness at the bottom 2 elements of the FE mesh. For the deterioration hotspots at the left and right midspans, the evolution of the elements' stiffness reduction over the lifespan of the bridge is described by employing the damage model of equation (\ref{Youngs_modulus_reduction_left}). $E^{(0)}$ is the initial undamaged value of the Young's modulus, and $D_1(t)$, $D_2(t)$ are the deterioration models (reduction of stiffness) employed for each location, as described by equation (\ref{deterioration}). The random variables of the deterioration models are summarized in Table \ref{table:Parameters}. According to \cite{Elingwood}, for this simple empirical deterioration model, a value of $B$=0.5 corresponds to diffusion-controlled damage processes. Therefore the mean values of $B_1$ and $B_2$ have been chosen equal to 0.5. The mean and coefficient of variation of the four uncertain parameters are chosen so that they result in a 1\% probability that $D(t=50)>$9 at the end of the lifespan.

	\begin{equation}
		E_j(t) = \frac{E^{(0)}}{(1+D_j(t))} = \frac{E^{(0)}}{(1+A_jt^{B_j})}, j = 1, 2
		\label{Youngs_modulus_reduction_left}
		\end{equation}
	
	\begin{table}[ht]
		\caption{Parameters of the stochastic deterioration model for corrosion.}
		\footnotesize
		\centering
		\begin{tabular}{cccc}\hline
			Parameters&Distribution&Mean&CV\\\hline
			$A_1, A_2$&Lognormal&0.506&0.4\\
			$B_1, B_2$&Normal&0.5&0.15\\\hline
		\end{tabular}
		\label{table:Parameters} 
	\end{table}
	
	\subsection{Synthetic monitoring data creation}
	\label{subsec: Synthetic}
\begin{figure}
		\centerline{
			\includegraphics[width=\textwidth]{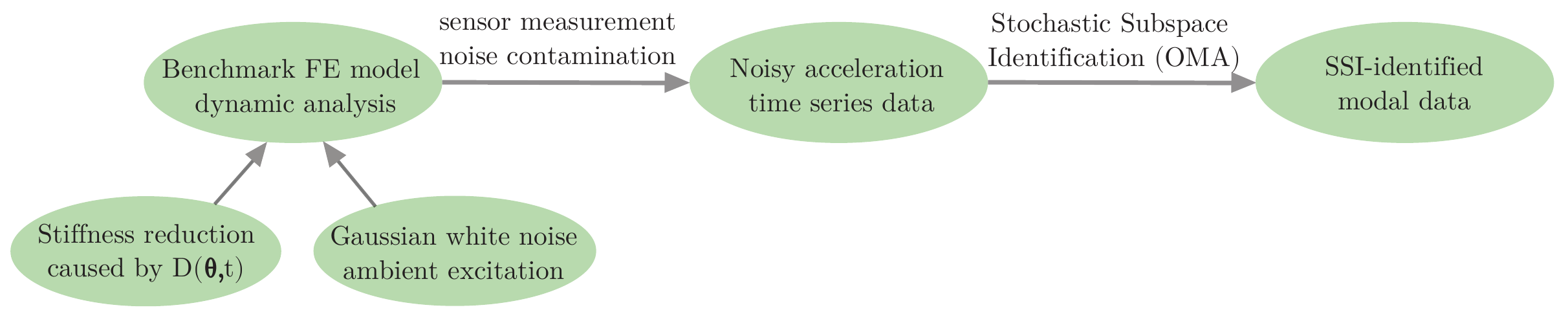}
		}
		\caption{Process for generating the SHM data}
		\label{f:synthetic}
	\end{figure}

	For the purpose of the VoI analysis framework presented in this paper, for every deterioration time instance at which we want to simulate a monitoring data set obtained from the deployed SHM system, the corresponding stiffness reduction is implemented in the FE benchmark model, a dynamic time history analysis is run and the ``true" vertical acceleration signals $\ddot{x}$ at the sensor locations (FE nodes) are obtained. The noise-free acceleration time series data set is contaminated with Gaussian white noise of 2\% root mean square noise-to-signal ratio, simulating a sensor measurement error. Subsequently the noisy accelerations $\tilde{\ddot{x}}$ are fed into an output-only operational modal analysis (OMA) scheme. Specifically, the stochastic subspace identification (SSI) \cite{SSI} algorithm is used to identify a set of the lower eigenvalues (squares of natural frequencies) and mode shapes. The data creation process can be seen in Figure \ref{f:synthetic}.
	
	\subsection{Continuous Bayesian model updating}
	\label{subsec: BMU results_chapter_6}
	Initially we demonstrate how the Bayesian framework performs in learning the parameters of the deterioration model on the basis of availability of the SHM modal data. In this work, the model predicting the eigenvalues and mode shapes for the Bayesian updating process is the same FE model as the one described in Section \ref{subsec: Synthetic} for the creation of the noise-contaminated synthetic data. Despite addition of artificial noise, adoption of the same model constitutes a so-called inverse crime \cite{inverse_crime}. This is the a built in feature of standard preposterior analysis. 
	
	We draw samples of the deterioration parameters, defining the evolution of sample deterioration curves. For each of these deterioration curves we create one monitoring history, i.e., we generate one set of OMA-identified modal data every year over the fifty years of the lifetime. In this simple example, the structural properties are not assumed influenced by environmental (temperature, humidity) and operation (non stationary effects due to traffic) variability. For this reason, we assume it suffices to utilize one estimate of the modal properties set per year. Using this data, we employ the sequential Bayesian deterioration model updating framework of Section \ref{sec:Bayes}. 
	
	The sequential Bayesian analysis framework requires a substantial number of evaluations of the likelihood function, implying multiple forward runs of the FE model. Within a VoI framework, Bayesian analysis must be performed numerous times. For this reason a VoI analysis can quickly become intractable. To enable the VoI analysis, we employ simple surrogate models to replace the structural FE model, which are described in the following two subsections.
	
	\subsubsection{Bridge system subject to scour deterioration - Global damage identification}
\begin{figure}
		\centerline{
			\includegraphics[width=\textwidth]{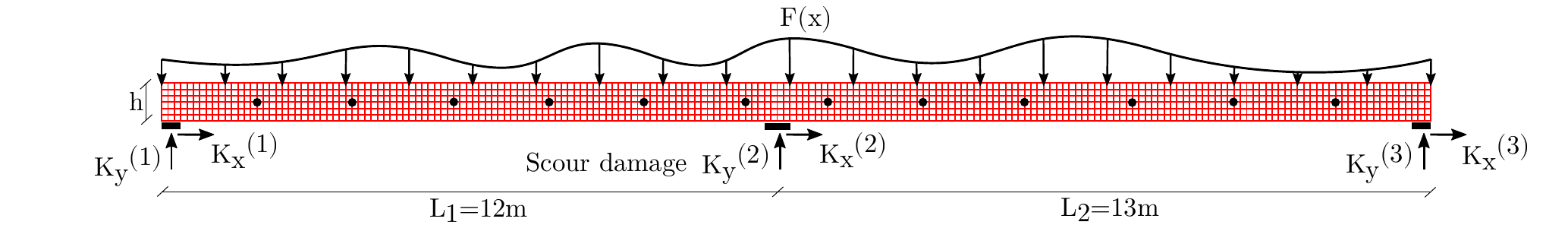}
		}
		\caption{Bridge system subject to scour damage}
		\label{f:benchmark_3}
	\end{figure}

	 In this assumed damage scenario we are interested in identifying damage in a global scale, for which use of the OMA-identified eigenvalue data alone may be sufficient. The benefit is that eigenvalue data can be successfully identified from an OMA procedure, even when only a rather small number of accelerometers are employed on the structure. The sensor placement that we assume here is the one corresponding to Figure \ref{f:benchmark_3}, with twelve employed sensors. This configuration is selected on the basis of engineering judgment, when seeking to identify the type of damage (local stiffness reduction) considered herein. Using the SSI algorithm, we identify the lower $N_m=6$ modes, which we then use for the updating. 
	
	We employ a surrogate model to replace the structural FE model for facilitating the Bayesian updating procedure. To this end, we create a fine uniform grid of values for $D(t)$; for each of these, we execute a modal analysis using the FE model and store the output eigenvalues. Eventually, we replace the modal analysis run of the structural FE model with a simple nearest neighbor lookup in the precomputed database.  
	
\begin{figure}
		\begin{subfigure}{.24\textwidth}
			\centering
			% include first image
			\includegraphics[width=1.\linewidth]{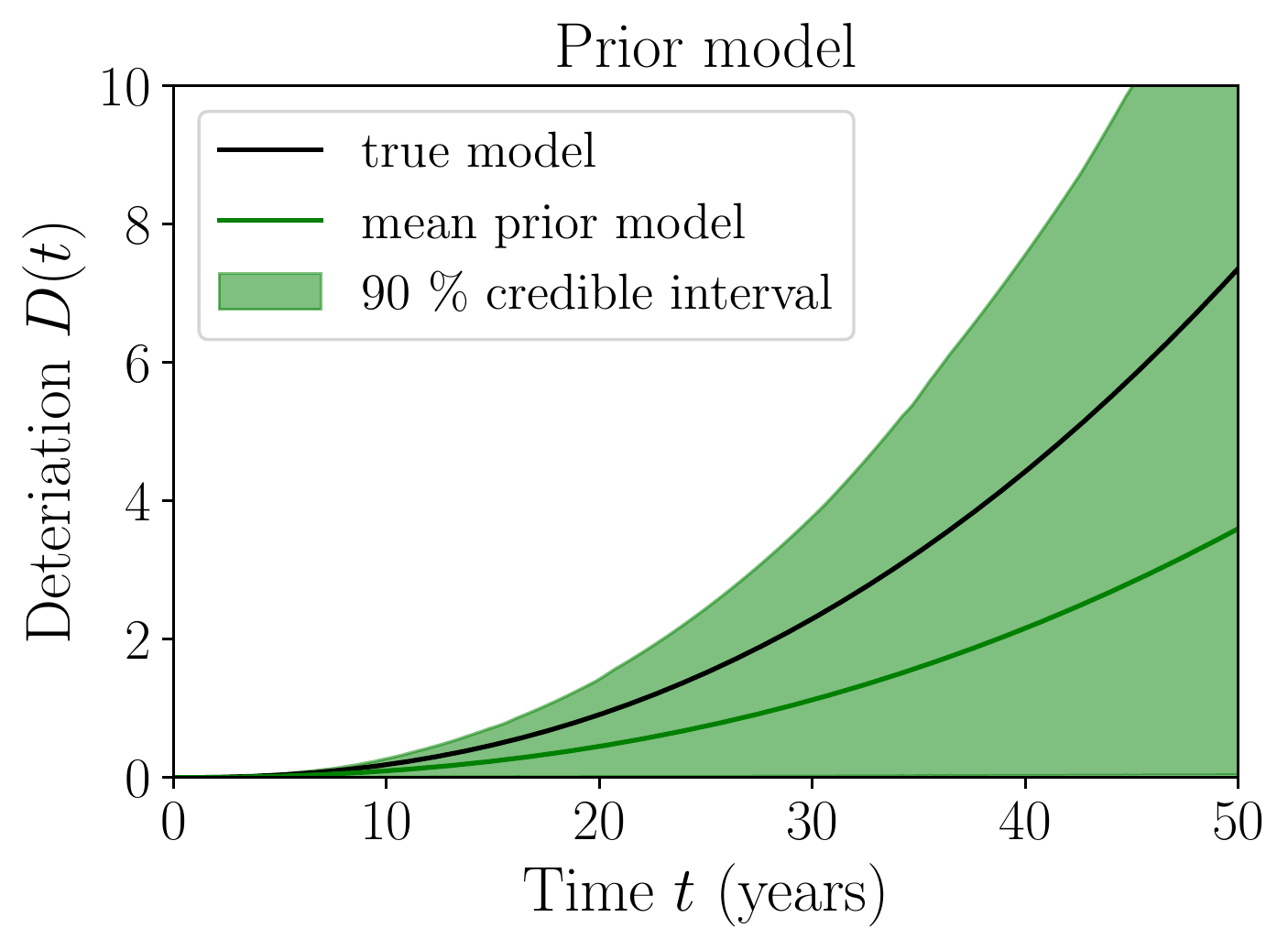}  
		\end{subfigure}
		\begin{subfigure}{.24\textwidth}
			\centering
			% include second image
			\includegraphics[width=1.\linewidth]{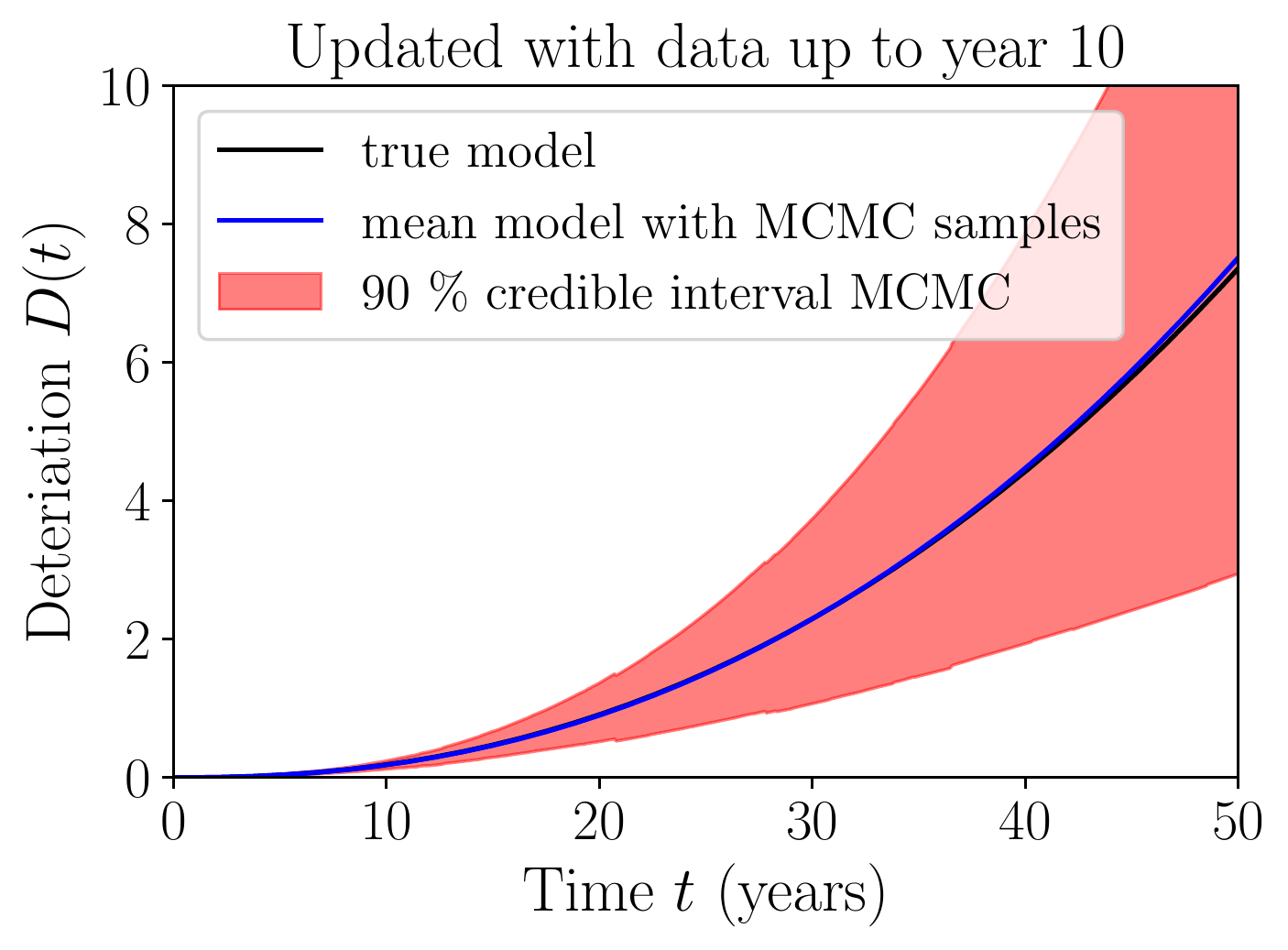}  
		\end{subfigure}
		\begin{subfigure}{.24\textwidth}
			\centering
			% include second image
			\includegraphics[width=1.\linewidth]{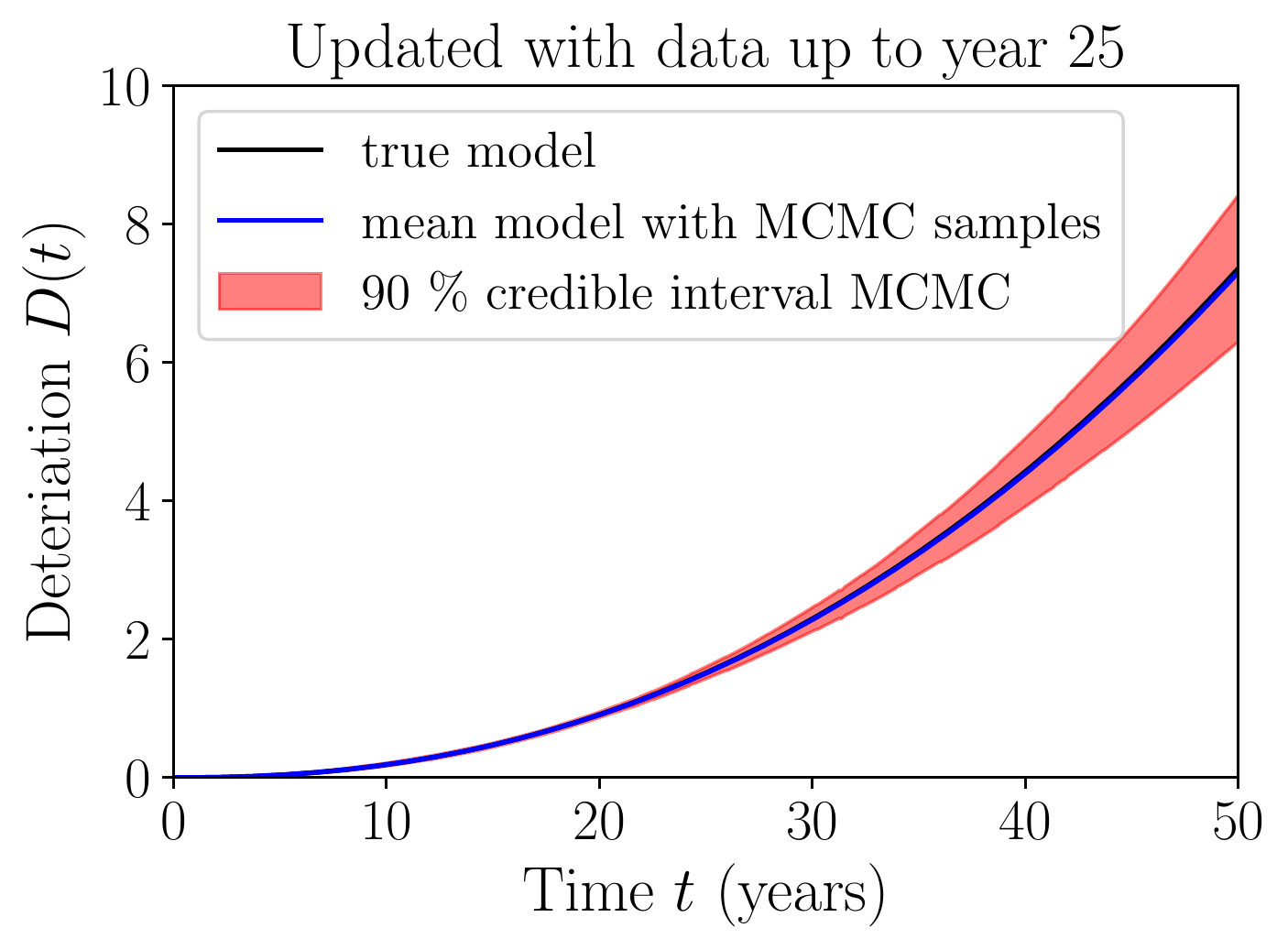}  
		\end{subfigure}
		\begin{subfigure}{.24\textwidth}
			\centering
			% include second image
			\includegraphics[width=1.\linewidth]{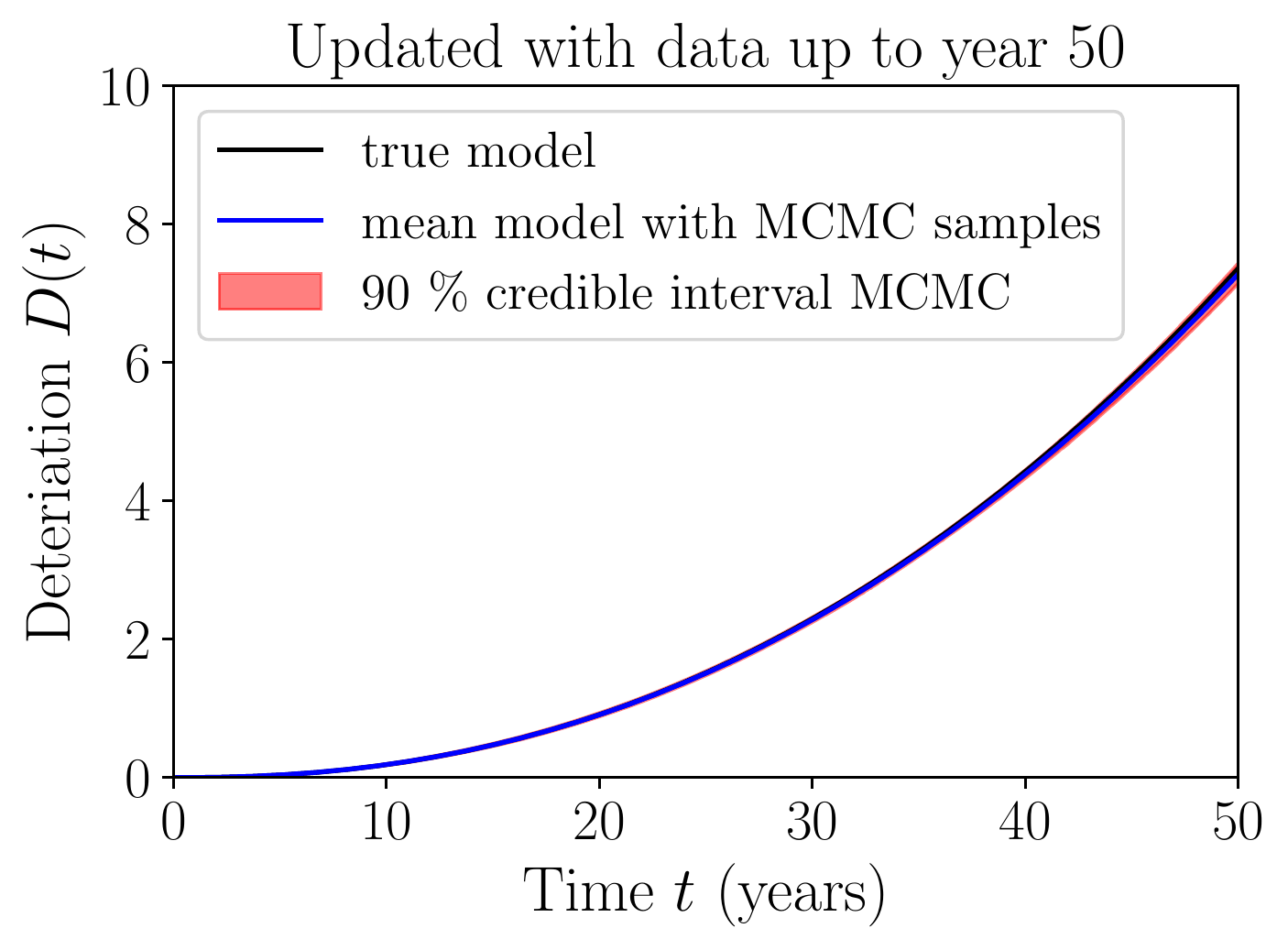}  
		\end{subfigure}
		\caption{Sequential Bayesian learning of the scour deterioration model}
		\label{f:updating_scour}
	\end{figure}

	For illustrating the data sampling and updating process, we assume a scenario where the underlying ``true" deterioration model corresponds to parameters values $A^*$=9.85$\times$10\textsuperscript{-4} and $B^*$ = 2.28. The ``true" deterioration curve can be seen in black in all the subfigures of Figure \ref{f:updating_scour}.
	
\begin{figure}
		\begin{subfigure}{.24\textwidth}
			\centering
			% include first image
			\includegraphics[width=1.\linewidth]{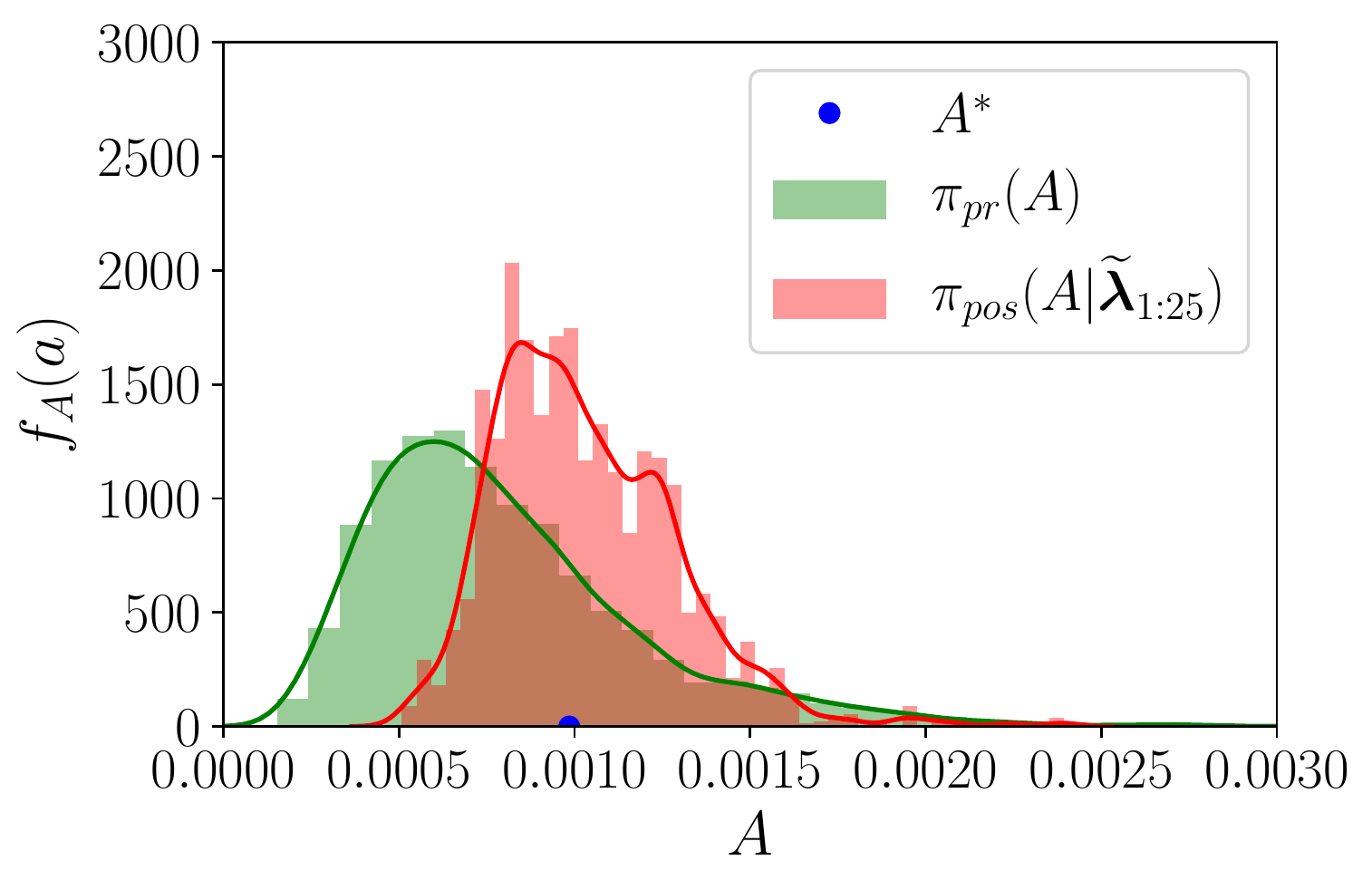}  
		\end{subfigure}
		\begin{subfigure}{.24\textwidth}
			\centering
			% include second image
			\includegraphics[width=1.\linewidth]{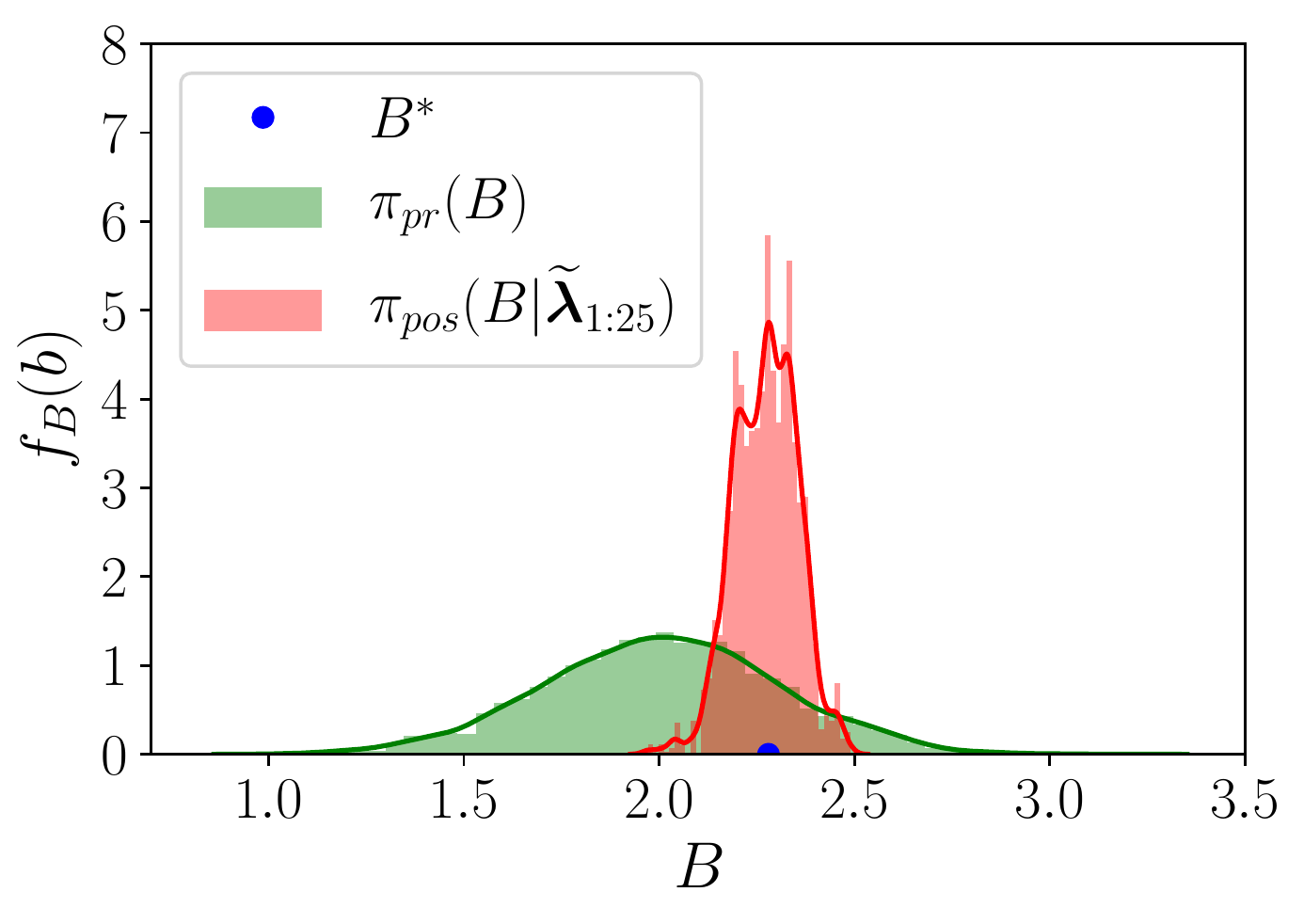}  
		\end{subfigure}
		\begin{subfigure}{.24\textwidth}
			\centering
			% include second image
			\includegraphics[width=1.\linewidth]{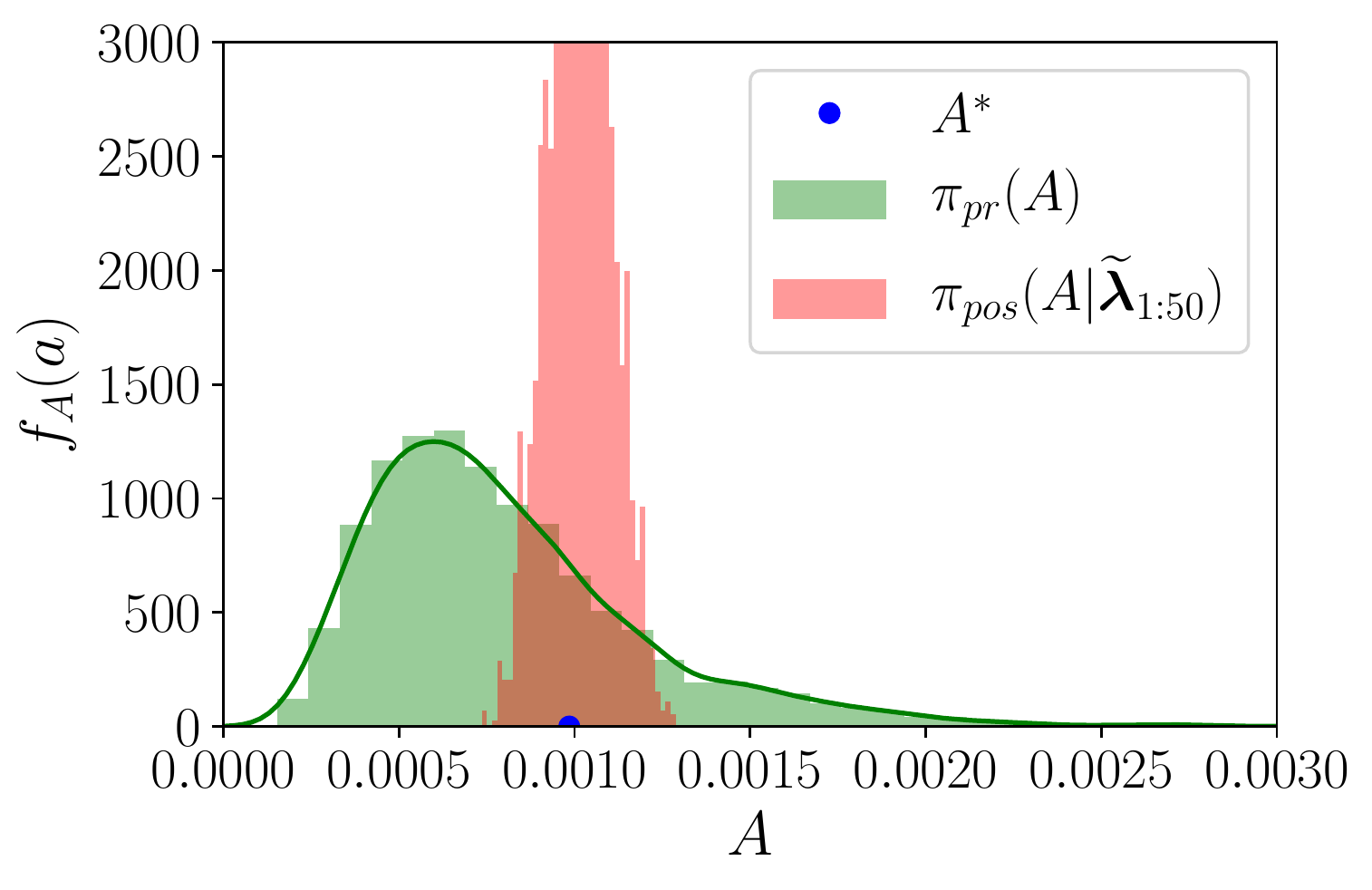}  
		\end{subfigure}
		\begin{subfigure}{.24\textwidth}
			\centering
			% include second image
			\includegraphics[width=1.\linewidth]{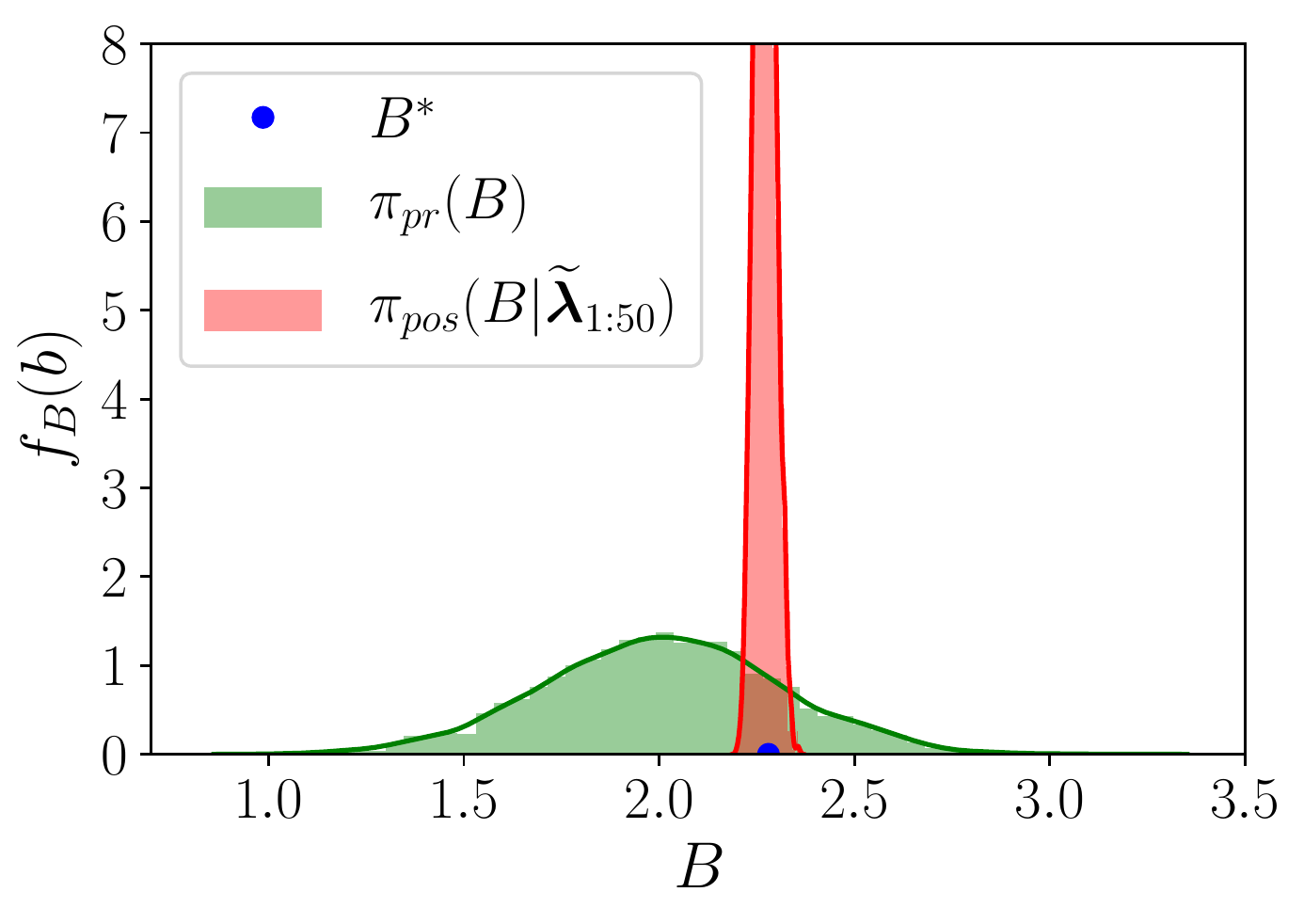}  
		\end{subfigure}
		\caption{Prior PDF and posterior PDF at years 25 and 50 for deterioration model parameters ($c_{\lambda m}=c_{\Phi m}=0.02$)}
		\label{f:parameters_learnt_scour}
	\end{figure}

	Figure \ref{f:parameters_learnt_scour} demonstrates how the distribution of the deterioration model parameters is updated by comparing the prior PDF of $A$ and $B$ with the posterior PDF of $A$ and $B$ at year 25 and year 50. For this analysis, both factors $c_{\lambda m}$ and $c_{\Phi m}$ are assumed equal to $0.02$, i.e. we assume that the total prediction error causes up to two percent deviation on the nominal model predicted values. 5000 MCMC samples are used for the Bayesian analysis at each time step. The posterior PDFs are given via a kernel density estimation using the 5000 posterior MCMC samples of the parameters. It is observed that using one SHM data set per year, the uncertainty in the deterioration model parameters gradually decreases, the PDFs become narrower and peak around the underlying ``true'' values for which the data was created.
	
	Figure \ref{f:updating_scour} contains the following: The mean estimated deterioration model together with its $90\%$ credible interval in the prior case, obtained via a MCS from the prior distribution of the uncertain parameters, is plotted in the left panel in green. In red we plot the posterior predictive mean models together with their $90\%$ credible intervals, which are estimated with posterior MCMC samples using monitoring data up to the three different time instances. For example in the second column, we use the monitoring data of the first ten years to obtain the posterior distribution $\pi_{\text{pos}}(\boldsymbol{\theta} \mid \boldsymbol{\widetilde{\lambda}}_{1:10},\boldsymbol{\widetilde{\Phi}}_{1:10})$, and then we use the posterior MCMC samples to predict the evolution of the deterioration model over the structural lifetime. We observe that already the data obtained during the first few years of the deterioration process (up to year 10) help in shifting the mean posterior model towards the underlying ``true'' deterioration curve, however the posterior uncertainty in the estimation is still relatively large. The posterior uncertainty is reduced significantly as more SHM modal data become gradually available (year 25, year 50).
	

	\subsubsection{Bridge system subject corrosion deterioration - Damage detection and localization}
	\label{subsubsec: Updating_corrosion}
	
	\begin{figure}
		\centerline{
			\includegraphics[width=\textwidth]{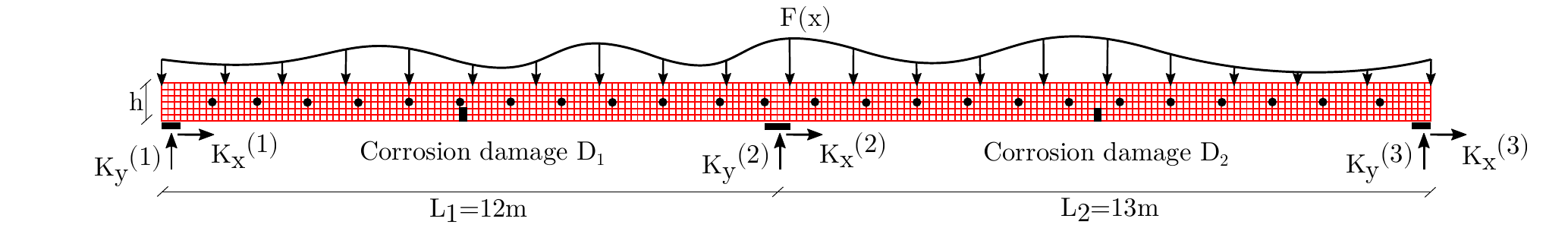}
		}
		\caption{Bridge system subject to corrosion damage in two locations}
		\label{f:benchmark_2}
	\end{figure}
	
	In the assumed scenario with two potential corrosion damage locations, the employed Bayesian model updating framework should be able to both detect and localize damage. Therefore both eigenvalue as well as mode shape displacement data should become available. As discussed in Section \ref{sec:Bayes}, a relatively large number of sensors is required for an accurate measurement and representation of the mode shape displacements. The sensor placement that we assume here is the one corresponding to Figure \ref{f:benchmark_2}, with 24 equally distributed accelerometers. By using a finite difference scheme, we can also obtain the mode shape curvatures, which are used instead of the mode shapes in the likelihood function, which seems to enhance the localization capabilities of the framework. Also in this case we identify the lower $N_m$=6 modes. 
	
	For defining a surrogate model, we create a two-dimensional grid of values for $D_1(t), D_2(t)$, and for each of the grid points we run a modal analysis with the FE model, and we store the output eigenvalues and mode shape vectors. Eventually we employ the following surrogates: For each of the eigenvalues, we fit a two-dimensional polynomial regression response surface model. For the mode shape displacement vector data, we replace the run of the structural FE model with a simple nearest neighbor lookup in the precomputed two-dimensional database. 
	 
\begin{figure}
		\begin{subfigure}{.24\textwidth}
			\centering
			% include first image
			\includegraphics[width=1.\linewidth]{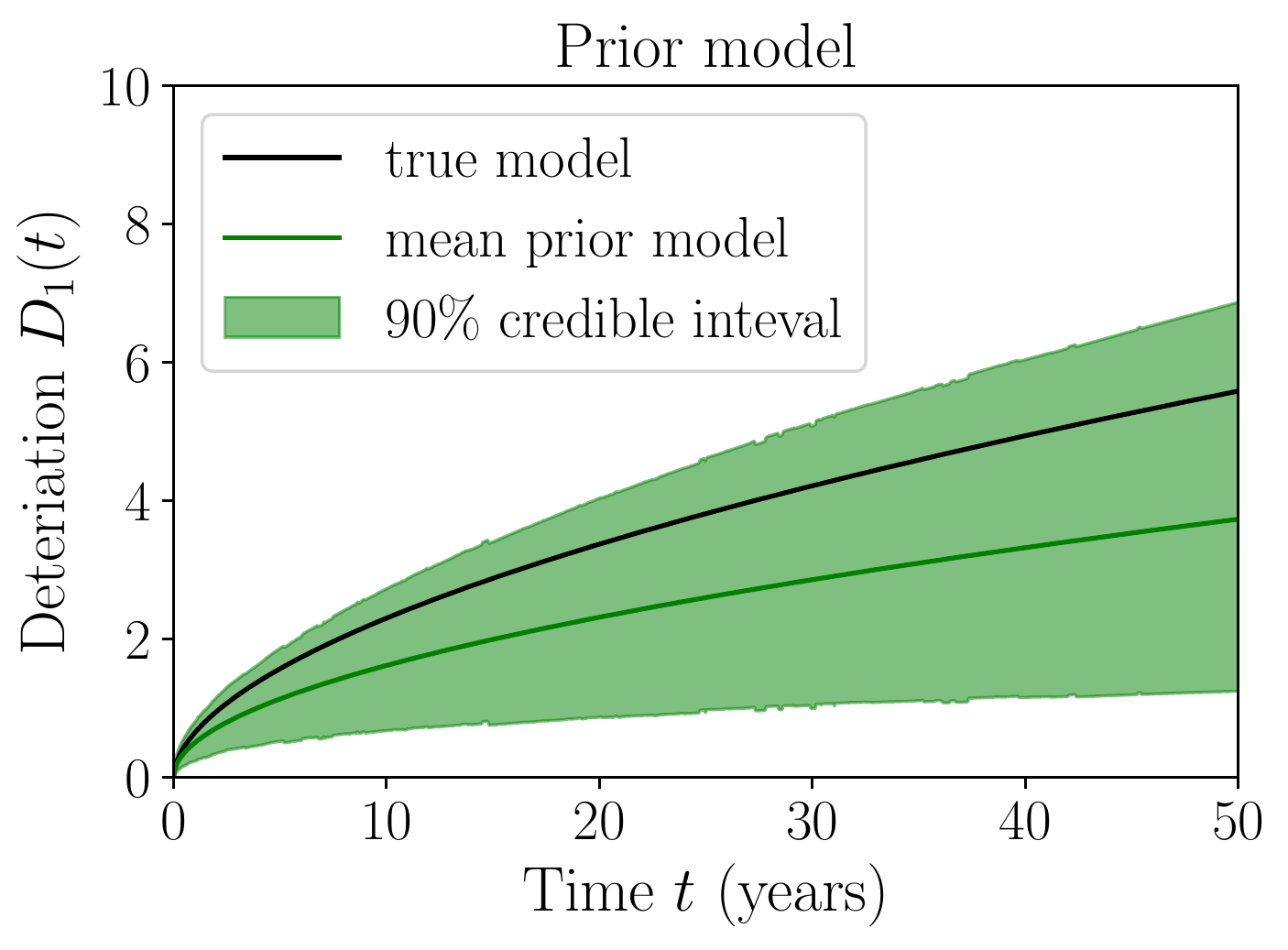}  
		\end{subfigure}
		\begin{subfigure}{.24\textwidth}
			\centering
			% include second image
			\includegraphics[width=1.\linewidth]{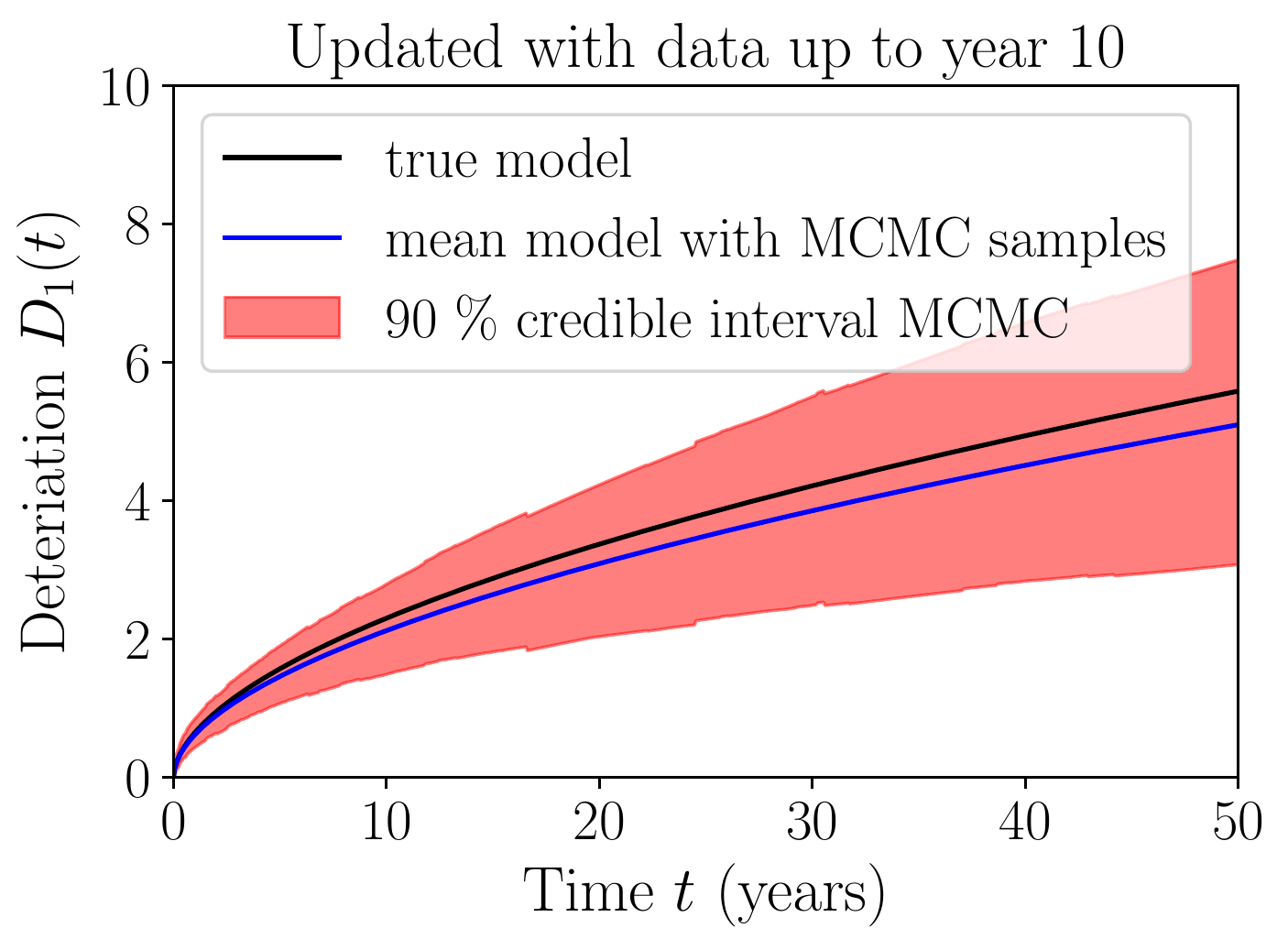}  
		\end{subfigure}
		\begin{subfigure}{.24\textwidth}
			\centering
			% include second image
			\includegraphics[width=1.\linewidth]{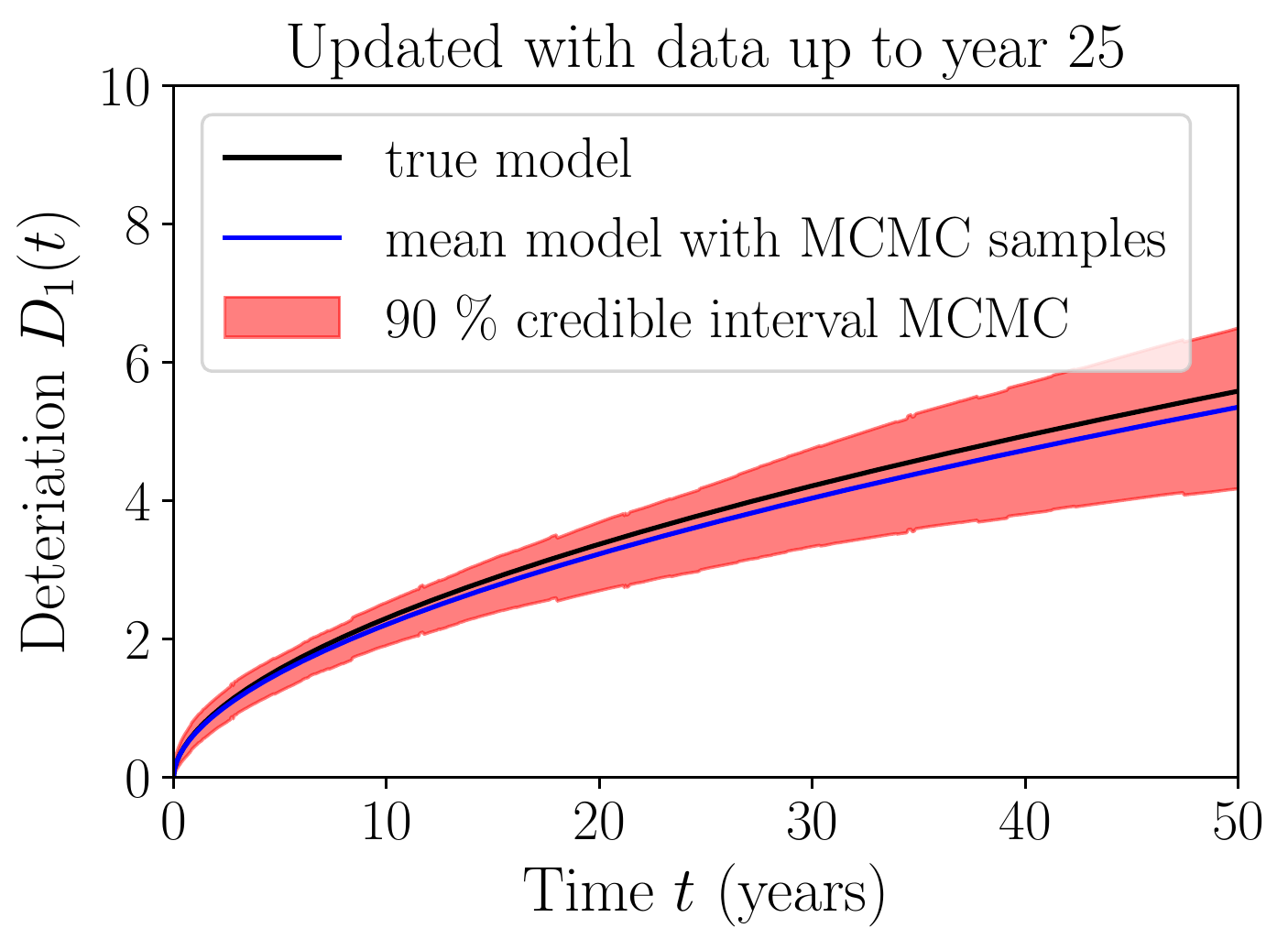}  
		\end{subfigure}
		\begin{subfigure}{.24\textwidth}
			\centering
			% include second image
			\includegraphics[width=1.\linewidth]{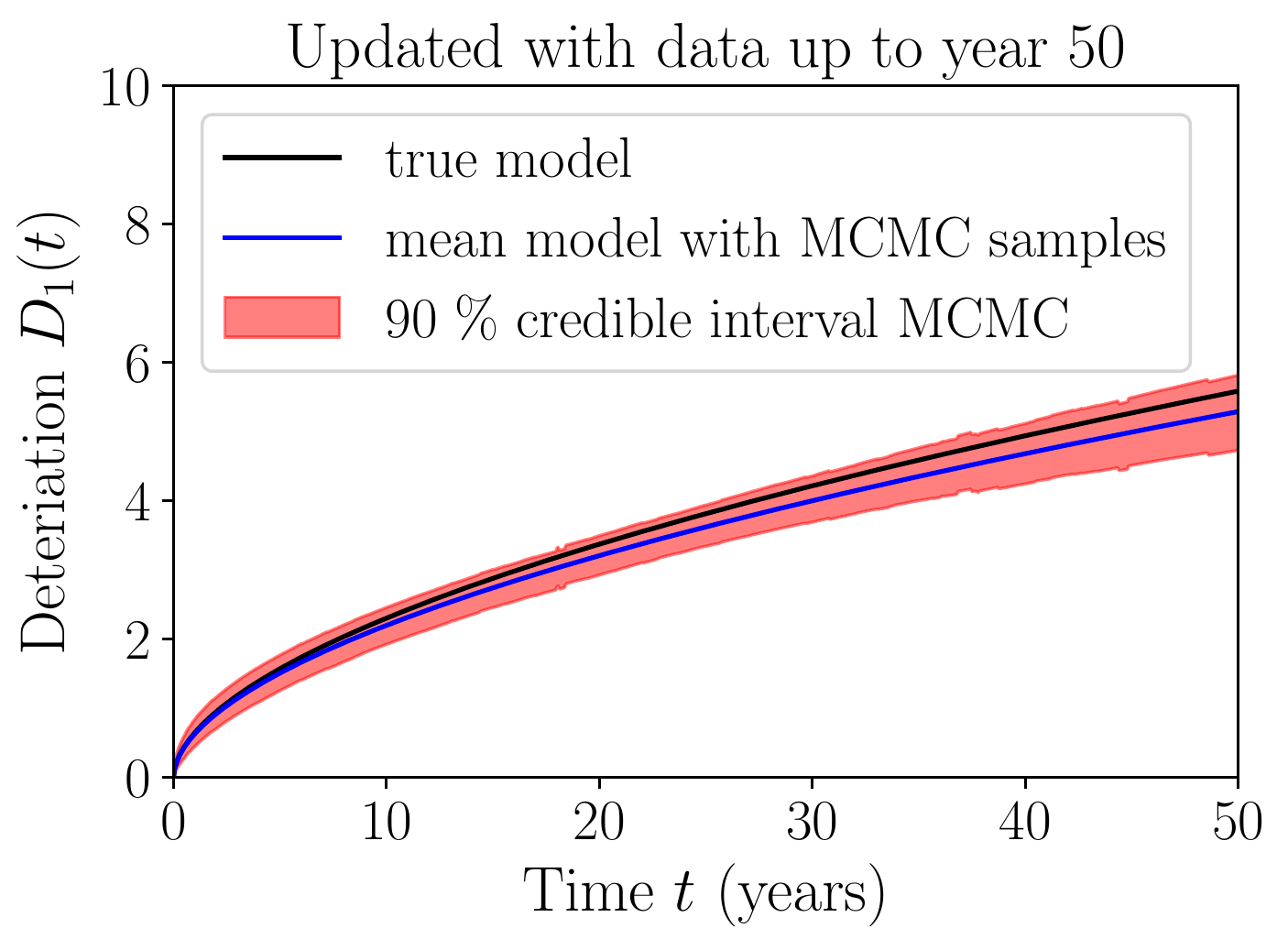}  
		\end{subfigure}
		\begin{subfigure}{.24\textwidth}
			\centering
			% include first image
			\includegraphics[width=1.\linewidth]{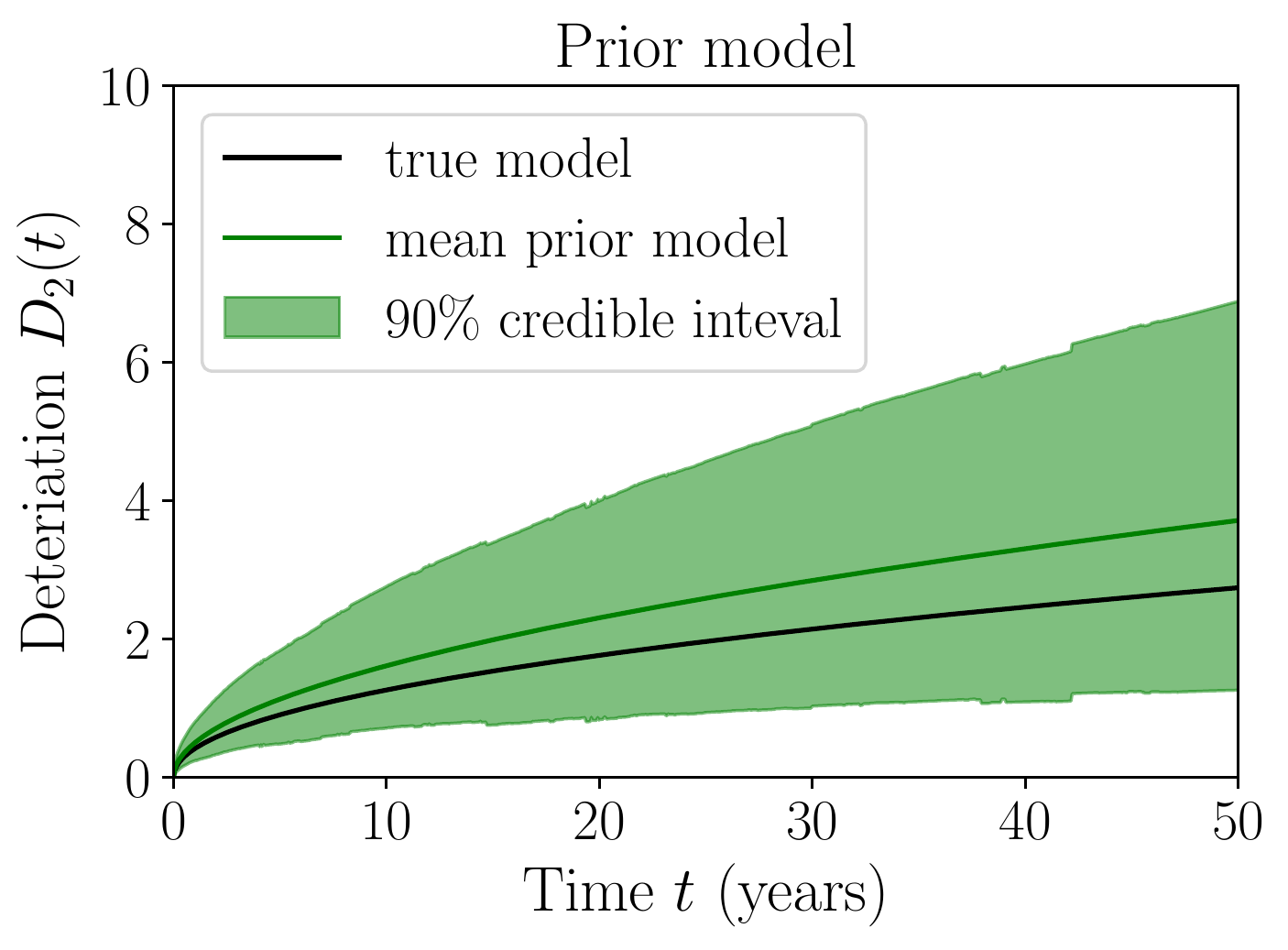}  
		\end{subfigure}
		\begin{subfigure}{.24\textwidth}
			\centering
			% include second image
			\includegraphics[width=1.\linewidth]{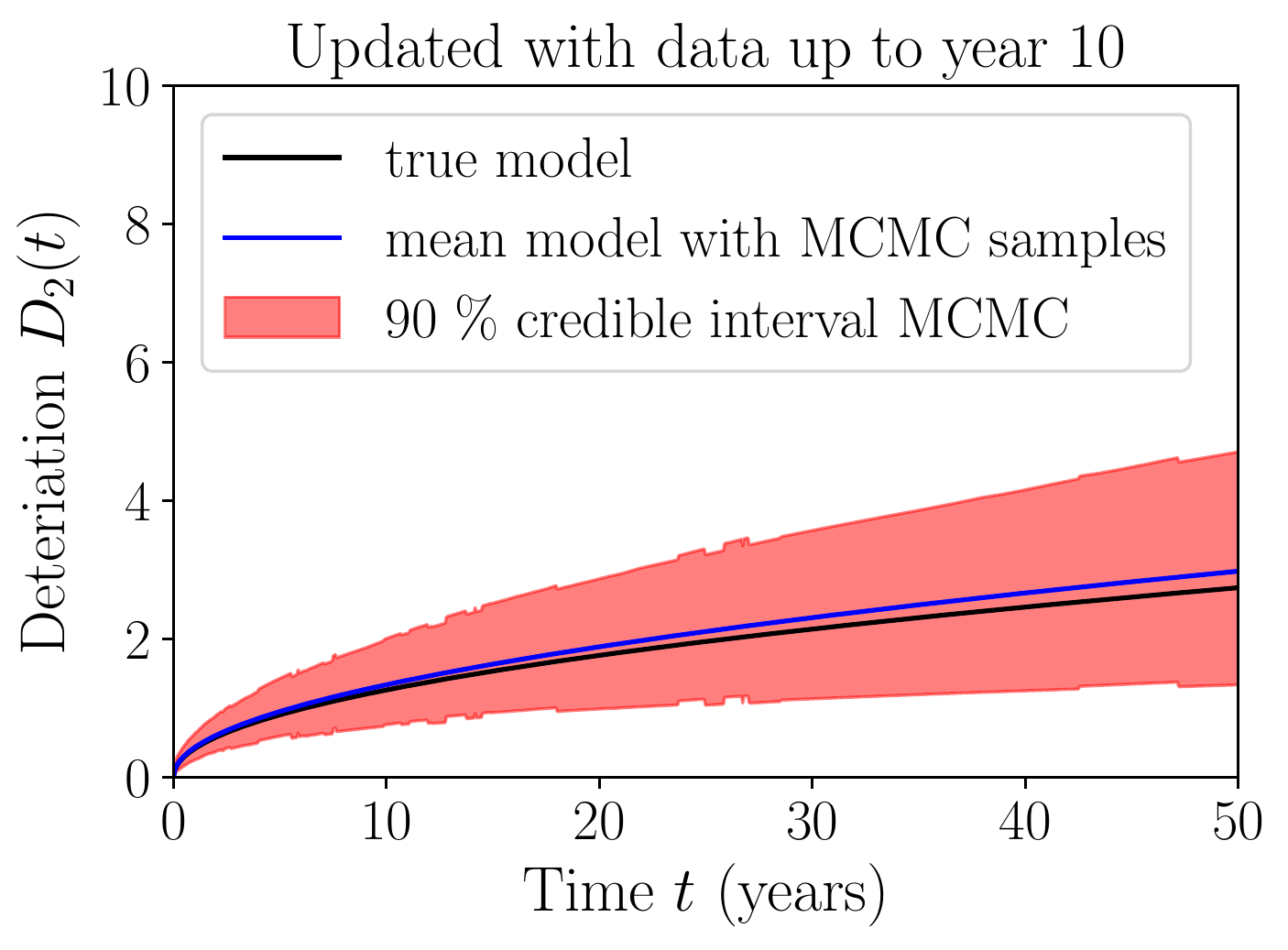}  
		\end{subfigure}
		\begin{subfigure}{.24\textwidth}
			\centering
			% include second image
			\includegraphics[width=1.\linewidth]{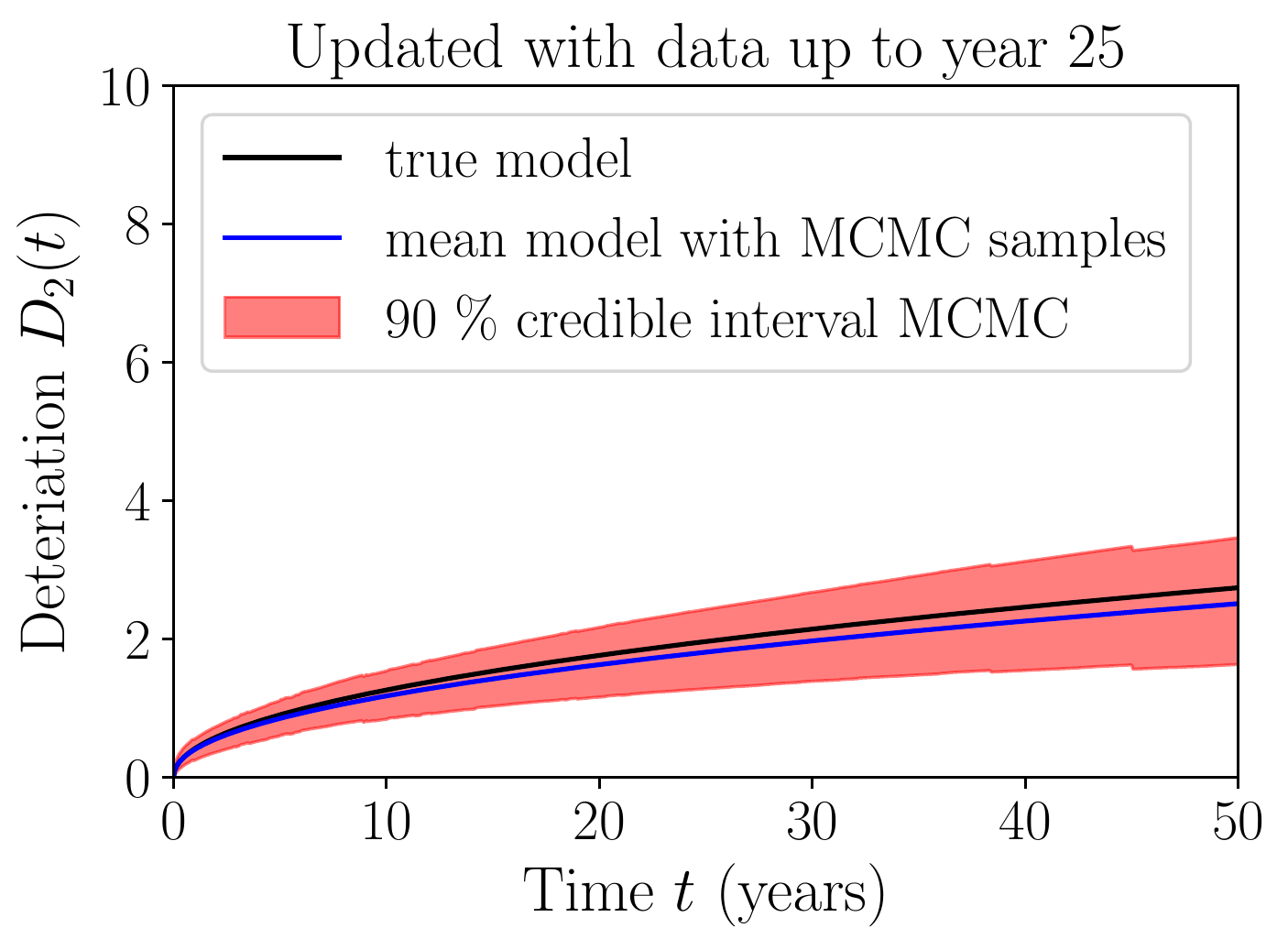}  
		\end{subfigure}
		\begin{subfigure}{.24\textwidth}
			\centering
			% include second image
			\includegraphics[width=1.\linewidth]{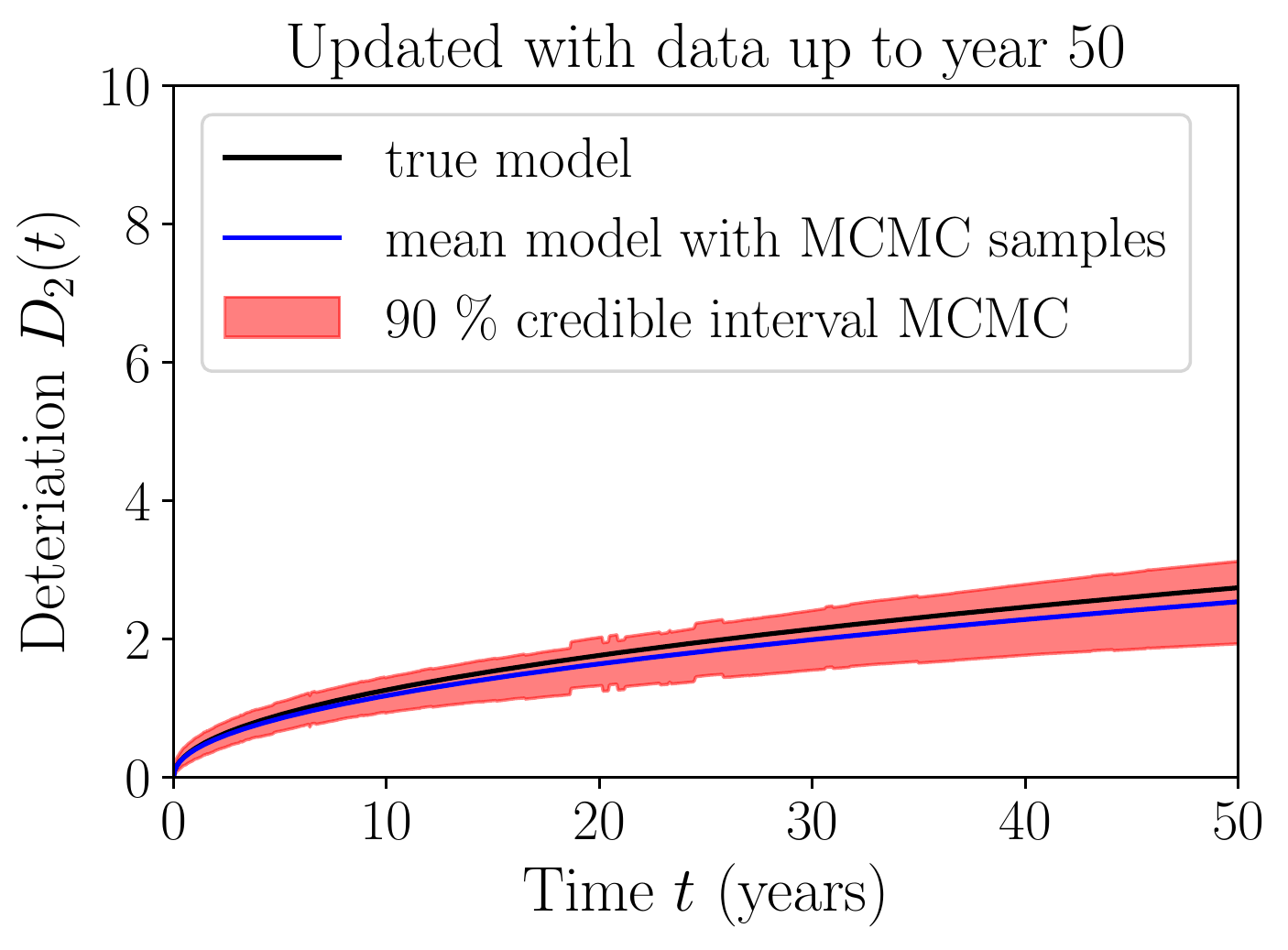}  
		\end{subfigure}
		\caption{Sequential Bayesian learning of the two corrosion deterioration models}
		\label{f:updating}
	\end{figure}

 	For illustration purposes, we draw one sample $\boldsymbol{\theta}^*$, which corresponds to the underlying ``true" deterioration parameter values $A_1^*=0.65$, $B_1^*=0.55$, $A_2^*=0.42$ and $B_2^*=0.48$ (``true" deterioration curves can be seen in black in all the subfigures of Figure \ref{f:updating}).

\begin{figure}[!ht]
		\begin{subfigure}{.24\textwidth}
			\centering
			% include first image
			\includegraphics[width=1.\linewidth]{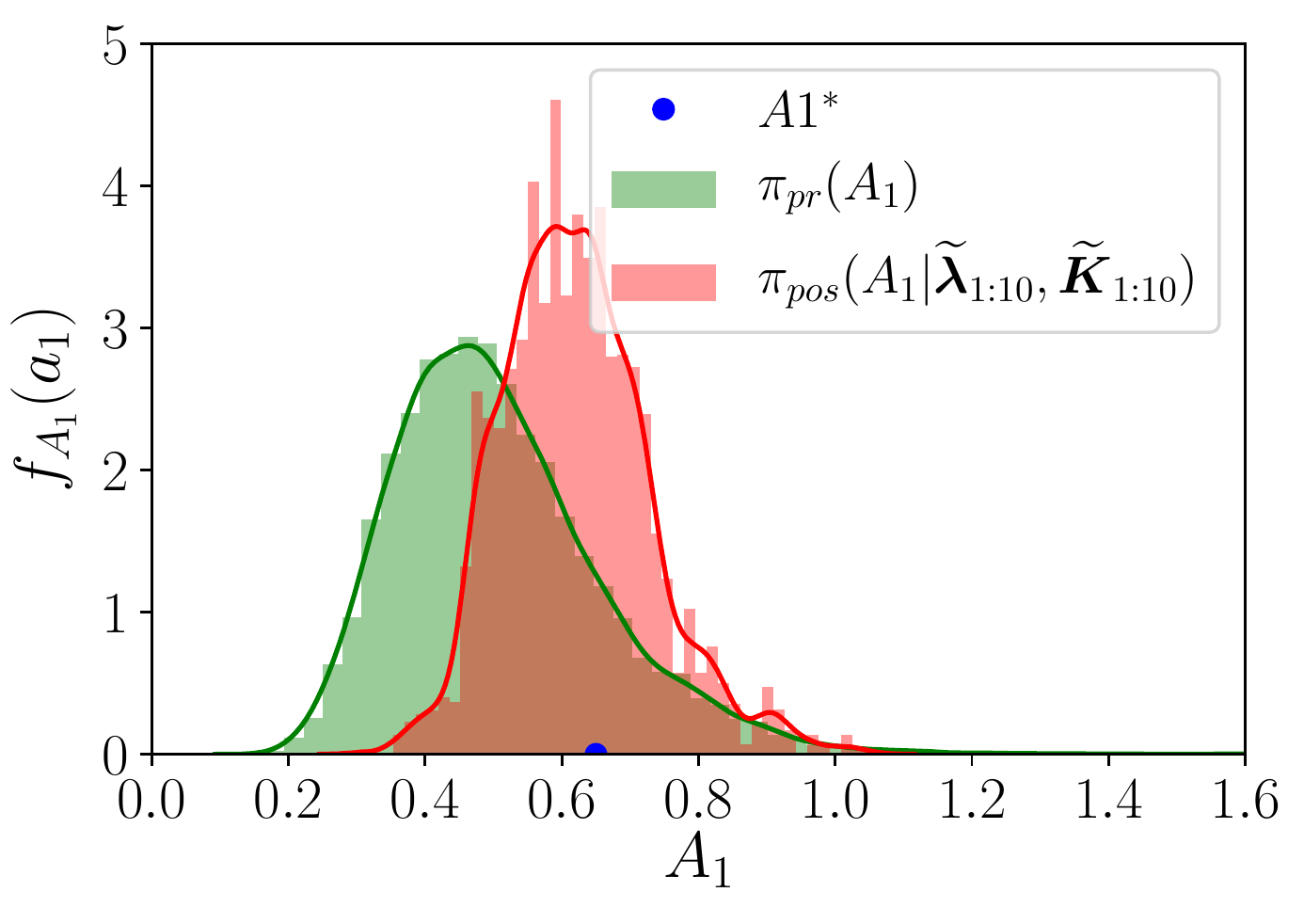}  
		\end{subfigure}
		\begin{subfigure}{.24\textwidth}
			\centering
			% include second image
			\includegraphics[width=1.\linewidth]{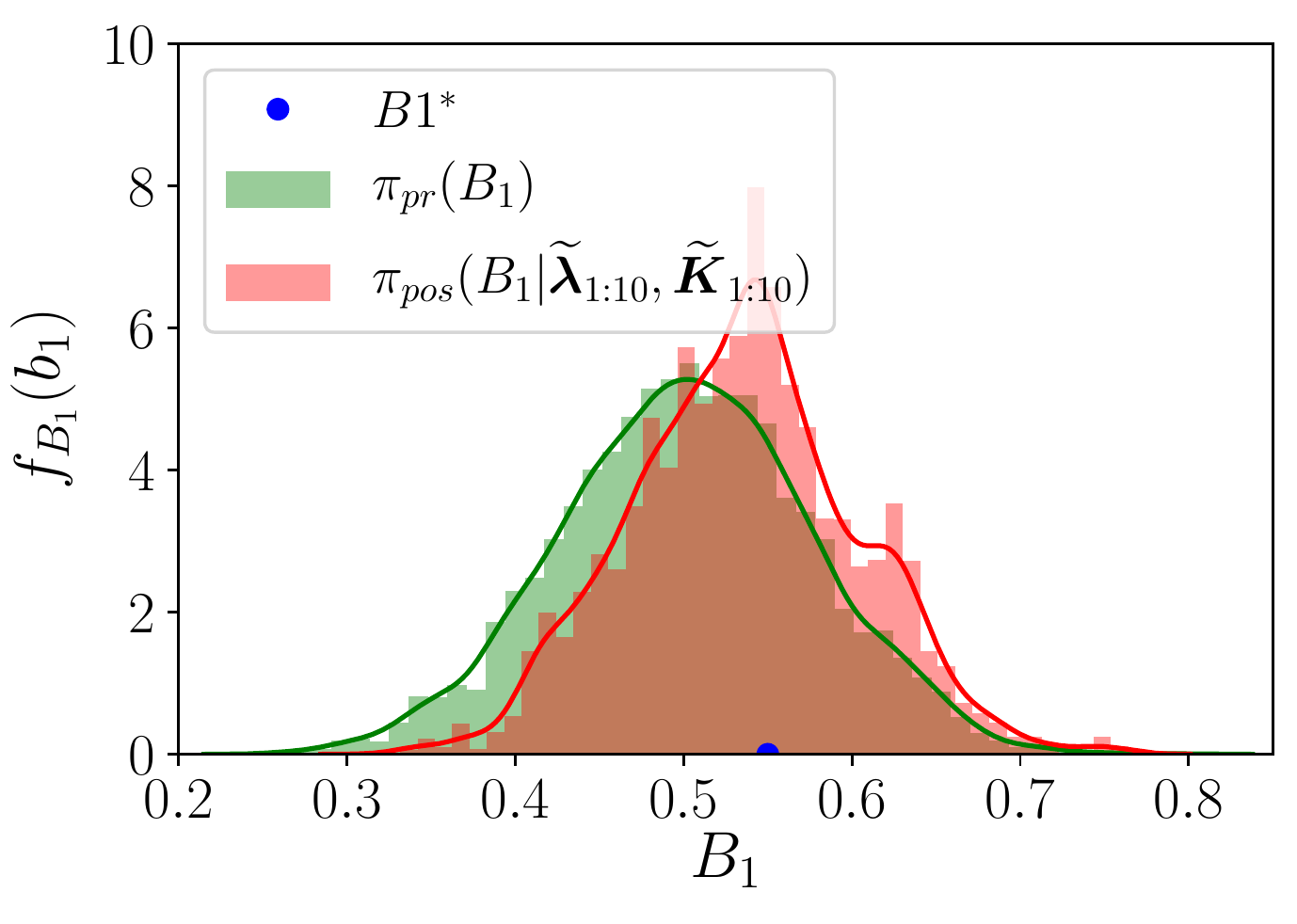}  
		\end{subfigure}
		\begin{subfigure}{.24\textwidth}
			\centering
			% include second image
			\includegraphics[width=1.\linewidth]{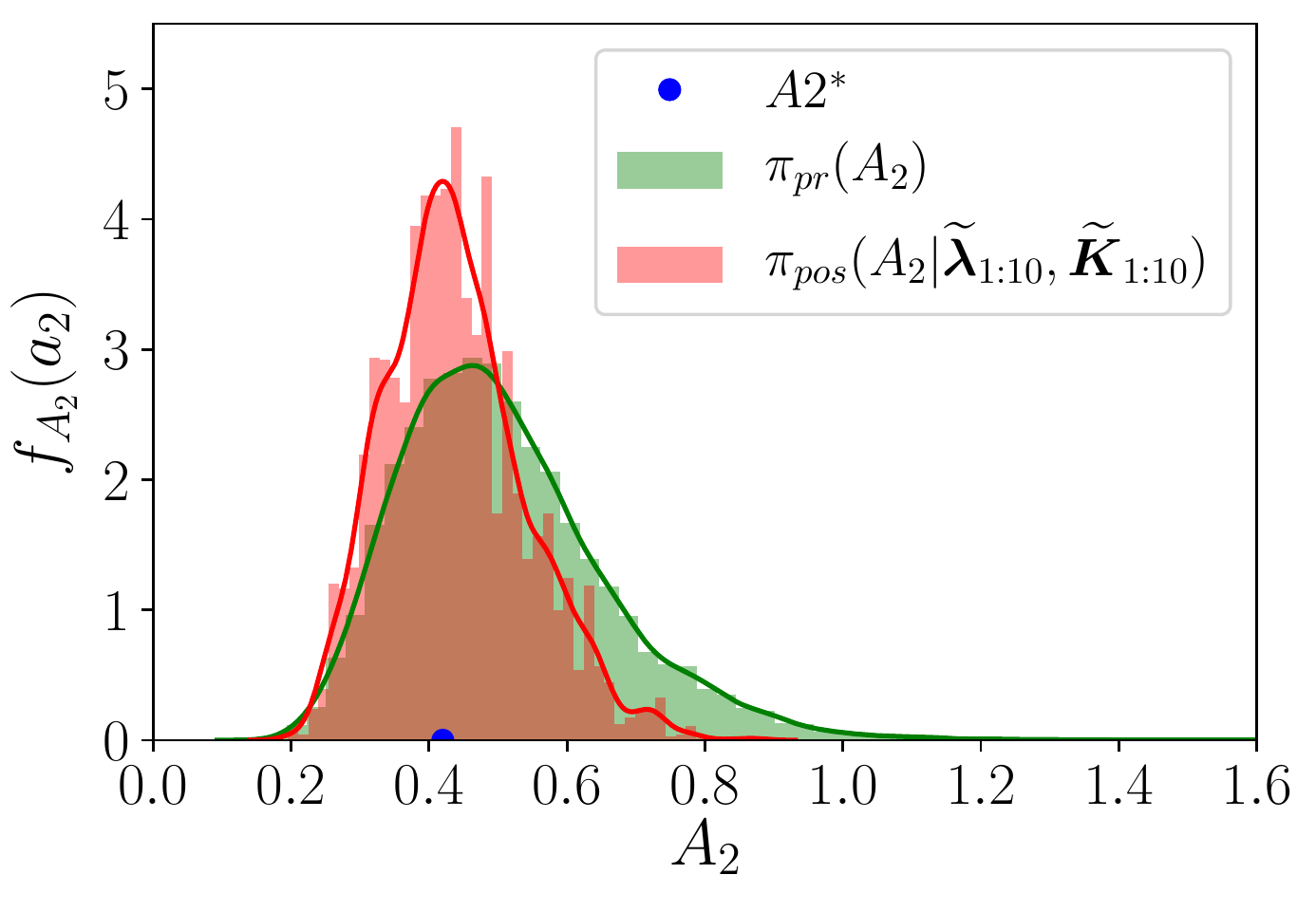}  
		\end{subfigure}
		\begin{subfigure}{.24\textwidth}
			\centering
			% include second image
			\includegraphics[width=1.\linewidth]{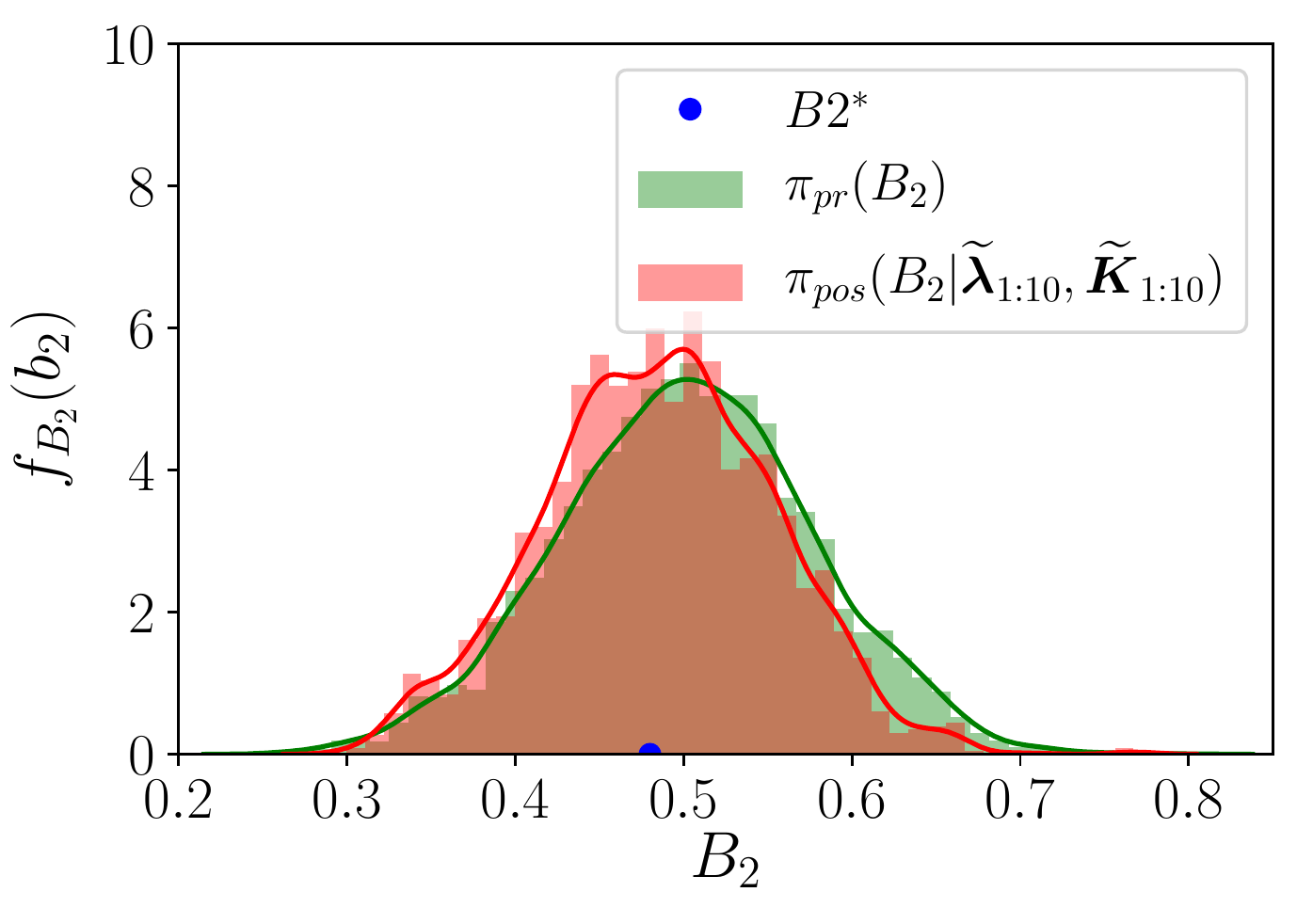}  
		\end{subfigure}
		\caption{Prior PDF and posterior PDF at year 10 for deterioration models parameters ($c_{\lambda m}=c_{\Phi m}=0.02$)}
		\label{f:parameters_learnt_1}
		\begin{subfigure}{.24\textwidth}
			\centering
			% include first image
			\includegraphics[width=1.\linewidth]{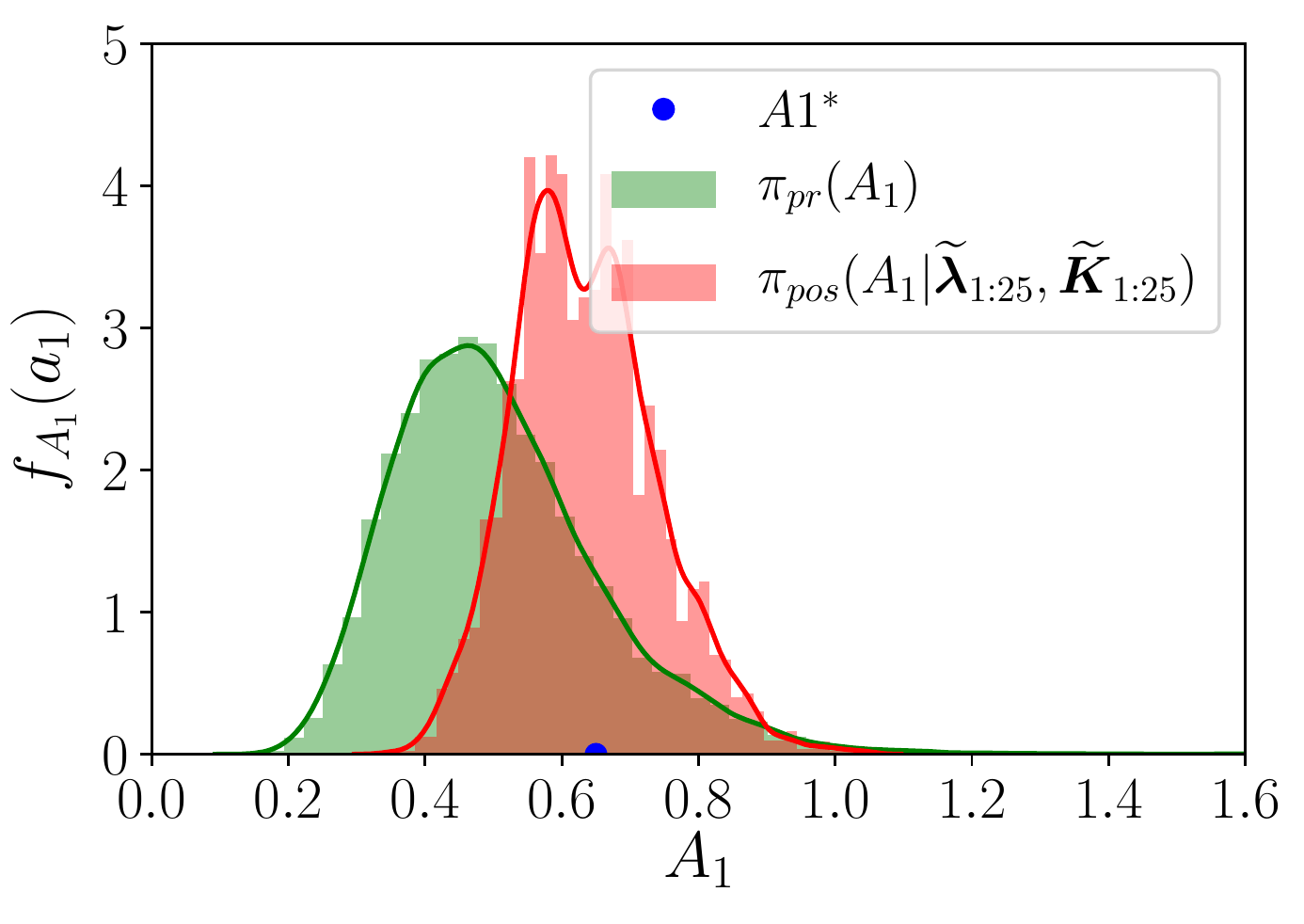}  
		\end{subfigure}
		\begin{subfigure}{.24\textwidth}
			\centering
			% include second image
			\includegraphics[width=1.\linewidth]{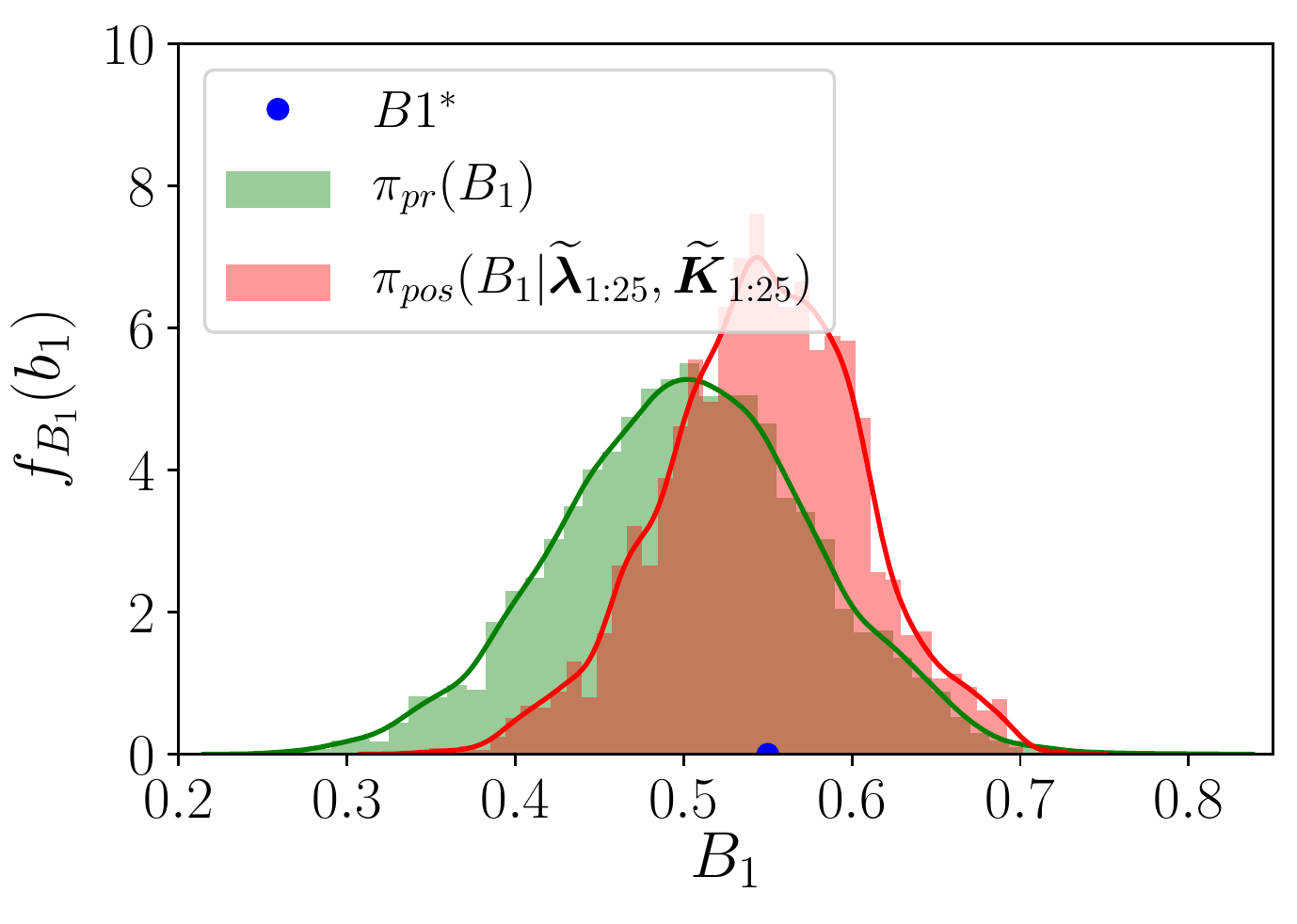}  
		\end{subfigure}
		\begin{subfigure}{.24\textwidth}
			\centering
			% include second image
			\includegraphics[width=1.\linewidth]{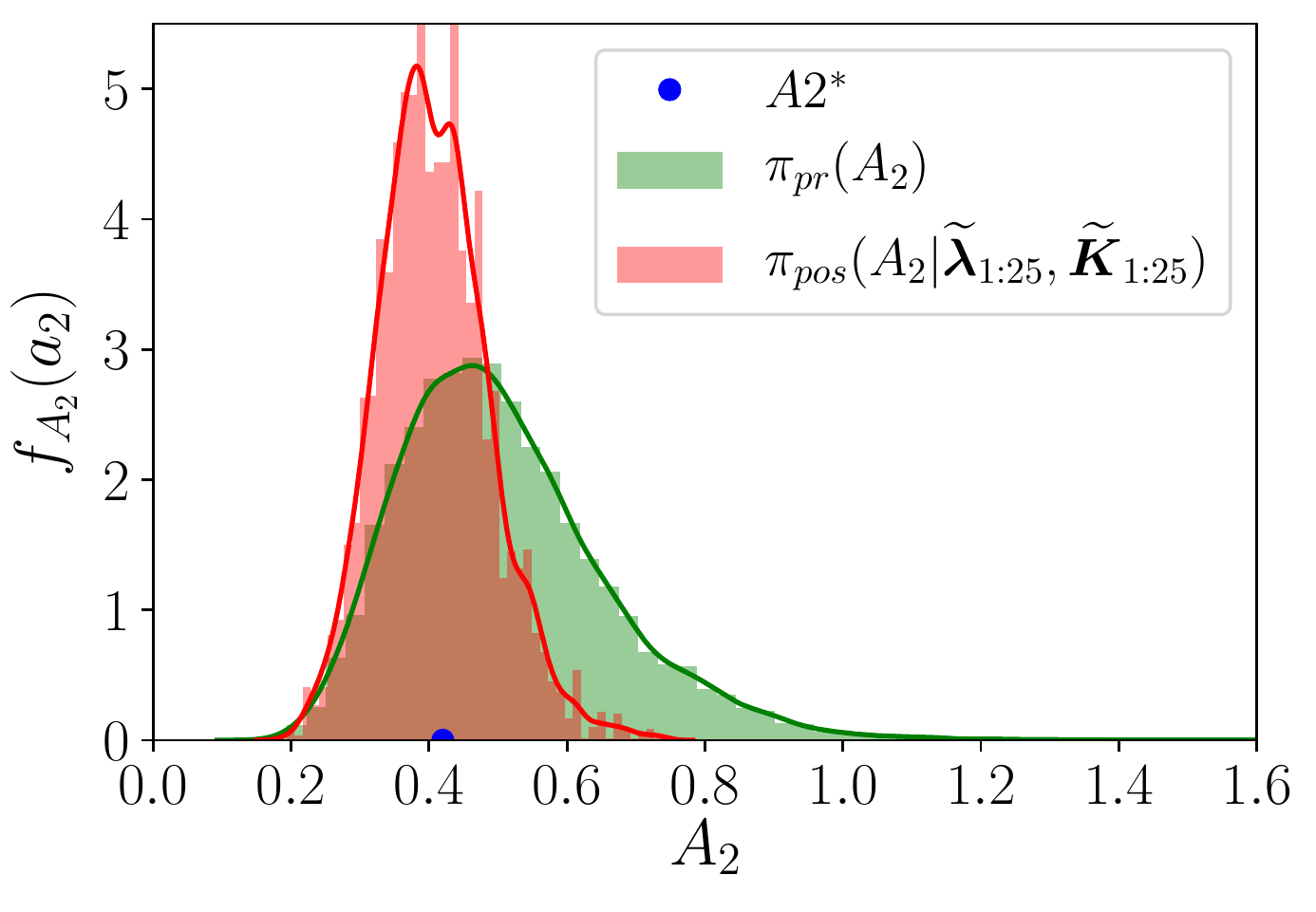}  
		\end{subfigure}
		\begin{subfigure}{.24\textwidth}
			\centering
			% include second image
			\includegraphics[width=1.\linewidth]{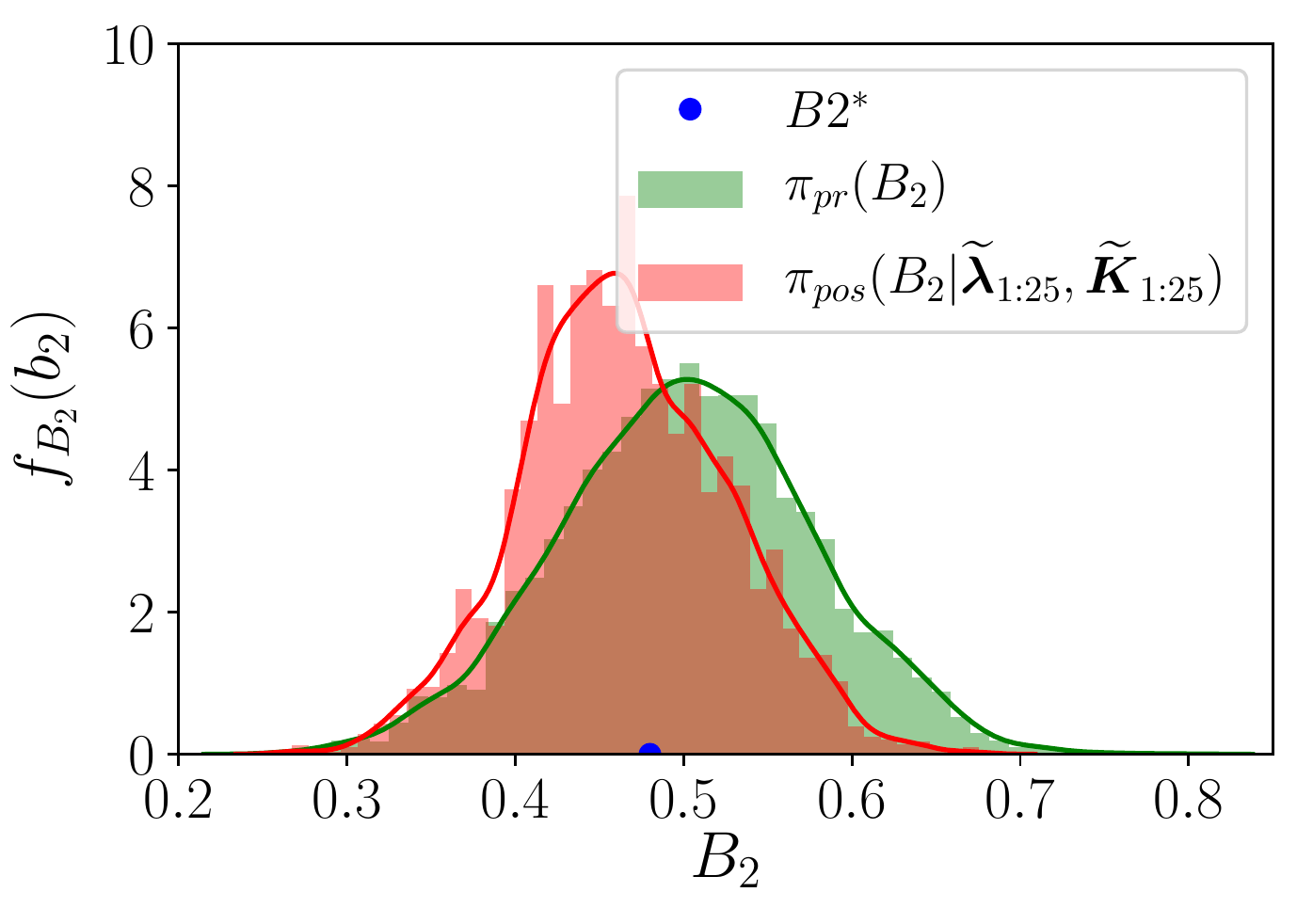}  
		\end{subfigure}
		\caption{Prior PDF and posterior PDF at year 25 for deterioration models parameters ($c_{\lambda m}=c_{\Phi m}=0.02$)}
		\label{f:parameters_learnt_2}
		\begin{subfigure}{.24\textwidth}
			\centering
			% include first image
			\includegraphics[width=1.\linewidth]{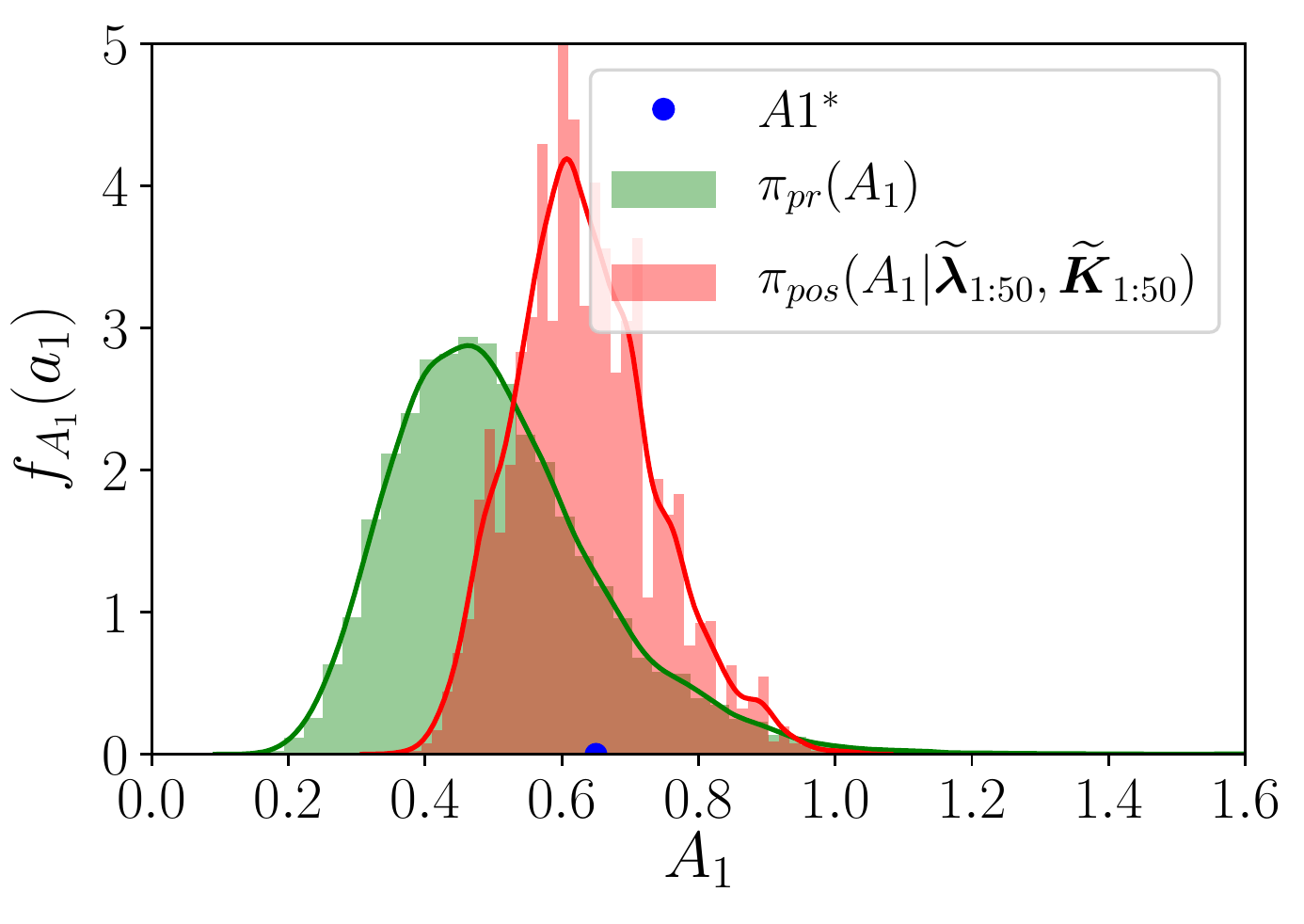}  
		\end{subfigure}
		\begin{subfigure}{.24\textwidth}
			\centering
			% include second image
			\includegraphics[width=1.\linewidth]{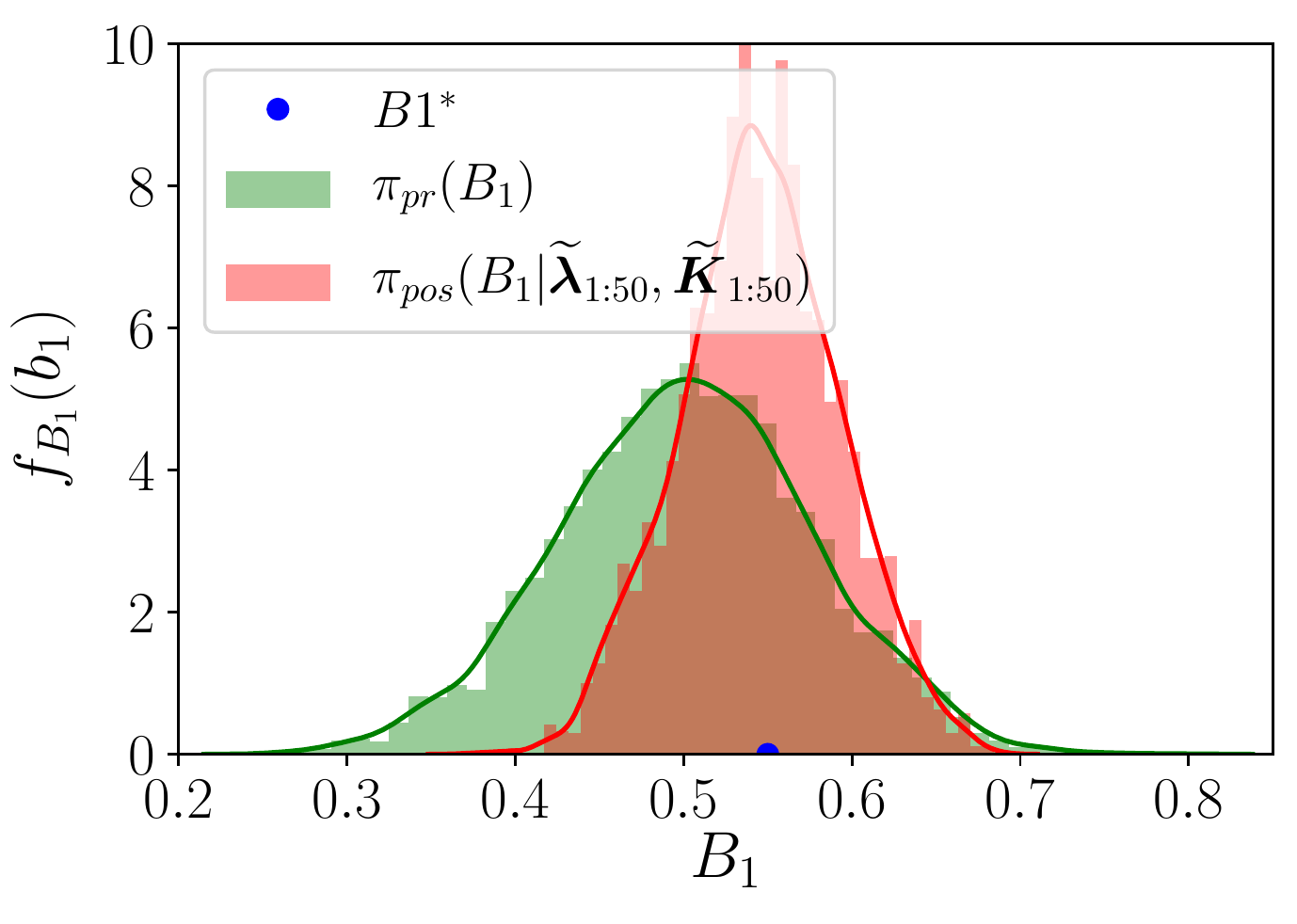}  
		\end{subfigure}
		\begin{subfigure}{.24\textwidth}
			\centering
			% include second image
			\includegraphics[width=1.\linewidth]{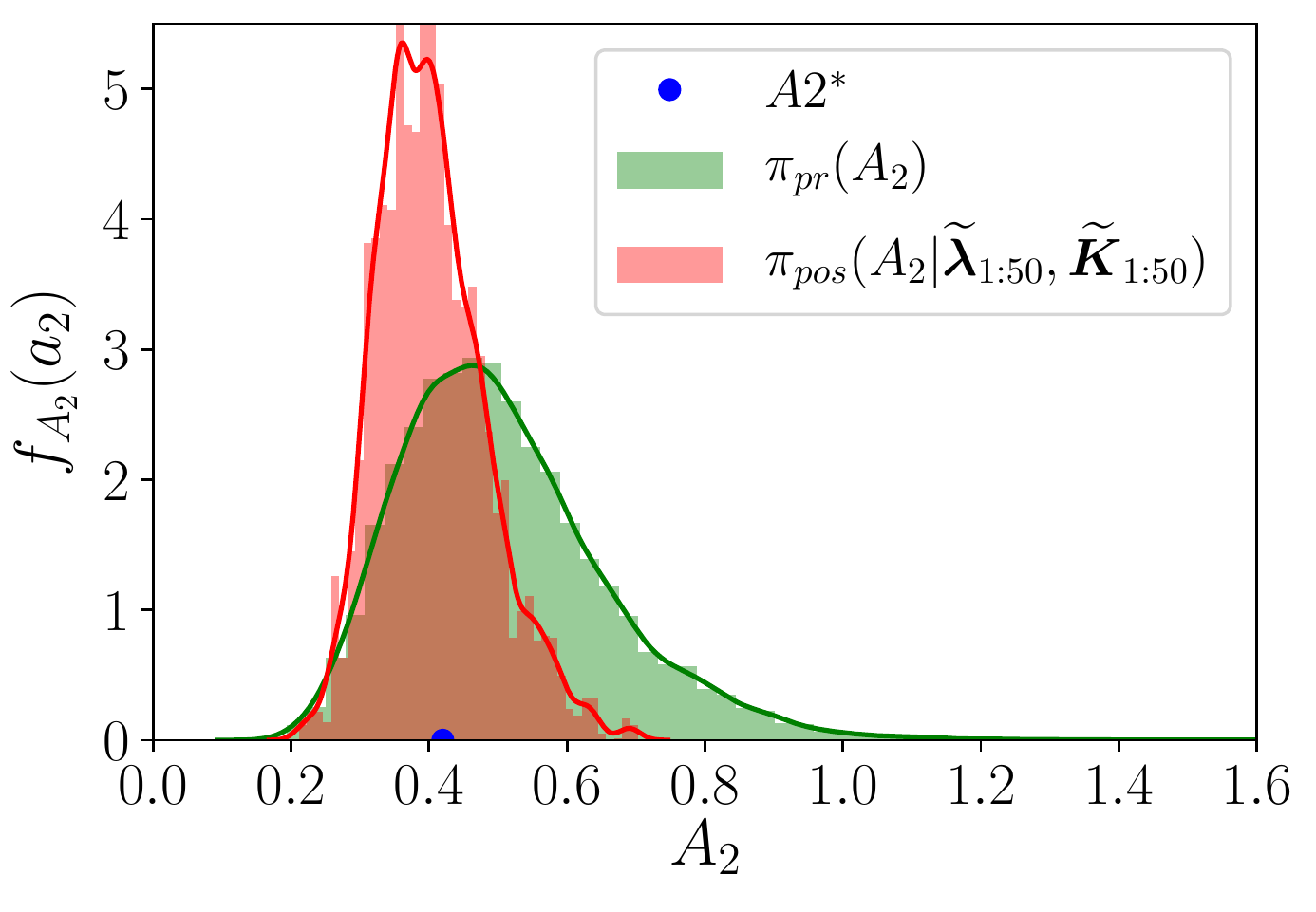}  
		\end{subfigure}
		\begin{subfigure}{.24\textwidth}
			\centering
			% include second image
			\includegraphics[width=1.\linewidth]{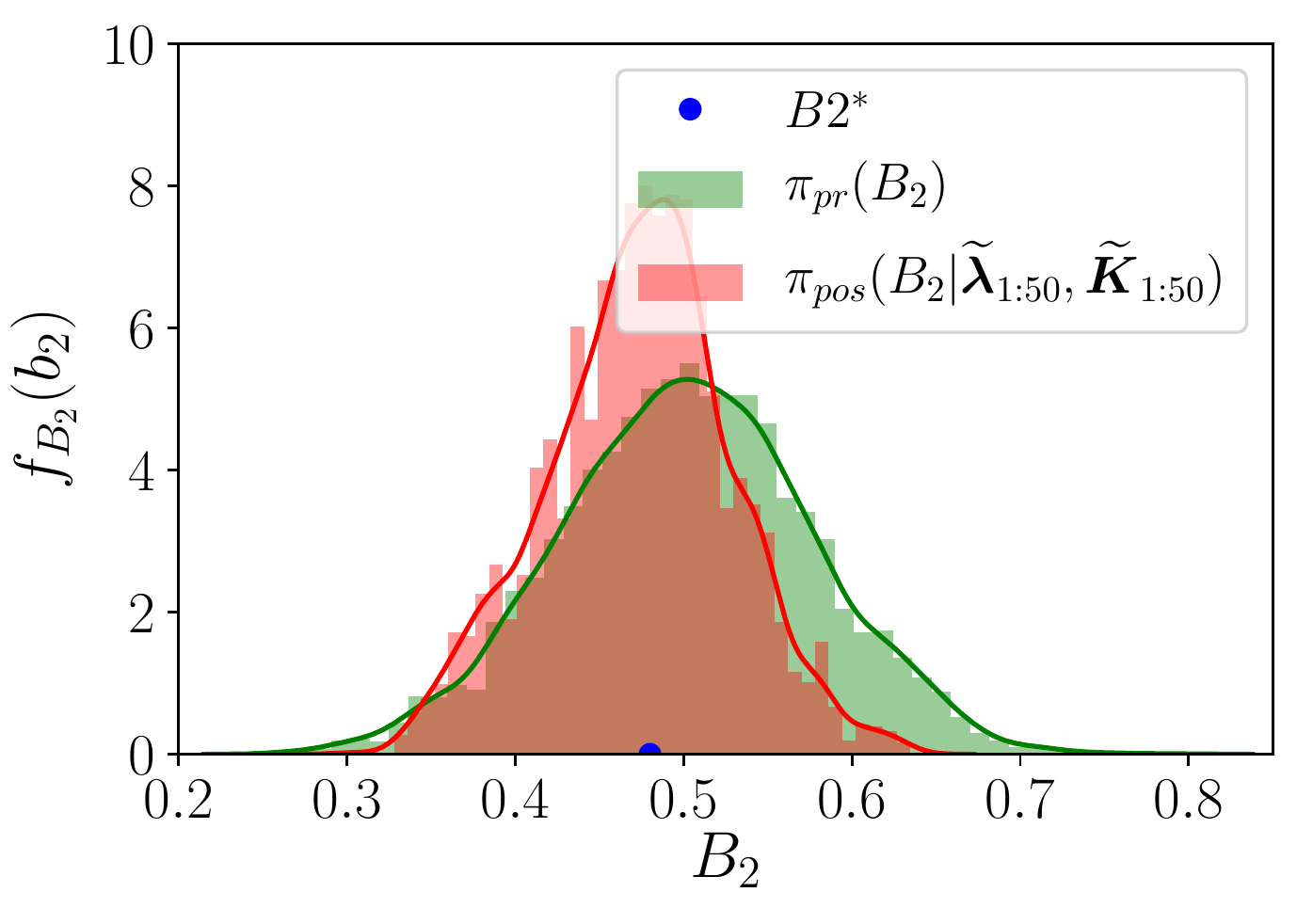}  
		\end{subfigure}
		\caption{Prior PDF and posterior PDF at year 50 for deterioration models parameters ($c_{\lambda m}=c_{\Phi m}=0.02$)}
		\label{f:parameters_learnt_3}
	\end{figure}

	Figures \ref{f:parameters_learnt_1}, \ref{f:parameters_learnt_2}, \ref{f:parameters_learnt_3} demonstrate how the distribution of the deterioration models' parameters is updated, by comparing the prior PDFs with the posterior PDFs at three different time instances. Both factors $c_{\lambda m}$ and $c_{\Phi m}$ in the likelihood function are assumed equal to $0.02$. In Figure \ref{f:updating}, we compare the underlying ``true" deterioration model with the deterioration model estimated using MCS in the prior case, and with the ones estimated with 5000 posterior MCMC samples at three different time instances.
	
\begin{figure}
		\begin{subfigure}{.24\textwidth}
			\centering
			% include first image
			\includegraphics[width=1.\linewidth]{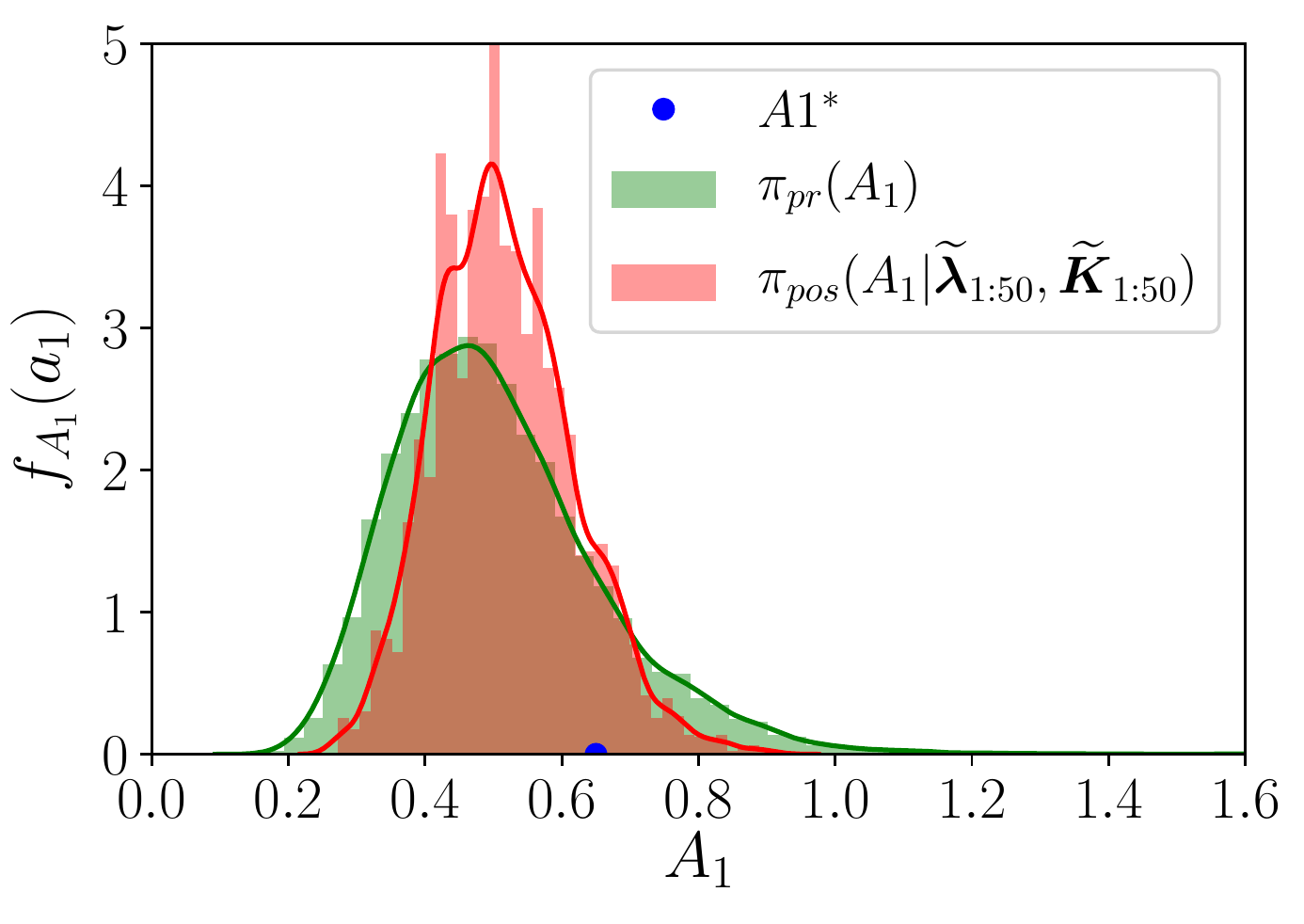}  
		\end{subfigure}
		\begin{subfigure}{.24\textwidth}
			\centering
			% include second image
			\includegraphics[width=1.\linewidth]{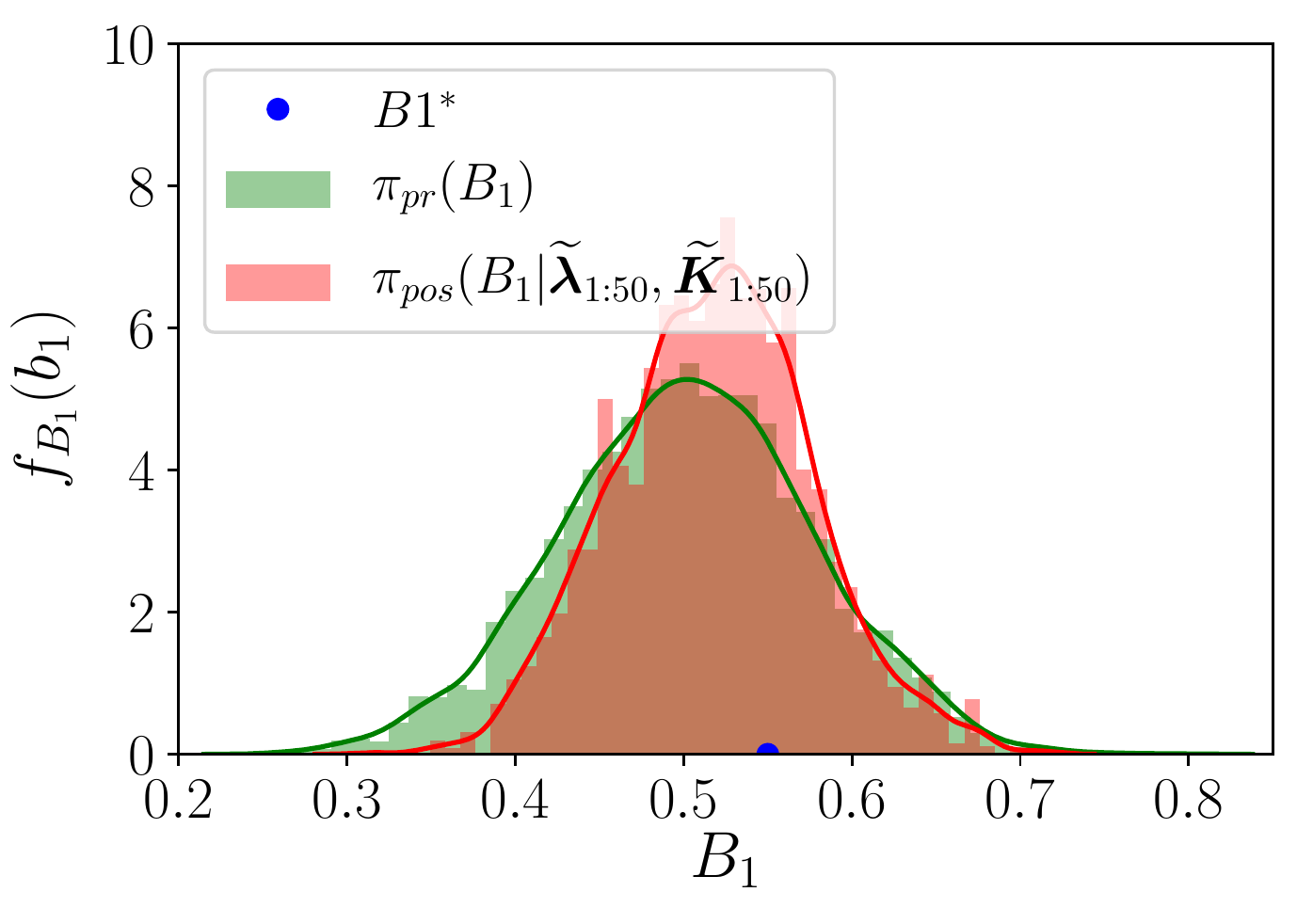}  
		\end{subfigure}
		\begin{subfigure}{.24\textwidth}
			\centering
			% include second image
			\includegraphics[width=1.\linewidth]{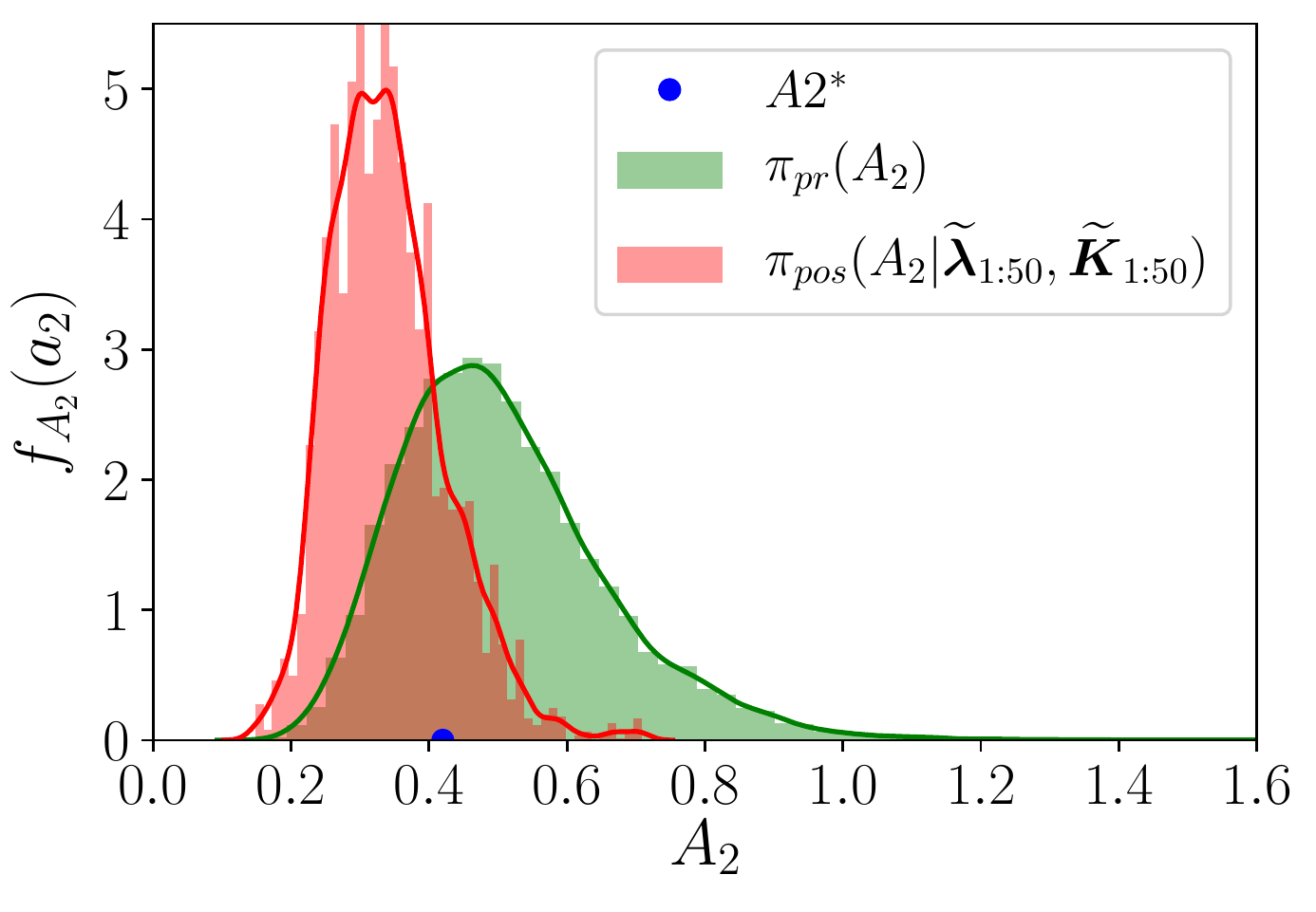}  
		\end{subfigure}
		\begin{subfigure}{.24\textwidth}
			\centering
			% include second image
			\includegraphics[width=1.\linewidth]{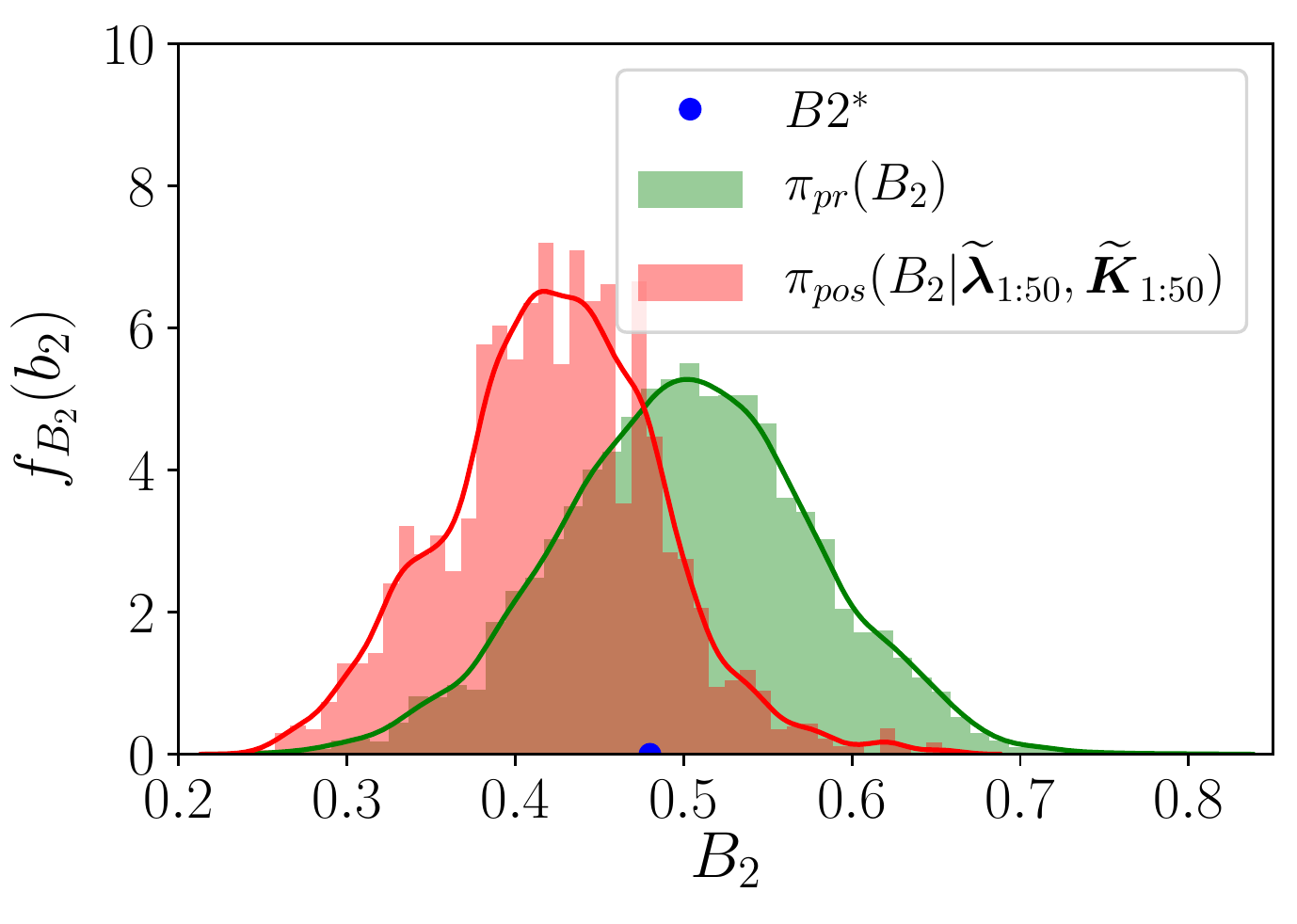}  
		\end{subfigure}
		\caption{Prior PDF and posterior PDF at year 50 for deterioration models parameters ($c_{\lambda m}=c_{\Phi m}=0.05$)}
		\label{f:parameters_learnt_cov_5}
	\end{figure}

	Section \ref{sec:Bayes} discusses the fact that quite often the choice of the magnitude of factors  $c_{\lambda m}$ and $c_{\Phi m}$ for constructing the likelihood function can be arbitrary, since usually very little is known about the magnitude of the total prediction error. Figure \ref{f:parameters_learnt_cov_5} attempts to demonstrate how crucial this choice can be for the results of the Bayesian updating, by performing it additionally for $c_{\lambda m}=c_{\Phi m}=0.05$. Comparing Figure \ref{f:parameters_learnt_cov_5} to Figure \ref{f:parameters_learnt_3} (both at year 50), it can be clearly observed that the posterior distribution of the deterioration model parameters that one learns is significantly affected by the choice of these factors.
	
	\subsection{Time-dependent structural reliability and its updating using monitoring data}
	%In Section \ref{subsec: SR_prior} we have described that there are cases when one can determine a deterministic capacity curve  $R(D(\boldsymbol{\theta},t))$ of the damaged structure for any realization of the deterioration $D(\boldsymbol{\theta},t)$. This will be shown in the following for both examples. 
	The uncertain demand acting on the structure is modeled by the maximum load in a one-year time interval with a Gumbel distribution (left subfigure of Figure \ref{f:capacity_Pr_F_scour}). The parameters of the Gumbel distribution are chosen such that the probability of failure in the initial undamaged state is equal to $10^{-6}$ and the coefficient of variation of the random load is $20\%$. 
	
	\subsubsection{Bridge system subject to scour deterioration}
\begin{figure}
		\begin{subfigure}{.33\textwidth}
			\centering
			% include first image
			\includegraphics[width=1.\linewidth]{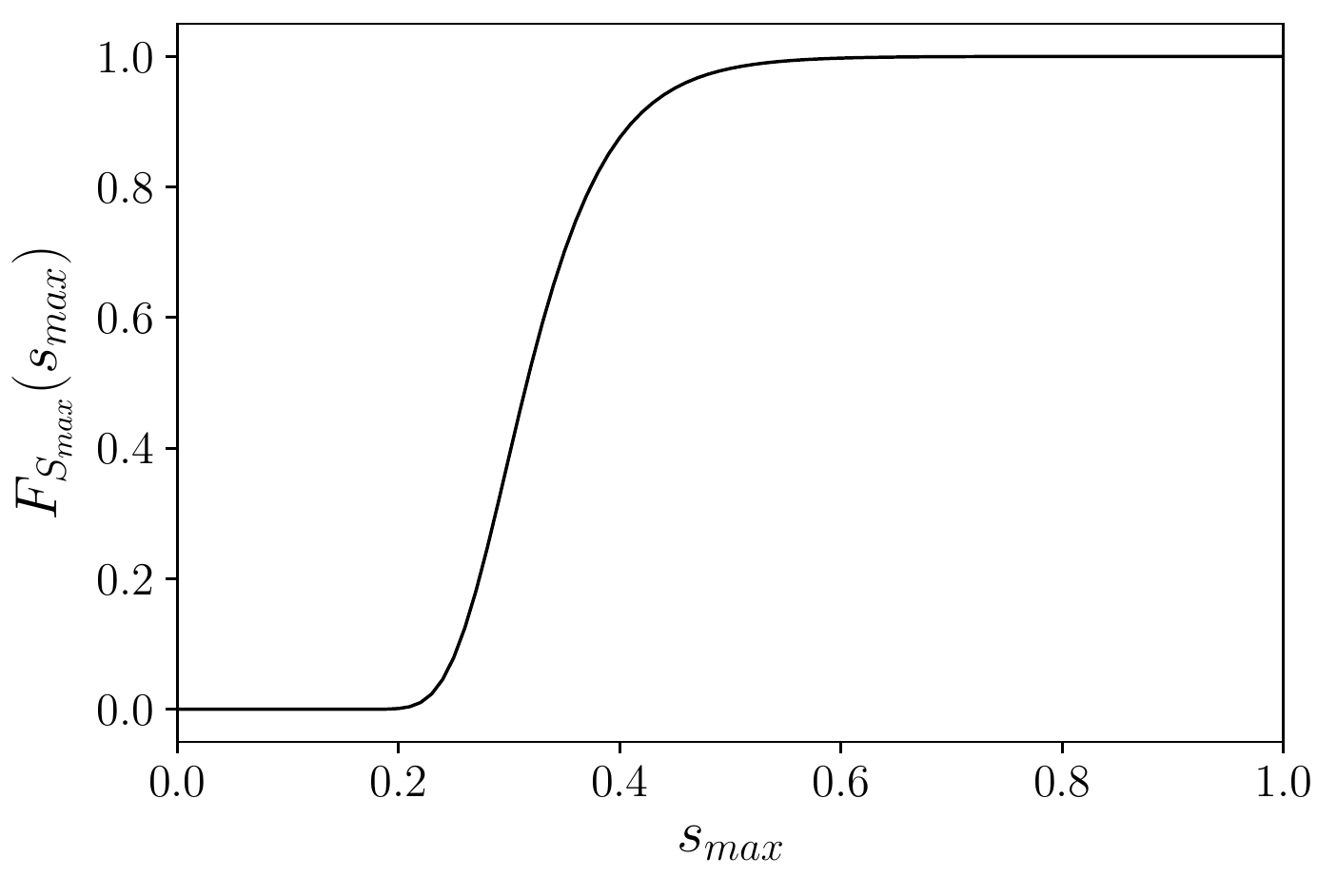}  
		\end{subfigure}
		\begin{subfigure}{.33\textwidth}
			\centering
			% include first image
			\includegraphics[width=1.\linewidth]{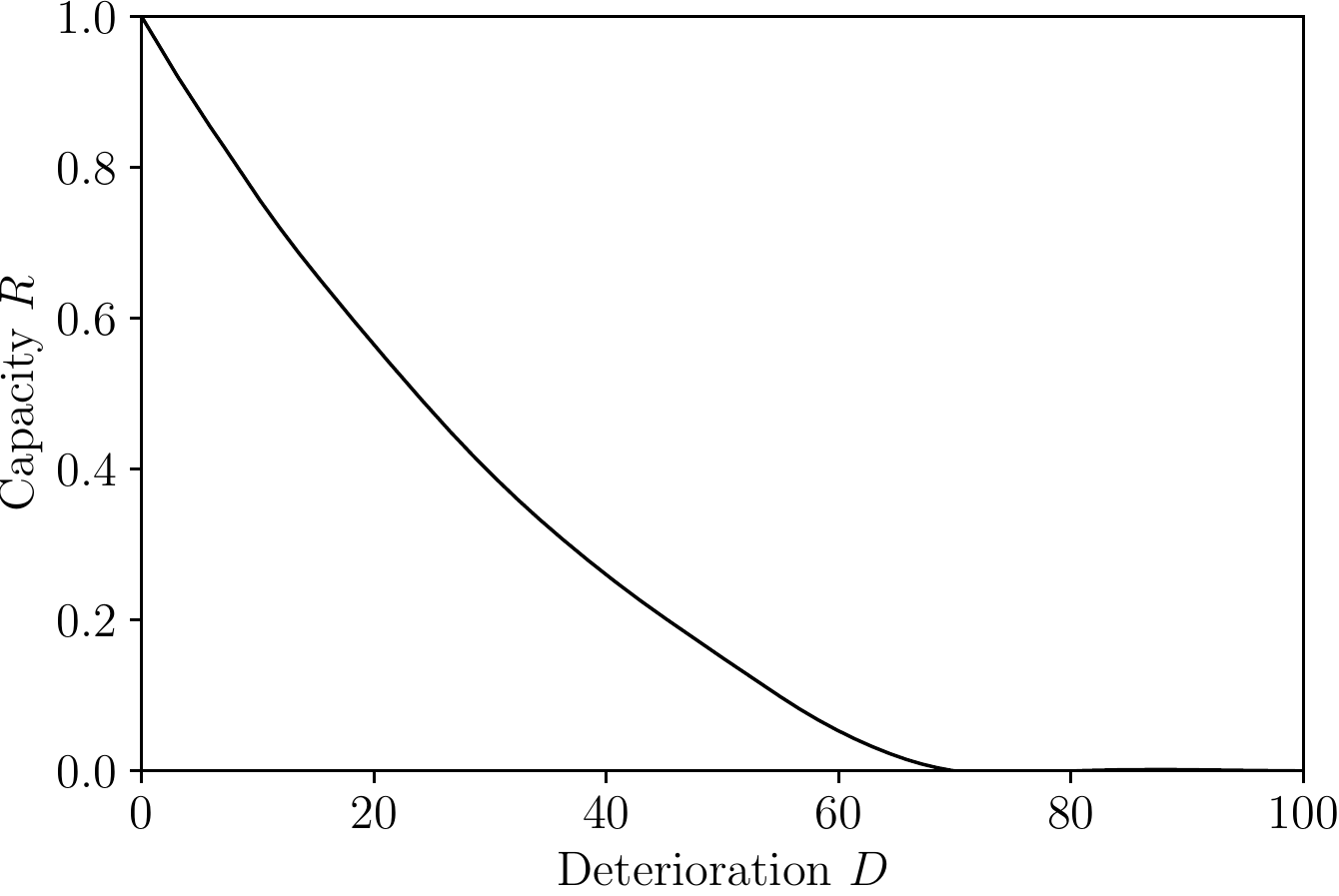}  
		\end{subfigure}
		\begin{subfigure}{.33\textwidth}
			\centering
			% include second image
			\includegraphics[width=1.\linewidth]{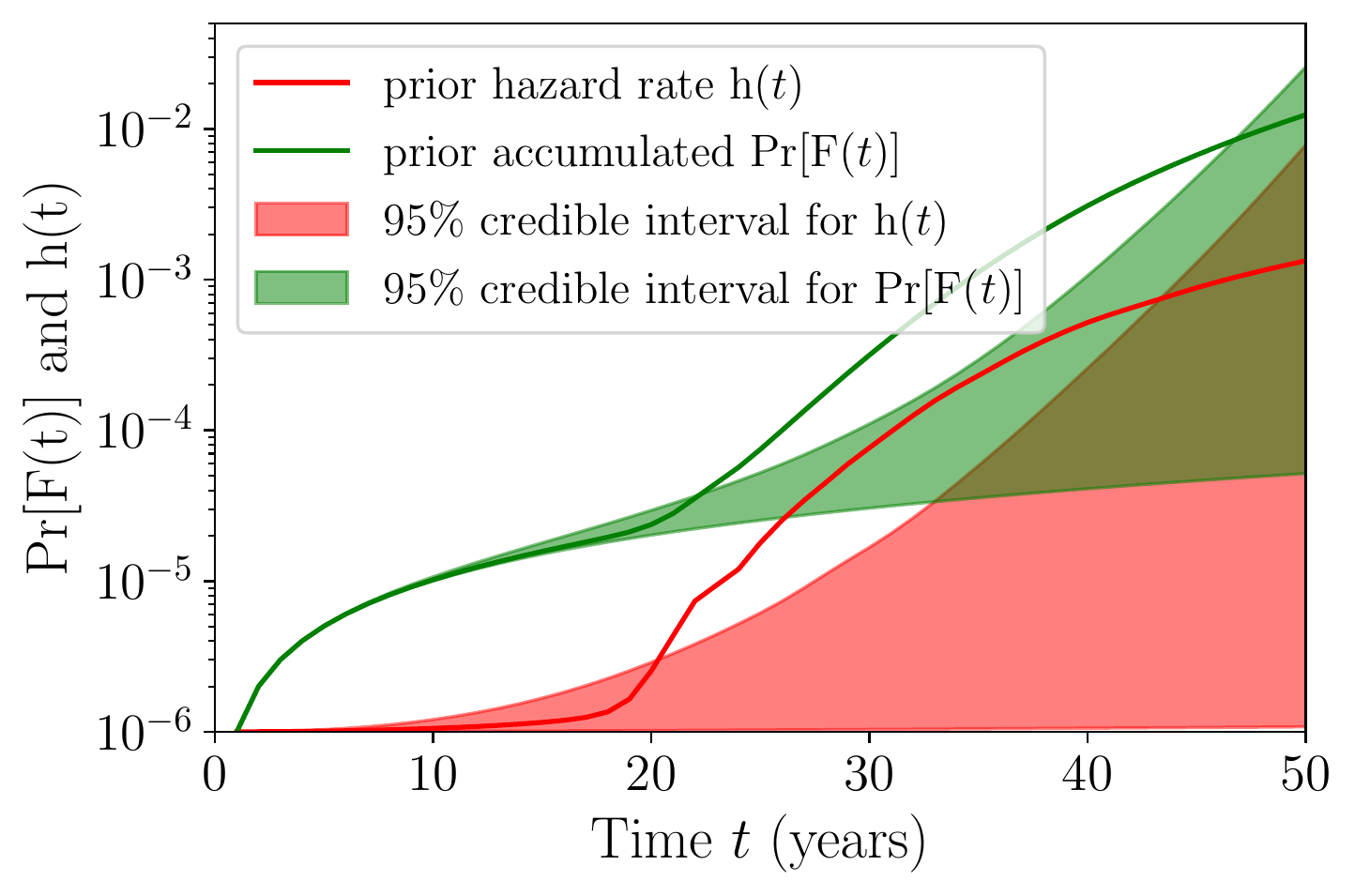}  
		\end{subfigure}
		\caption{left: CDF of the Gumbel distribution for the load with location $a_n = 0.0509$, scale $b_n= 0.297$. middle: Structural capacity in function of scour deterioration, right: Time-dependent structural reliability curves estimated with the prior model in the scour deterioration case.}
		\label{f:capacity_Pr_F_scour}
	\end{figure}

	The deterministic capacity curve $R(D(\boldsymbol{\theta},t))$ of the damaged structure for any realization of the scour deterioration $D(\boldsymbol{\theta},t)$ can be seen in the middle panel of Figure \ref{f:capacity_Pr_F_scour}. To determine this curve, we consider that when scour damage occurs in the middle support, the critical quantity that increases is the normal stress at the middle of the second, slightly longer, midspan. We create a fine one-dimensional grid of possible values as input for $D(\boldsymbol{\theta},t)$, for each of those we run a static analysis of our model, and we evaluate the loss of load bearing capacity of the structure relative to the initial undamaged state. 

	In the right panel of Figure \ref{f:capacity_Pr_F_scour} we plot the time-dependent accumulated probability of failure and the hazard function in the prior case, together with the 95\% credible intervals, estimated using $10^4$ prior samples. Because of the skewness of the assumed prior deterioration model, the mean estimated curves are not contained within the 95\% credible intervals.
	
	\subsubsection{Bridge system subject to corrosion deterioration}
\begin{figure}[ht]
		\begin{subfigure}{.33\textwidth}
			\centering
			% include first image
			\includegraphics[width=0.90\linewidth]{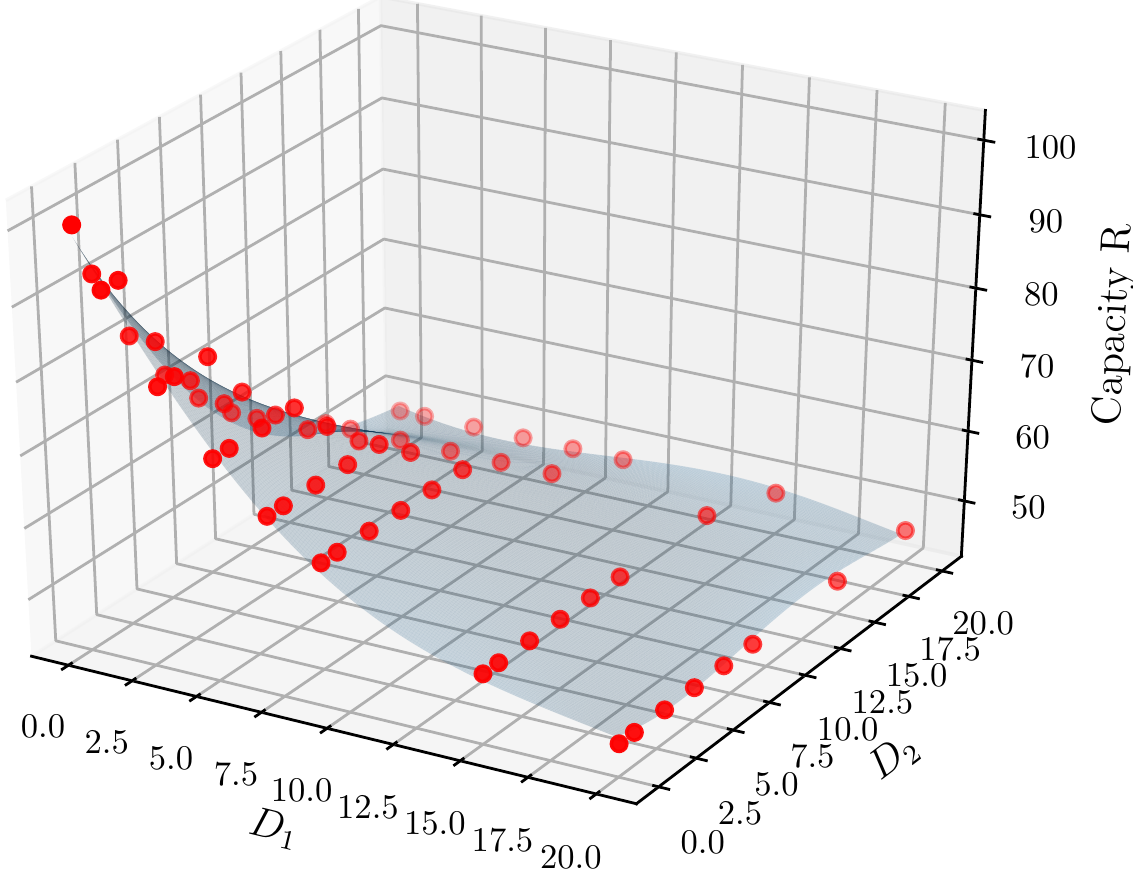}  
		\end{subfigure}
		\begin{subfigure}{.33\textwidth}
			\centering
			% include first image
			\includegraphics[width=1.05\linewidth]{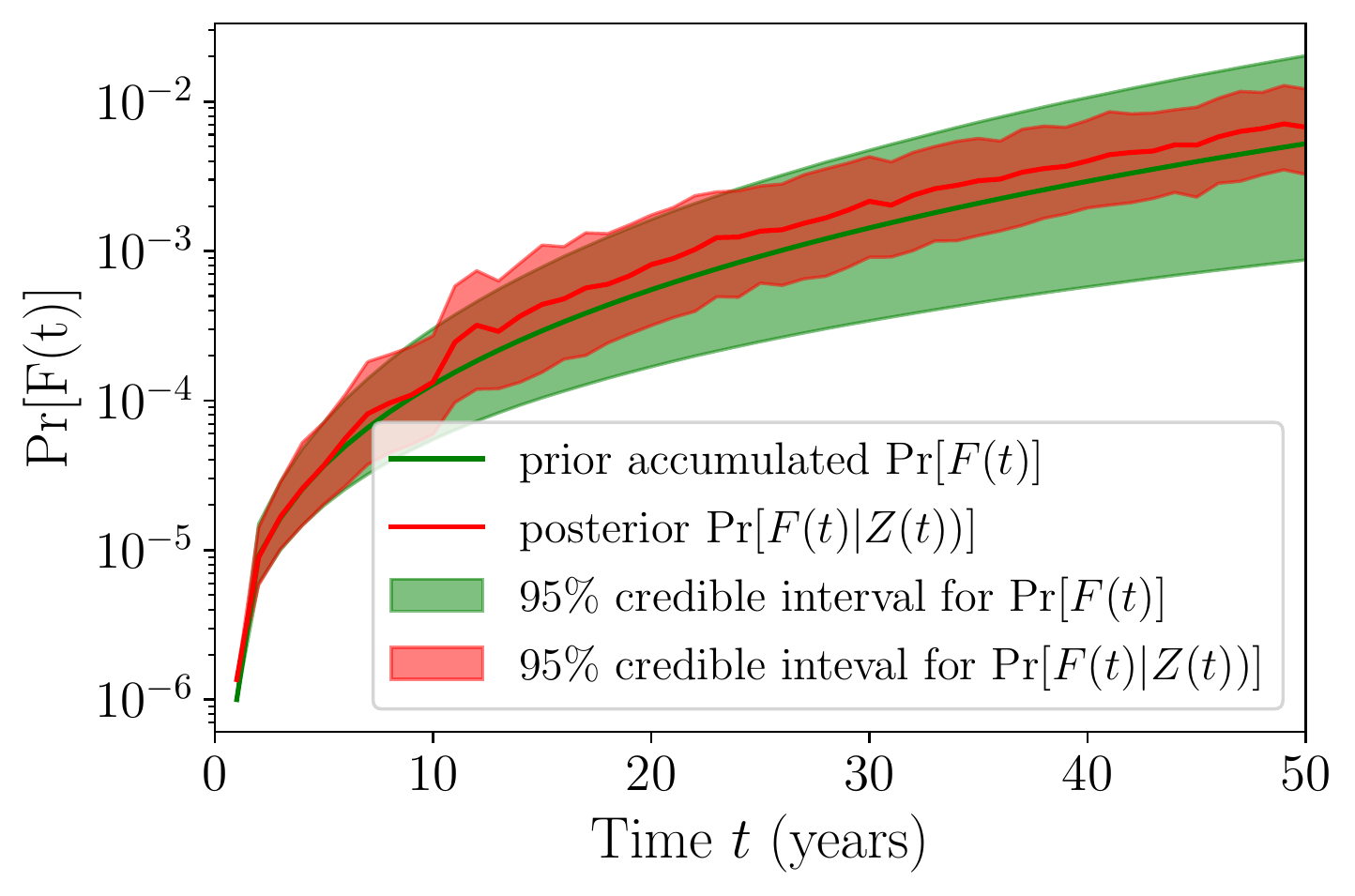}  
		\end{subfigure}
		\begin{subfigure}{.33\textwidth}
			\centering
			% include second image
			\includegraphics[width=1.05\linewidth]{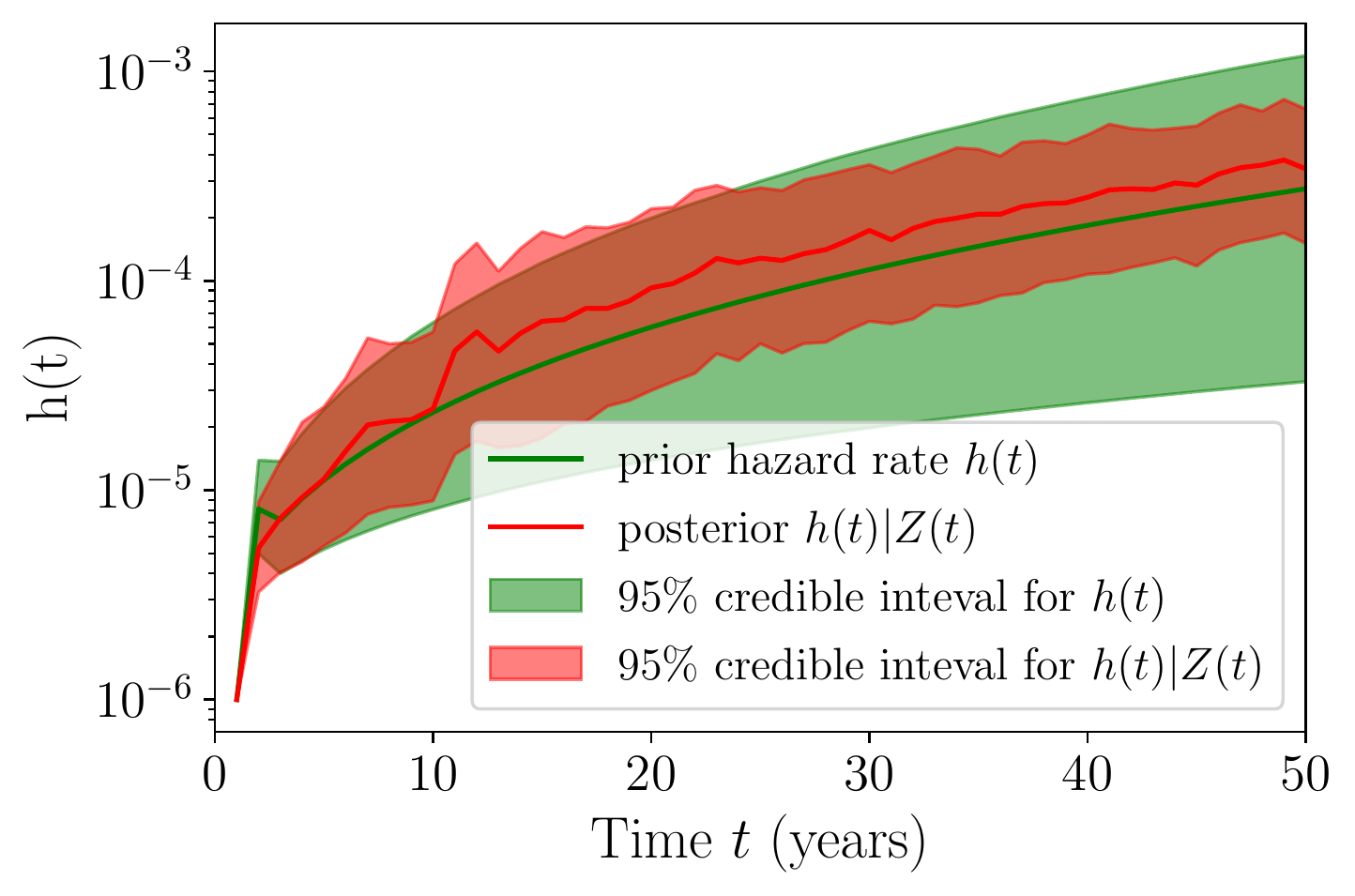}  
		\end{subfigure}
		\caption{left: Polynomial regression response surface for structural capacity in function of corrosion deterioration, right: time dependent structural reliability curves in the prior/posterior corrosion case.}
		\label{f:SR_corrosion_example}
	\end{figure}

	When damage (stiffness reduction) occurs in the elements at the bottom of each midspan, the quantity that increases critically are the normal stresses at the top of each midspan. We create a two-dimensional grid of possible values of the two corrosion deteriorations $D_1$ and $D_2$.  For each of those possible combinations, we run a static analysis with our model, and we evaluate the loss of load bearing capacity of the bridge structure relative to the undamaged state. Eventually we fit a two-dimensional polynomial regression response surface curve that describes $R(\boldsymbol{D}(\boldsymbol{\theta},t))$; it can be seen in the left panel of Figure \ref{f:SR_corrosion_example}. 
	
	As presented in Section \ref{subsec:SR_updating}, learning the parameters of the deterioration models, and the reduction of the uncertainty in their estimation through the sequential acquisition of SHM modal data, affects the estimation of the time-dependent structural reliability. In Figure \ref{f:SR_corrosion_example}, we plot in green the accumulated probability of failure and the hazard rate of the bridge structure in the case of using the prior deterioration model, and we compare it with the red plots of the accumulated probability of failure and the hazard rate conditional on the continuous monitoring data (1 data set per year), which correspond to the underlying ``true'' deterioration models described by $A_1^*=0.65$, $B_1^*=0.55$, $A_2^*=0.42$ and $B_2^*=0.48$. The prior estimates are obtained with 5000 Monte Carlo samples following equations (\ref{MCS}), (\ref{hazard_prior}). The posterior estimates are obtained via equations (\ref{accumulated_posterior}), (\ref{hazard_posterior}) using 5000 MCMC samples at each time step. The $95\%$ credible intervals are computed using the Monte Carlo prior samples in the prior case, and the MCMC posterior samples in the posterior case. It is observed that the uncertainty in the estimation of the structural reliability is reduced in the posterior case. This reduction of the uncertainty and the updated estimate of the structural reliability form the basis for the VoI analysis.
	
	\subsection{VoI analysis}
	\label{subsubsec: VOI_results}
	The VoI is computed with equation (\ref{VOI}) following the Bayesian preposterior decision analysis framework presented in Section \ref{sec:LCC}. The expected total life-cycle costs, $\boldsymbol{\text{E}}_{\boldsymbol{\theta}}[C_{\text{tot}} \mid w]$ in the prior case and $\boldsymbol{\text{E}}_{\boldsymbol{\theta}, \boldsymbol{Z}}[C_{\text{tot}}|w]$ in the preposterior case, are both computed with MCS. In the preposterior case, as already explained in Section \ref{subsec: Preposterior_cost}, the system state space $\boldsymbol{\theta}$ and the monitoring data space $\boldsymbol{Z}$ are jointly sampled. %This means that for each system state sample $\boldsymbol{\theta}$, one full set of accelerations over the lifetime has to be created, and subsequently processed by the SSI algorithm which identifies the modal data $\boldsymbol{Z}$. 
	For each full history of modal data, one sequential Bayesian posterior analysis has to be performed, which is a costly procedure by itself. It is clear that such an analysis can be very computationally expensive, therefore some considerations on the available computational budget, and how to distribute it, have to be made in advance. The computational cost of the VoI analysis is approximately proportional to the number of MCS samples used in the expected life-cycle cost computation and the necessary corresponding synthetic monitoring data creation, and by the computational cost of the employed method for performing the sequential Bayesian updating.

	For our investigation we assume $\hat{c}_F = 10^7$\euro{}, and for the repair cost $\hat{c}_R$ we investigate different ratios $\frac{\hat{c}_R}{\hat{c}_F}=[10^{-1},10^{-2},10^{-3}]$, and for each of those we calculate the VoI. The discount rate is taken as $r=2\%$.
	
	The solution to the stochastic life-cycle optimization problem of equation (\ref{optimization}) is performed through an exhaustive search among a large discrete set of values of the heuristic parameter (the threshold at which a repair is performed).
	
	\subsubsection{VoI results for bridge system subject to scour deterioration}
	
	For this example, we draw  1000 samples of $\boldsymbol{\theta}$, which are used in both prior and preposterior analysis. In the preposterior case, for each $\boldsymbol{\theta}$ sample we create one continuous set of identified modal data $\boldsymbol{Z}_{1:50}$. For the 1000 different sequential Bayesian analyses that have to be performed, we employ the adaptive MCMC algorithm. For the estimation of the different posterior accumulated probabilities of failure in equation (\ref{accumulated_posterior}), 2000 posterior MCMC samples are used.

	Tables \ref{table:LCC_scour} and \ref{table:LCC_SHM_scour} summarize the results of the life cycle optimization, documenting the optimal value of the heuristic parameter $w^*$, and the optimal expected total life cycle costs that correspond to $w^*$. Table \ref{table:LCC_scour} documents also the optimal time for a repair action in the prior case. This is not documented in Table \ref{table:LCC_SHM_scour}, since in the preposterior case there is not one single optimal $t_{repair}$ value, but $t_{repair}$ varies for each sample $\boldsymbol{\theta}$ and the corresponding monitoring history. 
	
	Table \ref{table:VOI_scour} documents the resulting VoI values that we obtain with the 1000 samples via equation (\ref{VOI}) for the three different cost ratios, while Table \ref{table:VPPI_scour} reports the VPPI values obtained via equation (\ref{VPPI}), related to the hypothetical case when we learn perfectly the condition of the structure from the SHM system. We also include the CV of the mean VoI, VPPI estimates, which quantifies the uncertainty in the estimates obtained via MCS. In cost ratio cases for which the optimal action in the prior case is not to perform any repair, the VoI estimate has a quite large variability. This is because the samples in the preposterior analysis that lead to a different optimal $t_{repair}$ than in the prior case are only a few, which is an indication that a larger number of Monte Carlo samples or more efficient sampling techniques (e.g. importance sampling) should be used to reduce the variance. It is important to take into account that equation (\ref{VPPI}) for computing the VPPI can easily be solved even for a very large number of MC samples, which would reduce the variability of the estimate shown here. 
	
	For all the cost ratio cases, the VoI is positive, which indicates a potential benefit of installing an SHM system on the deteriorating bridge structure. It is interesting to compare the obtained VoI values to the VPPI values. We observe that in this example the VoI from SHM extracted via Bayesian model updating is close to optimal, as it provides almost the full VPPI value.
	
	\begin{table}
		\caption{Results of preposterior Bayesian decision analysis for the scour example}
		\begin{subtable}{.5\textwidth}
			\footnotesize
			\centering
			\caption{Life-cycle optimization in the prior case.}
			\begin{tabular}{cccc}
				$\frac{\hat{c}_R}{\hat{c}_F}$&$w_0^*$&$\boldsymbol{\text{E}}[C_{\text{tot}}|w_0^*]$&$t_{repair}$\\\hline
				$10^{-1}$&$\ge2\times10^{-3}$&45395&no repair\\
				$10^{-2}$&$\ge2\times10^{-3}$&45395&no repair\\
				$10^{-3}$&2$\times10^{-5}$&5924 &year 31\\\hline
			\end{tabular}
			\label{table:LCC_scour} 
		\end{subtable}
		\begin{subtable}{.5\textwidth}
			%\vspace{-5pt}
			%\begin{table}[!ht]
			\footnotesize
			\centering
			\caption{Life-cycle optimization in the preposterior case.} 
			\begin{tabular}{cccc}
				$\frac{\hat{c}_R}{\hat{c}_F}$&$w_{mon}^*$&$\boldsymbol{\text{E}}[C_{\text{tot}}|w_{mon}^*]$\\\hline
				$10^{-1}$&2.1$\times10^{-2}$&12552\\
				$10^{-2}$&1.2$\times10^{-3}$&3125\\
				$10^{-3}$&9.9$\times10^{-5}$&1109\\\hline
			\end{tabular}
			\label{table:LCC_SHM_scour} 
		\end{subtable}
		%\vspace{-5pt}
		\begin{subtable}{.5\textwidth}
			\footnotesize
			\centering
			\caption{Value of information (VoI)}
			\begin{tabular}{ccc}
				$\frac{\hat{c}_R}{\hat{c}_F}$&VoI (CV)\\\hline
				$10^{-1}$&32843 (0.34)\\
				$10^{-2}$&42270 (0.30)\\
				$10^{-3}$&4815 (0.02)\\\hline
			\end{tabular}
			\label{table:VOI_scour} 
		\end{subtable}
		\begin{subtable}{.5\textwidth}
			\footnotesize
			\centering
			\caption{Value of partial perfect information (VPPI)}
			\begin{tabular}{ccc}
				$\frac{\hat{c}_R}{\hat{c}_F}$&VPPI (CV)\\\hline
				$10^{-1}$&35013 (0.23)\\
				$10^{-2}$&42717 (0.21)\\
				$10^{-3}$&4918 (0.02)\\\hline
			\end{tabular}
			\label{table:VPPI_scour} 
		\end{subtable}
	\end{table}
	
	\subsubsection{VoI results for bridge system subject to corrosion deterioration}
	\label{subsubsec: VoI_results_corrosion}
	
	For this second example, we draw  2000 samples of $\boldsymbol{\theta}$, which are used in both prior and preposterior analysis. In the preposterior case, for each $\boldsymbol{\theta}$ sample we create one continuous set of identified modal data $\boldsymbol{Z}_{1:50}$. For the 2000 different sequential Bayesian analyses that have to be performed, we employ the Laplace approximation method of Section \ref{subsubsec: Laplace} for the solution, which introduces an approximation error in the posterior solution, especially in the initial years, when the data set is not so large, yet is computationally much faster than an MCMC solution. For the estimation of the posterior accumulated probability of failure in equation (\ref{accumulated_posterior}), 10000 samples are drawn from the approximate multivariate Gaussian posterior distribution.
	
	The computed VoI and VPPI estimates can be seen in Table \ref{table:VOI_corrosion}. We observe that the VoI is 0 in the case when the costs have a ratio $\frac{\hat{c}_R}{\hat{c}_F}=10^{-1}$, which means that one does not get any benefit from the data obtained from the SHM system. This is related to the fact that, for this cost ratio, the optimal decision is to not perform a repair action in the lifespan of the bridge, in both the prior and all the preposterior samples, since at all time steps the cost of a repair is much larger than the the risk of failure cost. For the cost ratio $\frac{\hat{c}_R}{\hat{c}_F}=10^{-2}$, we observe that the VoI from SHM extracted via Bayesian model updating is not optimal, as it does not provide the full VPPI value, but $51\%$ of this value, while for the cost ratio $\frac{\hat{c}_R}{\hat{c}_F}=10^{-3}$ it only provides $22\%$ of the VPPI value.
	
	\begin{table}
	\caption{Results of preposterior Bayesian decision analysis for the corrosion example}
	\begin{subtable}{.5\textwidth}
		\footnotesize
		\centering
		\caption{Life-cycle optimization in the prior case.}
		\begin{tabular}{cccc}
			$\frac{\hat{c}_R}{\hat{c}_F}$&$w_0^*$&$\boldsymbol{\text{E}}[C_{\text{tot}}|w_0^*]$&$t_{repair}$\\\hline
			$10^{-1}$&$\ge2.8\times10^{-4}$&$26792$&no repair\\
			$10^{-2}$&$\ge2.8\times10^{-4}$&$26792$&no repair\\
			$10^{-3}$&$2\times10^{-5}$&$9308$&year $8$\\\hline
		\end{tabular}
		\label{table:LCC} 
	\end{subtable}
	\begin{subtable}{.5\textwidth}
		%\vspace{-5pt}
		%\begin{table}[!ht]
		\footnotesize
		\centering
		\caption{Life-cycle optimization in the preposterior case.}
		\begin{tabular}{cccc}
			$\frac{\hat{c}_R}{\hat{c}_F}$&$w_{mon}^*$&$\boldsymbol{\text{E}}[C_{\text{tot}}|w_{mon}^*]$\\\hline
			$10^{-1}$&$\ge9\times10^{-3}$&$26792$\\
			$10^{-2}$&$7.5\times10^{-4}$&$25334$\\
			$10^{-3}$&$1.83\times10^{-5}$&$9200$\\\hline
		\end{tabular}
		\label{table:LCC_SHM} 
	\end{subtable}
	%\vspace{-5pt}
	\begin{subtable}{.5\textwidth}
		\footnotesize
		\centering
		\caption{Value of information (VoI)}
		\begin{tabular}{ccc}
			$\frac{\hat{c}_R}{\hat{c}_F}$&VoI (CV)\\\hline
			$10^{-1}$&0\\
			$10^{-2}$&1458 (0.42)\\
			$10^{-3}$&108 (0.08)\\\hline
		\end{tabular}
		\label{table:VOI} 
	\end{subtable}
	\begin{subtable}{.5\textwidth}
		\footnotesize
		\centering
		\caption{Value of partial perfect information (VPPI)}
		\begin{tabular}{ccc}
			$\frac{\hat{c}_R}{\hat{c}_F}$&VPPI (CV)\\\hline
			$10^{-1}$& 132 (0.30)\\
			$10^{-2}$& 2871 (0.22)\\
			$10^{-3}$& 497 (0.05)\\\hline
		\end{tabular}
		\label{table:VPPI} 
	\end{subtable}
	\label{table:VOI_corrosion}
	\end{table}

	At a glance, one could claim by looking at the VPPI values, that for both constructed examples, the upper limit to how much value the SHM information can have for supporting the single repair decision is small, since this value does not contain the cost of the SHM system itself. In principle, if the cost of obtaining this information, i.e. the cost of the SHM system (installation, maintenance etc.), is higher than the VPPI, then no further investigation into a VoI analysis would make sense. The resulting VPPI and VoI estimates for these simplified example case studies are affected by the fact that only a single repair action case is explored.

	\subsubsection{VoI results - Sensor placement study}
	The purpose of this section is to demonstrate that the presented VoI analysis can be employed as a formal decision analysis tool for performing various parametric studies related to different choices in designing the SHM system and performing the Bayesian model updating procedure.
	
	One critical choice when designing an SHM system is the number and position of the sensors to be employed on the structure. One could employ the proposed VoI analysis to perform optimal sensor placement studies for a deteriorating structural system. Each sensor arrangement choice will result in a VoI value, and the choice which leads to the highest VoI would be the preferred one.
	
	Herein we demonstrate this with the use of the second example of the bridge system subject to corrosion deterioration at two locations. For the decision problem, we now fix the cost of failure to $\hat{c}_F = 10^7$\euro{} and the cost of repair to $\hat{c}_R = 3.5\times10^4$\euro{}. We consider the following two different arrangements of the sensors: i) 24 uniformly distributed accelerometers along the structure, ii) 12 uniformly distributed accelerometers along the structure. In both cases the VoI analysis is performed by drawing 1000 samples of $\boldsymbol{\theta}$.
	
	It becomes evident that in the case that the structure is subjected to deterioration at two different damage locations, the number of sensors and consequently the quality of the mode shape displacement or curvature information that one obtains clearly affects the BMU results and therefore leads to a notable difference in the heuristic-based life-cycle optimization and the VoI result that we obtain.
	
	\begin{table}[ht]
	\caption{Parametric study for the effect of the number of sensors on VoI result}
	\begin{subtable}{1\textwidth}
		\footnotesize
		\centering
		\caption{Life-cycle optimization in the prior case.}
	    \begin{tabular}{cc}
	        $w_0^*$&$t_{repair}^*$\\\hline
	        $6.1\times10^{-5}$&21\\\hline
	    \end{tabular}
	    \vspace{5pt}
		\label{table:LCC_prior_parametric} 
	\end{subtable}
	\begin{subtable}{.5\textwidth}
		\footnotesize
		\centering
		\caption{Life-cycle optimization in the preposterior case.}
		\begin{tabular}{cc}
			sensors&$w_{mon}^*$\\\hline
			24&$1\times10^{-4}$\\
			12&$3.2\times10^{-4}$\\\hline
		\end{tabular}
		\label{table:LCC_opt_parametric} 
	\end{subtable}
	\begin{subtable}{.5\textwidth}
		\footnotesize
		\centering
		\caption{Effect of number of sensors on the VoI}
	    \begin{tabular}{c|cccc}
		    VPPI (CV)&sensors&VoI (CV)&$\frac{\text{VoI}}{\text{VPPI}}$\\\hline
		    7681 (2.6\%)&24&4614 (5.3\%)&60\%\\
		    &12&2711 (15\%)&35\%\\\hline
	    \end{tabular}
		\label{table:VOI_parametric} 
	\end{subtable}
	\label{table:VOI_corrosion_parameter_study}
	\end{table}
	
	\section{Concluding remarks}
	This paper investigates the quantification of the VoI yielded via adoption of SHM systems acting in long-term prognostic mode for cases of deterioration. A preposterior Bayesian decision analysis for quantifying the VoI, specifically tailored for application on an employed numerical benchmark structural model is presented. The modeling of the acquired SHM data is done in a realistic way, following a state-of-the-art operational modal analysis procedure. The data is used within a Bayesian model updating framework, implemented in a sequential setting, to continuously update the uncertain structural condition, which subsequently leads to the updating of the estimate of the structural reliability. A heuristic-based solution to the simplified decision problem  is provided for finding the optimal time to perform a single repair action, which might be needed during the lifetime of the structure. We discuss specific computational aspects of a VoI calculation. The VoI analysis requires the integration over the monitoring data, which are here modeled in a realistic way, adding an extra computationally expensive layer in the analysis. In addition to the VoI solution, an upper limit to the VoI through the value of partial perfect information is also provided, related to hypothetical situations of perfect knowledge on the system condition. It should be noted that the resulting VoI estimates are affected by the fact that only a single repair action case is explored. In the present exemplary analysis, we do not take into account dependence on varying environmental effects (e.g dependence on temperature). The VoI analysis and results presented herein focus on demonstrating, for the first time, a VoI analysis on the full SHM chain, from data acquisition to utilization of a structural model for the purpose of the updating and reliability calculation. 
	
	\section*{Declaration of competing interest}
	The authors declare that they have no known competing financial interests or personal relationships that could have appeared to influence the work reported in this paper.
	
	\section*{Acknowledgments}
	The authors would like to gratefully acknowledge the support of the TUM Institute for Advanced Study through the Hans Fischer Fellowship.
	
%	\section*{References}
	\bibliography{mybibfile}
	
\end{document}